\documentclass[ amsmath,amssymb, aps]{revtex4-1}

\usepackage{soul}
\usepackage{cancel}
\usepackage{caption}
\usepackage{listings}
\usepackage{relsize}
\usepackage{graphicx}
\usepackage{amsmath}
\usepackage{filecontents}
\DeclareFontFamily{U}{mathx}{}
\DeclareFontShape{U}{mathx}{m}{n}{<-> mathx10}{}
\DeclareSymbolFont{mathx}{U}{mathx}{m}{n}
\DeclareMathAccent{\widehat}{0}{mathx}{"70}
\DeclareMathAccent{\widecheck}{0}{mathx}{"71}

\newcommand{\argmax}{\text{argmax}}
\newcommand{\x}{\mathbf{x}}
\newcommand{\n}{\overrightarrow{\mathbf{n}}}
\renewcommand{\selectlanguage}[1]{}

\newcommand{\ta}[1]{{\color{magenta} #1}}

\newcommand{\ce}[1]{\ensuremath{#1}}

\begin{document}

\title{Chemically reactive thin films: dynamics and stability}

\author{Tilman Richter}
\affiliation{
 Helmholtz Institute Erlangen-N\"urnberg for Renewable Energy,
Forschungszentrum J\"ulich, Cauerstr.~1, 91058 Erlangen, Germany
}%
\author{Paolo Malgaretti}
\affiliation{
p.malgaretti@fz-juelich.de\\
 Helmholtz Institute Erlangen-N\"urnberg for Renewable Energy,
Forschungszentrum J\"ulich, Cauerstr.~1, 91058 Erlangen, Germany
}
\author{Thomas M. Koller}
\affiliation{Department of Chemical and Biological Engineering and Department of Physics, Friedrich-Alexander-Universität Erlangen-Nürnberg, Cauerstr.~1, 91058 Erlangen}
\author{Jens Harting}
\affiliation{Helmholtz Institute Erlangen-N\"urnberg for Renewable Energy,
Forschungszentrum J\"ulich, Cauerstr.~1, 91058 Erlangen, Germany \\
Department of Chemical and Biological Engineering and Department of Physics, Friedrich-Alexander-Universität Erlangen-Nürnberg, Cauerstr.~1, 91058 Erlangen}


\begin{abstract}
\subsection*{Significance statement}
The yield of catalytic reactions within small liquid droplets or thin liquid films differs much from their bulk counterparts due to the large surface-to-volume ratio provided by the liquid interface. A liquid interface triggers a set of phenomena ranging from enhanced surface diffusion and Marangoni flows to effective surface interactions that have been shown to enhance the yield by orders of magnitude. Employing a simple theoretical model, we show that the inhomogeneous density of reactants and products due to the chemical reactions can trigger Marangoni flows that affect the spatial distribution of the catalyst. These entangled dynamics provide new means to prevent/enhance the rupture of thin films, generate non-homogeneous films and non-spherical droplets and induce non-steady states characterized by stationary and traveling waves.
\subsection*{Abstract}
Catalyst particles or complexes suspended in liquid films can trigger chemical reactions leading to inhomogeneous concentrations of reactants and products in the film. We demonstrate that the sensitivity of the liquid film's gas-liquid surface tension to these inhomogeneous concentrations strongly impacts the film stability. Using linear stability analysis, we identify 
novel scenarios in which the film can be either stabilized or destabilized by the reactions. Furthermore, we find so far unrevealed rupture mechanisms which are absent in the chemically inactive case.
The linear stability predictions are confirmed by numerical simulations, which also demonstrate that the shape of chemically active droplets can depart from the spherical cap and that unsteady states such as traveling and standing waves might appear. 
Finally, we critically discuss the relevance of our predictions by showing that the range of our selected parameters is  well accessible by typical experiments. 
\end{abstract}

\maketitle

In the last decade it has been shown that the dynamics of chemical reactions confined within microdroplets and thin films substantially differ from their bulk counterpart~\cite{wei_accelerated_2020}. For example, chemical reactions within droplets occur at rates that are up to $\simeq 10-200$ faster than their bulk counterparts~\cite{Fallah2014,Bain2016,Li2018}. 
The physical mechanisms responsible for such a speed-up have not been identified yet and indeed they may be multiple at the same time. In fact, when confined within small droplets ($\lesssim 100 \mu$m) or thin films~\cite{shi_interfacial_2022}, chemical reactions can be affected by different physical phenomena induced by the liquid-liquid interface and even by the three-phase contact line. In these conditions, the chemical reaction may be favored by local surface charges that may enhance the reaction rate~\cite{buch_water_2007,Petersen2008,ruiz-lopez_molecular_2020}. At the same time transport phenomena such as enhanced interface diffusion~\cite{mondal_enhancement_2018} and even Marangoni flows (due to the inhomogeneous surface tension induced by evaporation or varying concentration of chemicals) can lead to a significant speedup of the reactions. In particular, catalytic nanoparticles within water droplets suspended in oil have been shown to induce Marangoni flows since the reaction products act like weak surfactants~\cite{singh_interface-mediated_2020}. 

Similarly, thin films have been exploited to enhance chemical reaction yields. For example, model chemical reactions within thin films have shown a $\sim 100$ fold enhancement in the conversion efficiency~\cite{Wei2018}. At the same time, thin films and catalytically active species in the form of dissolved complexes or nanoparticles, are used in Supported Liquid Phase catalysis~\cite{zhao_developments_2006,zeng_silica-supported_2015,alsalahi_rhodium-catalyzed_2021}. Within such a framework, the liquid is deposited on a solid support, and molecular catalyst complexes are dissolved in the liquid phase~\cite{schorner_gasphase_2021, scholten_transition_2012}. 
In particular, the latter cases typically involve molecular or particulate catalysts dispersed in the liquid film. Accordingly, the reaction does not proceed uniformly throughout the film and it can lead to local variations in reactant and product concentrations, resulting in solutal Marangoni flows~\cite{masoud_reciprocal_2014, dominguez_effective_2016,jafari_kang_forward_2020,squarcini_inhomogeneous_2020,imamura_modeling_2021}.  
Recently, a novel twist has been represented by Supported Ionic Liquid Phase (SILP) catalysis~\cite{steinruck_ionic_2015,Haumann_book}.
The concept combines the attractive properties of ionic liquids (ILs), including their tunable physiochemical properties and low sorption, with the benefits of homogeneous and heterogeneous catalysis. Furthermore, the low thickness of the IL film, in order of several nanometers, reduces mass transport limitations. This allows to perform a wide range of industrially relevant reactions such as hydrogenation, dehydrogenation, or hydroformylation~\cite{hatanaka_continuous_2021, marinkovic_fifteen_2019, gu_ionic_2009, riisager_very_2005}. However, the stability of IL films within nanopores remains only partially understood~\cite{knapp_impact_2009, frosch_wetting_2023}.

All the above-mentioned examples clearly show that the overall reaction yield strongly depends on the interplay between the coupled dynamics of the catalysts and the liquid interface. 
However, so far, the study of the stability of thin liquid films has, with a few exceptions~\cite{pereira_interfacial_2007, pereira_dynamics_2007, bender_thin_2017}, focused mainly on the ``passive'' case, i.e., in the absence of chemical reactions.

Here, we propose a new model capturing the response of thin liquid films to the chemical reactions taking place therein.  
Our theoretical analysis leads to three main predictions. First, the steady-state inhomogeneous distribution of chemicals induced by the chemical reactions may stabilize an otherwise unstable thin film or break it into droplets much smaller than the passive counterpart.  Second, depending on the partial accumulation of catalyst at the liquid-gas or solid-liquid interface the chemical reactions can either enhance the rupture by several orders of magnitude or lead to a non-monotonous dependence of the rupture time on the net density of reactants. Third, eventually, the system may attain either a droplet state, whose shape may significantly depart from the spherical cap or even a film with non-uniform height.
These complex dynamics are due to the interplay between the gradients in the Laplace and disjoining pressure, induced by variations in the film height, and the Marangoni flows generated by the inhomogeneous concentrations of reactants and products. 
Our predictions show that all these effects can be controlled by tuning the effective interactions between the catalysts and the interfaces (liquid and solid) as well as by tuning the solubility of reactants and products in the thin film.
Remarkably, all these effects can be observed even for weak sensitivity of the liquid-gas surface tension to the composition of the liquid phase (less the $3\%$) which is well within the range of values that have been determined experimentally in the case of \ce{CO_2} or \ce{Ar} dissolved in ionic liquids~\cite{zhai_influence_2023}.

\begin{figure}[hh]
    \centering
    \includegraphics[width=\linewidth]{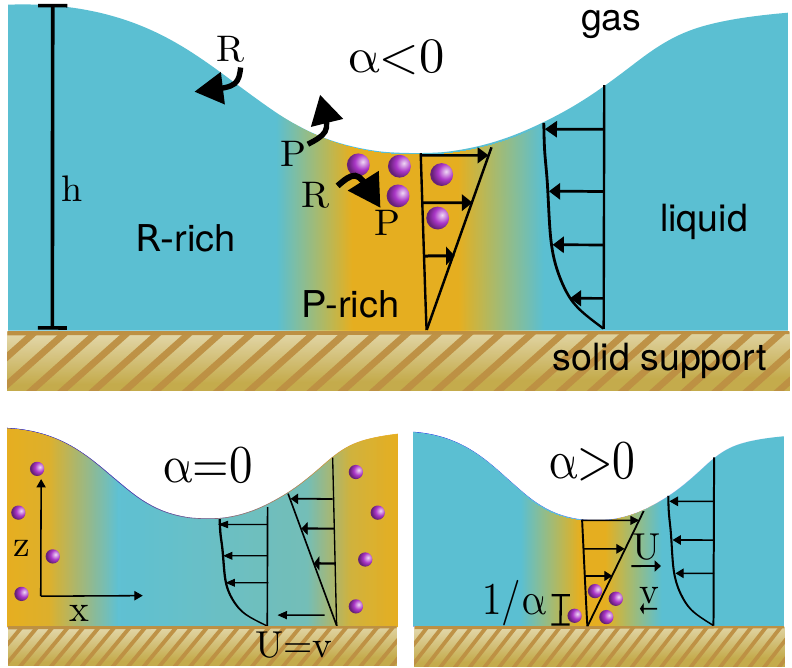}
    \caption{Sketch of the modeled setup. The blue and yellow regions are the liquid film enriched with reactant $R$ and product $P$, respectively. The purple dots represent the catalytic particles. Their distribution upon varying the parameter $\alpha$ is shown in the sub-figures. The expected contributions to the flow field are depicted with arrows. $U$ represents the flux associated to the liquid height $h$ as used in~\eqref{eq:TFE-0} and $v$ represents the advection of catalytic particles as defined in~\eqref{eq:fick-0}.}
    \label{fig:sketch} 
\end{figure}

\section*{Results}
\subsection*{Model}
The dynamics of thin reactive films are quite complex since they involve many physical processes that are all entangled. Focusing on the case of thin films or shallow droplets (i.e. with contact angles $\theta \ll \pi/2$) allows for a simple theoretical description to gain insight into the relevant physical phenomena and to capture the essentials of the underlying dynamics. As such, we can exploit the separation in length scales along the transverse (local film height) and the longitudinal directions. 
This separation allows us to assume that the liquid pressure $p$ is homogeneous along the direction normal to the solid substrate, i.e.~$p(\x,z) =p(\x)$~\cite{oron_long-scale_1997, craster_dynamics_2009}. Therefore, the problem can be simplified by integrating along the transverse direction leading to the following relevant quantities, namely, the local height of the film $h(\x)$ and the densities of catalysts, $\rho(\x)$, reactants, $\rho_R(\x)$, and products, $\rho_P(\x)$ integrated along the direction normal to the solid substrate. 
Accordingly, the local height of the film is governed by the thin film equation (TFE)~\cite{oron_long-scale_1997, craster_dynamics_2009}
\begin{align}
\label{eq:TFE-0}
    \partial_t h(\x)\!\! =\!\! \nabla\!\cdot\! \left[\dfrac{M(h(\x))h(\x)}{\mu}\nabla p(\x)\! -\! \dfrac{h^2(\x) + 2 bh(\x)}{2\mu}\nabla \gamma(\x) \right]. 
\end{align}
Here, $M(h(\x))$ is a height-dependent mobility (see Secs.~\ref{sec:model} and \ref{sec:inhomogenous_solutes} in the supporting information) and $\mu$ is the dynamic viscosity. $\gamma(\x)$ is the local liquid-gas surface tension (from now on referred to solely as surface tension) that is assumed to depend on the concentration of reactants, $\rho_R(\x)$ and products, $\rho_P(\x)$,
\begin{align}
\label{eq:surface_tension-0}
    \gamma(\x)=& \gamma_0 -\Gamma_P\dfrac{l}{h(\x)}\rho_P(\x)-\Gamma_R\dfrac{l}{h(\x)}\rho_R(\x). 
\end{align}
We remark that the pressure $p(\x)$ accounts for both the Laplace pressure as well as for the disjoining pressure (see Secs.~\ref{sec:model} and \ref{sec:inhomogenous_solutes} in the supporting information). 
For what concerns the dynamics of the reactant and products, we do not account for any effective interactions with the liquid or solid interface and, due to their fast diffusion (granted by their small size),  we assume that their density along the transverse direction is homogeneous. To account for the impact of the varying height of the film on their dynamics we follow the standard Fick-Jacobs scheme~\cite{zwanzig_diffusion_1992,Reguera2001,Malgaretti2013,Malgaretti2023} according to which their integrated densities, $\rho_{P,R}(\x)$ obey the following advection-diffusion equations~\cite{thiele_note_nodate, thiele_gradient_2013, xu_variational_2015}:
\begin{align}
    & \begin{split}\label{eq:reactant-0}
    \partial_t \rho_R =& \nabla \cdot \left( D_R \nabla \rho_R - v_{P} \rho_R + D_R \beta \rho_R \nabla\mathcal{F}_{0}\right) \\
    &- \frac{\omega \rho\rho_R}{h}  - \sigma_R^\uparrow \dfrac{l}{h}\rho_R + \rho_{R,Res}\sigma_R^\downarrow.
    \end{split}\\
    & \begin{split}\label{eq:product-0}
    \partial_t \rho_P =& \nabla \cdot \left( D_P \nabla \rho_P - v_{R} \rho_P + D_P \beta \rho_P \nabla\mathcal{F}_{0}\right) \\
    &+ \frac{\omega \rho\rho_R}{h}  - \sigma_P^\uparrow \dfrac{l}{h}\rho_P.
    \end{split}
\end{align}
Here, $D_{R,P}$ are the effective diffusion coefficients of the reactants/products (estimated from the Stokes-Einstein relation), $\sigma_R^\uparrow$ denotes the rate at which the reactant is transferred from the liquid film to the surrounding, $\rho_{R,Res}$ represents the concentration of reactants in the gas reservoir, and $\sigma_R^\downarrow$ stands for the rate at which the reactant enters the liquid film. 
The parameter $l$ is a molecular length scale that we introduce to account for the fact that only products located in the upmost molecular liquid layer, of thickness $l$, are prone to dissolve into the surrounding gas. 
The chemical reaction occurs with rate $\omega$ whenever the reactants are in contact with the catalyst described by its density profile $\rho$. 
The effective velocity $v_0$ accounts for the integrated advective flow (see Secs.~\ref{sec:model} and \ref{sec:inhomogenous_solutes} in the supporting information) whereas the effective potential gradient, $\nabla \mathcal{F}_0=-k_BT\dfrac{\nabla h}{h}$ (calculated in the supplemental material) accounts for the entropic force on the chemical species induced by the inhomogeneous film height. 

For catalysts, it is well known that, within thin films, they may accumulate at the liquid~\cite{steinruck_surface_2011} or solid~\cite{shylesh_situ_2013} interface, or they may be homogeneously distributed along the film height. Accordingly, we model their effective interactions with the liquid interfaces as
\begin{align}
\label{eq:potential-0}
    \beta\xi(z,h(\x)) = \begin{cases}
    \alpha z & \text{for }\alpha> 0\\
    0 & \text{for }\alpha = 0\\
    \alpha (z-h(\x))& \text{for }\alpha< 0.
    \end{cases}
\end{align}
Here $\beta:=1/k_BT$ with Boltzmann-constant $k_B$ and absolute temperature $T$.
This potential reflects different scenarios for the distribution of the catalyst within the liquid film along the vertical z-direction in the form of an exponential behavior with a characteristic length scale of $1/\alpha$. For $\alpha>0$ the catalyst is distributed towards the solid substrate while for $\alpha<0$ it is attracted to the liquid-vapor interface. For $\alpha = 0$, the catalyst is homogeneously distributed along the vertical direction. See Fig.~\ref{fig:sketch} for an illustration of the three cases.
As for the chemical species, in order to integrate along the film thickness we follow the Fick-Jakobs~\cite{zwanzig_diffusion_1992,Reguera2001,Malgaretti2013} approximation which, within the length scale separation that we are considering, assumes that the density factorizes in 
\begin{subequations}
\label{eq:factorise_rho}
    \begin{align}
    \tilde{\rho}(\x,z) &= \rho(\x) \tilde{\rho}_z(z,h(\x)),\\     
    \tilde{\rho}_z (z,h(\x)) &=\dfrac{e^{-\beta \xi(z,h(\x))}}{\int_0^{h(\x)} e^{-\beta\xi(z,h(\x))}}.
\end{align}
\end{subequations}
Accordingly, integrating the advection-diffusion equation along the transverse direction leads to
\begin{align}
    \label{eq:fick-0}
    \partial_t \rho(\x) &= \nabla \cdot\left( D\nabla \rho(\x) - v_\alpha(\x)\rho(\x) + D\beta \rho(\x) \nabla \mathcal{F}(\x)\right),
\end{align}
where $D$ is the effective diffusion coefficient of the catalyst in the liquid film (estimated via the Stokes-Einstein relation). The effective potential in \eqref{eq:fick-0} 
is defined as~\cite{zwanzig_diffusion_1992,Reguera2001,Malgaretti2013}
\begin{align}
\label{eq:potential_F}
    \mathcal{F}(\x)= - \dfrac{1}{\beta} \ln \left( \int_0^{h(\x)} e^{-\beta \xi(z,h(\x))} dz \right) 
\end{align}
that indeed is the local equilibrium free energy of the catalyst. The energy $\xi$ in \eqref{eq:potential_F} defines the vertical distribution of the catalyst and is given in Sec.~\ref{sec:model} of the supporting information. 
The effective velocity $v_\alpha(\x)$ in \eqref{eq:fick-0} is related to the integrated flow of catalyst across the transverse direction and it is sensitive to the effective interactions of the catalysts with the interfaces (see supporting information).

\subsection*{Stability}

 \begin{figure}
     \centering
     \includegraphics[width=\linewidth]{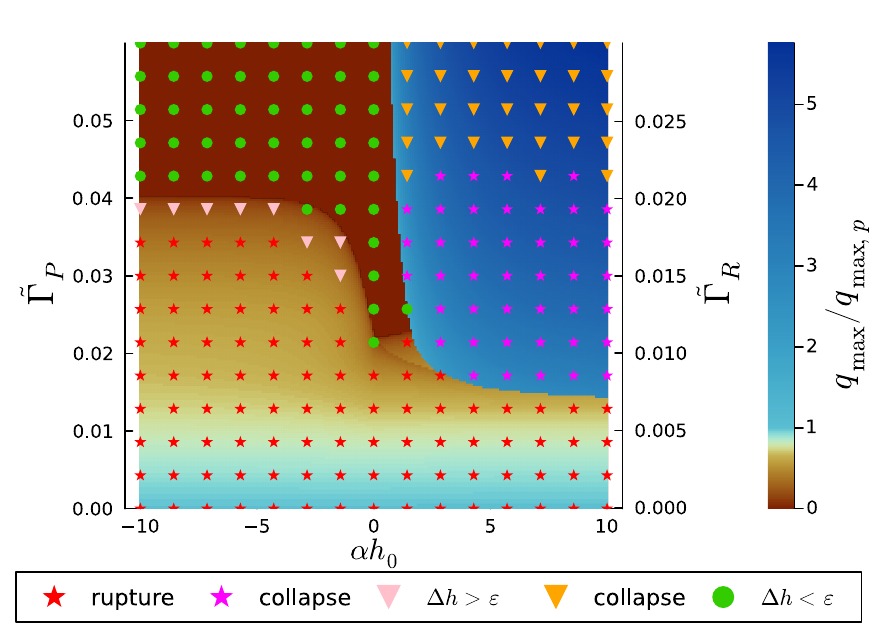}
     \caption{Representation of the fastest growing wave mode/vector of the studied thin film system in terms of the percentage changes in the vapor-liquid surface tension due to the presence of the reaction product, $\tilde{\Gamma}_P:=\Gamma_P\rho_{0,P}l/(h_0\gamma_0)$, and reactant $\tilde{\Gamma}_R:=\Gamma_R\rho_{0,R}l/(h_0\gamma_0)$, for varying concentration distribution of the catalyst particles in vertical direction, $\alpha h_0$. The colormap in the background indicates the wavenumber of the fastest growing mode of the system obtained from \eqref{eq:linear_soloution}. The dots in the foreground give the eventual state of a numerical solution of Eqs.~(\ref{eq:TFE-0}-\ref{eq:product-0}) and \eqref{eq:fick-0} after $10^{9}$ iterative time steps. Stars represent simulation results where the initially closed film forms droplets. Triangles represent still closed films with excitation $\delta h$ higher than a certain threshold $\epsilon$, while green dots correspond to an excitation $\delta h$ below that threshold $\epsilon$. The colors differentiate between simulations that are driven by the accumulation of catalyst (magenta and orange), called the collapse regime, or by the passive dynamics of the thin film (red and pink). The parameters and properties are the same as listed in the Material and Methods section. The symbols and colors used for the numerical simulation agree with the ones used in Fig.~\ref{fig:simulations}.}
     \label{fig:stability_portrait}
 \end{figure}

Thin liquid films of a homogeneous height $h_0<h_{crit}$ are known to be unstable, i.e. small fluctuations will lead to film rupture and to the formation of droplets.
This instability is mainly due to attractive van der Waals forces between the surrounding gas atmosphere and the solid substrate~\cite{vrij_possible_1966,doi_soft_2013} which in the current approach are captured by the disjoining pressure $\gamma_0 (1-\cos\theta)f(h)$.
By standard linear stability analysis about the homogeneous state $h(\x)=h_0$ it is possible to predict the wavenumber of the fastest growing mode of the TFE \eqref{eq:TFE-0} for $\gamma=\gamma_0 \equiv const.$ to be~\cite{rauscher_spinodal_2008, fetzer_thermal_2007, zitz_lattice_2021,PSH22}
 \begin{align}
 \label{eq:q_max_p}
    q_{\max,p}=\sqrt{\dfrac{1}{2}f'(h_0)(1-\cos \theta_0)}. 
\end{align}
Accordingly, even for non-zero contact angles, $\theta$, a thin liquid film with a uniform height $h_0$ is  meta-stable when $f'(h_0) \rightarrow 0$ i.e., when the potential $f$ in the disjoining pressure (see Eq.~\eqref{eq:f_pressure}
in the supporting information) becomes negligible. 
For liquids like water, this happens when $h_0$ is tens of nanometers~\cite{doi_soft_2013}. In the ongoing discussion, we assume that the film is thin enough, such that $f'(h_0)>0$, as this is the relevant scenario for many technological applications including SILP catalysis~\cite{riisagera_supported_2006}.

Employing a linear stability analysis (see Materials and Methods), the relevant dimensionless numbers can be identified. These are the relative changes in surface tension caused by the presence of the product, $\tilde{\Gamma}_P$, and by the reactants $\tilde{\Gamma}_R$, as well as the height of the film in units of the ``screening length'' $\alpha h_0$. For $\Gamma_P=\Gamma_R=0$ the chemical reactions have no impact on the film stability.
As a reference case, we consider a water film of an initial height $h_0=10 \text{nm}$. The contact angle is set to $\theta_0=\pi/9=20^\circ$. In the following, if not mentioned otherwise, 
we assume that the sorption rates of the chemicals are  equal to the reaction rate at the catalyst. We will discuss the implications of different choices later on. All parameters used are reported in the Materials and Methods section.

Fig.~\ref{fig:stability_portrait} shows the stability portrait of a film that is initially prepared with a homogeneous thickness $h_0$, and catalyst $\rho_0$, reactants $\rho_{R,0}$, and product $\rho_{P,0}$ densities for different transverse distribution of catalysts ($\alpha h_0$, x-axis) and surface tension sensitivity to chemicals (y-axis) where for convenience, we have set $\tilde{\Gamma}_P=2\tilde{\Gamma}_R$. On the top of the numerical solutions (points), Fig.~\ref{fig:stability_portrait} reports also the wavenumber of the fastest growing mode $q_{max}$ normalized by the value $q_{\max,p}$ given by \eqref{eq:q_max_p} obtained in the absence of chemicals or when chemicals do not affect the surface tension $\Gamma_P=\Gamma_R=0$ (color code). 
In particular, five different scenarios can be identified: (1) the film is stabilized by the chemical reactions (green dots and dark brown background), (2) the film breaks as "passive" meaning in a very similar as if there was not chemical reaction happening (red stars and light brown/cyan), (3) a cross-over region between the former two where the simulation time and system size are insufficient to observe film rupture (pink triangles),  (4) the film breaks with much shorter wavelengths as compared to the "passive" case (magenta stars and blue background) and (5) short wavelength excitement but no film-rupture (orange triangles and blue background).

\begin{figure}
    \centering
    \includegraphics[width=\linewidth]{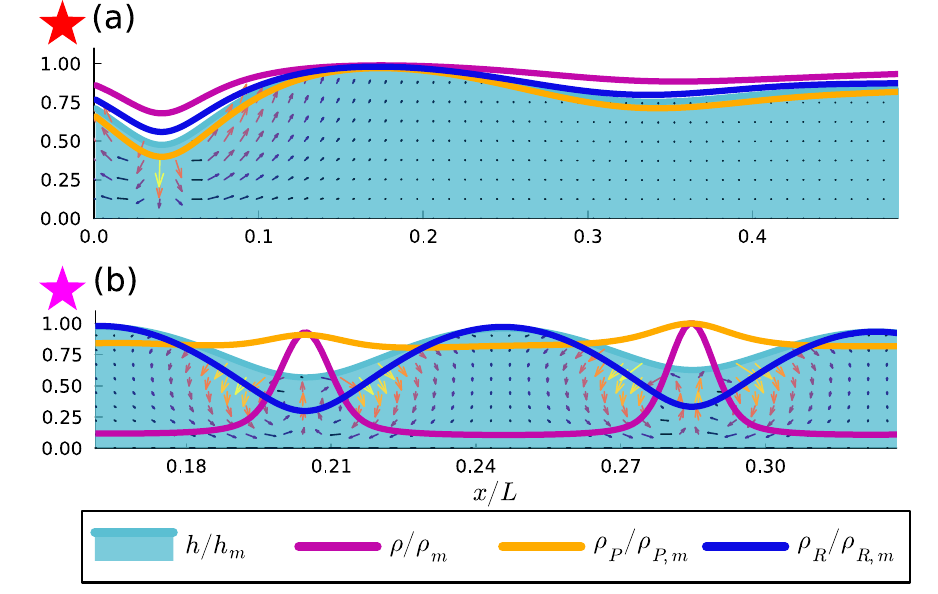}
    \caption{The height $h$, catalyst density $\rho$, product density $\rho_P$ and reactant density $\rho_R$ normalized by their respective maxima $h_m$, $\rho_m$, $\rho_{P,m}$ and $\rho_{R,m}$ for a snapshot shortly before film-rupture of the spinodal regime (a) and the collapse regime (b).  The flow field is interpolated from surface tension and pressure gradient $\nabla\gamma, \nabla p$ assuming incompressibility $\partial_x u + \partial_z v=0$. Arrows indicate the direction of the flow and their color the magnitude of the local velocity. The lengths of the arrows do not scale the same among the two insets, i.e.~the velocity field in (a) is exaggerated 10 times more than in (b). The simulations reported here correspond to the magenta and red stars in Fig.~\ref{fig:simulations}.}
    \label{fig:flow_field}
\end{figure}

The linear stability analysis indicates that when the catalyst particles are preferentially at the vapor-liquid interface ($\alpha h_0 \leq 0$), the thin film is stable (green dots) when 
\begin{align}
\tilde{\Gamma}_P+ \tilde{\Gamma}_R>\frac{\Gamma_{c}(\rho_{0,P}+\rho_{0,R})l}{(h_0\gamma_0)},    
\label{eq:cond-0}
\end{align}
where $\Gamma_{c}(\rho_{0,P}+\rho_{0,R})l/(h_0\gamma_0) =1-\cos\theta_0$ is the threshold value that is determined by the reduction in the surface tension due to both reactants and products and closely resembles the amount of surfactants needed to stabilize an otherwise unstable film~\cite{thiele_note_nodate}.  When \eqref{eq:cond-0} is not fulfilled (red stars) 
the thin film is unstable and the fastest growing mode is characterized by $q_{\max,p}>q_{\max}>0$. Now, in the same way as in the passive case, the homogeneous system is not stable which leads to the rupture of the liquid film. The resulting droplets in this case are larger than in the passive case. Fig.~\ref{fig:flow_field} (a) shows an example of a simulated flow field in this regime leading to film rupture. 

When the catalytic particles are homogeneously distributed ($\alpha\approx 0$), significantly less catalytic activity is required to stabilize the system, as can be seen in Fig.~\ref{fig:stability_portrait}. This is because, for a homogeneous distribution, both advection and diffusion enforce a strong correlation between the film height and the catalyst density, i.e.~$\rho \propto h$. For advection this can be seen from the effective mobilities of the catalyst, $M_p,M_\gamma$ (see Eq.~\eqref{eq:mob-cat}
in the supporting information), and film height, $M,(h+2b)/2$ (see \eqref{eq:TFE-0}). The mobilities become pairwise equal, $M_p=M$ and $M_\gamma= (h+2b)/2$ and this causes the catalytic particles to move along with the liquid. Concerning diffusion, for $\alpha=0$, $\nabla \mathcal{F} = - k_BT \nabla h / h$ and thus, at equilibrium, $\rho$ is  proportional to $h$. This proportionality stemming from both diffusion and advection, $\rho\propto h$, leads to enhanced consumption of reactant (and catalysis of products) proportional to $h$. In this way, variations in film height are suppressed by locally decreasing the surface tension due to the chemical reactions. Such behavior is already captured by the LSA as terms with the prefactors $M_p$ or $M_\gamma$ have a counterpart with prefactor $M$ or $(h+2b)/2$ canceling each other out (see supporting information, Sec.~\ref{sec:inhomogenous_solutes}).

Finally, for $\alpha> 0$, the catalysts are accumulated at the solid-liquid interface. Considering small surface tension effects $\tilde{\Gamma}_P, \tilde{\Gamma}_R\ll 1$ we observe that the wavenumber of the fastest growing mode is reduced as compared to the passive case wavenumber $q_{\max},q_{\max,p}$ (cyan to brown background in Fig.~\ref{fig:stability_portrait}). This stems from the reduction in surface tension by the presence of the chemicals.
In contrast, for larger values of 
$\tilde{\Gamma}_P, \tilde{\Gamma}_R$ 
the wavenumber of the fastest growing mode becomes larger than the passive wave number, $q_{\max}>q_{\max,p}$ (dark blue background in Fig.~\ref{fig:stability_portrait}), indicating the excitation of short wavelengths.  The mechanism underlying this phenomenon can be explained as follows: Under the assumptions that $\alpha\gg 0$ and $\Gamma_P,\Gamma_R\gg 0$ (i.e.~the catalytic particles are highly accumulated near the solid substrate, and the surface tension is lower in the product-rich phase than in the reactant rich phase, see Fig.~\ref{fig:sketch}), a high local concentration of catalyst at a point in space $\x$ locally reduces surface tension. This reduction leads to the creation of surface stress, inducing a Marangoni shear flow $\nabla\gamma (h+2b)/(2\mu)$ away from the area of high catalyst concentration. Consequently, the film height at that location decreases, forming a corrugated vapor-liquid interface with a radius of curvature $R$. This corrugation generates a Laplace pressure $\gamma/R$, resulting in a parabolic-shaped flow $\nabla p M(h)/\mu$ towards the area of high catalyst concentration. These two contributions, along with the disjoining pressure, sum up to form the height-averaged velocity $U$ (compare with Eq.~\eqref{eq:thin_sheet} in the supporting information
). According to \eqref{eq:fick-0}, the catalyst particles experience an effective flow with the velocity $v=\nabla \gamma M_\gamma /\mu+ \nabla p M_p /\mu $, while the film height $h$ is affected by the flow with the velocity $U$. Because the Marangoni 
flow and the pressure gradient-driven flow have distinct flow profiles (as depicted in Fig.~\ref{fig:sketch}), it is possible that $U$ and $v$ have different directions. Consequently, the film corrugation persists, since the flow related to $U$ transports liquid away from the regions of high concentration, while the pressure-driven flow related to $v$ accumulates more catalyst at that location. This effect is illustrated by the the flow field of an exemplary simulation in this regime as shown in the lower inset of Fig.~\ref{fig:flow_field}. The presence of such rollers or vortices in thin liquid films in the presence of Marangoni flows has been observed before for thermal gradients~\cite{de_gennes_capillarity_2004}.
Similar phenomena have been observed in the context of the Keller-Segel model~\cite{herrero_chemotactic_nodate}, in whose spirit we refer to this effect as a ``collapse.''

\begin{figure}
    \centering
    \includegraphics[width=\linewidth]{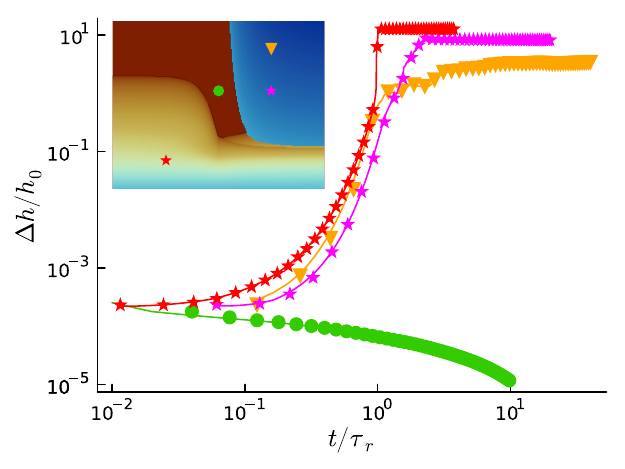}
    \caption{
    The maximum height $\Delta h /h_0=(\max h-\min h)/h_0$ difference in time plotted for four exemplary simulations, showing the different possible behaviors of the system. All simulations are started in the perturbed homogeneous state $h=h_0+ \epsilon \mathcal{N}_0^1, \rho=\rho_0, \rho_A=\rho_{0,A}, \rho_B=\rho_{0,B}$. 
    In the inset, we show the position of those simulations in the stability diagram Fig.~\ref{fig:stability_portrait} as colored symbols. For every simulation, a video is available in the supporting information. In the simulation shown with green dots, $\tau_r=\infty$ thus we choose an arbitrary value instead, to show its evolution in the same plot as the other three. For the axis labels of the inset see Fig.~\ref{fig:stability_portrait}. The symbols and colors used here match the ones in Fig.~\ref{fig:stability_portrait}.}
    \label{fig:simulations}
\end{figure}

 Upon raising $\alpha$  above $\alpha =0$ there is a jump discontinuity from $q_{\max}=0$ to $q_{\max}\gg q_{\max,p}$ (Fig.~\ref{fig:stability_portrait}, brown to blue background). Also when raising $\tilde{\Gamma}_P+\tilde{\Gamma}_R$  a jump discontinuity from smaller $q_{\max}$ to higher $q_{\max}$ can be observed at roughly $\tilde{\Gamma}_P+\tilde{\Gamma}_R\approx0.015+0.0075$ (Fig.~\ref{fig:stability_portrait}, light brown background to blue background). Analyzing the eigenvalues of the linearized system (refer to Fig.~S5 in the supplemental material), we observe that the two physical effects connected with the disjoining pressure and the collapse, are associated with local maxima in dispersion relation $\eta_{m}(q)$ (see Materials and Methods). The observed jump in the fastest-growing modes can be explained by the global maximum switching from one local maximum to the other.

Finally, for $\tilde{\Gamma}_P+ \tilde{\Gamma}_R=0$, where no effect of the product and reactant concentration on the surface tensions is given, we find $q_{\max}=q_{\max,p}$, as expected.


\subsection*{Rupture}

After reporting on the onset of the instability, we now turn to the dynamics leading to the breakage of the film. 
Fig.~\ref{fig:simulations} shows the time evolution of the height difference $\Delta h=\max h -\min h$  for some representative cases. 
When the thin film is stabilized by the chemical reactions (green points) or the chemical reactions are almost irrelevant (red stars) the dynamics of the system towards the steady state are relatively simple (see Fig.~\ref{fig:simulations}). This is confirmed by the fact that the time evolution obtained by the numerical simulations agrees well with the linear solution \eqref{eq:linear_soloution}: the amplitude of the dominant wavelength ($1/q_{\max}$) approaches exponentially its steady-state value. Only as the system approaches the steady state, the linear solutions \eqref{eq:linear_soloution} become less accurate.  
At variance, for red and pink stars, the chemical reactions are more relevant and indeed the time evolution cannot be captured fully by the linearized equations as for the simpler cases. 

Initially, the amplitudes grow as predicted by \eqref{eq:linear_soloution}, constructing the wavemode with wavenumber $q_{\max}$. Yet, at a certain point in time, this growth halts, and the full numerical solution no longer aligns with \eqref{eq:linear_soloution}. In such cases, the rupture of the film cannot occur at such small wavelengths ($1/q_{\max}\ll 1/q_{\max,p}$). Instead, the system's maxima merge, giving rise to larger dominant wavelengths that are not predicted by the LSA. An example of such a simulation can be observed in Fig.~S2 as well as in the video \lstinline|sim_3.mp4| in the supplemental material.

For  $\tilde{\Gamma}_P+\tilde{\Gamma}_R > \Gamma_c(\rho_{0,P}+\rho_{0,R}) l/(h_0\gamma_0)$, (orange down-triangle) we also observe the initial growth of the dominant wavelength $1/q_{\max}$, that is followed by a coarsening unifying the maxima of the concentration of catalyst $\rho$. In this situation, the surface tension $\gamma$ is very low and Young's law of wetting predicts a contact angle of $\theta=0$. Thus no film rupture occurs even though certain wavelengths of the system are excited by the dynamic interaction of chemical reactions and surface tension.



To quantitatively capture the dependence of the rupture dynamics on the chemical reactions we define the rupture time $\tau_r$ as the earliest time when the film height somewhere becomes as small as the precursor height $h^*$,
\begin{align}
\label{eq:rupture_time_def}
    h(\x,\tau_r) = h^*,
\end{align}
which captures the time it takes the homogeneously initialized thin film to undergo rupture and evolve into droplets. A good estimate on $\tau_r$ is provided by the linearized solution \eqref{eq:linear_soloution}
\begin{align}
\label{eq:rupture_time_0}
    \tau_r = \frac{1}{\eta_m(q_{\max})}\ln \frac{h_0}{\epsilon}.
\end{align}

\begin{figure}
    \centering
    \includegraphics[width=\linewidth]{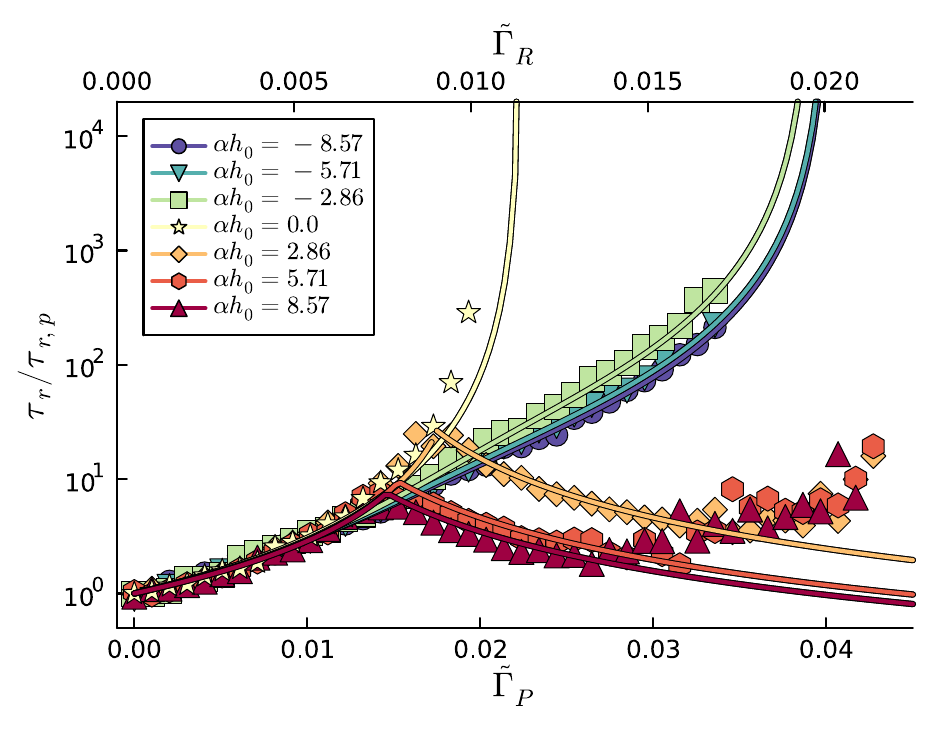}
    \caption{The rupture time of a liquid film containing catalytic particles normalized by the passive case rupture time $\tau_{r,p}$ of a thin liquid film is plotted as a function of the percentage change of surface tension at a mean concentration of product $\tilde{\Gamma}_P$ or reactant $\tilde{\Gamma}_R$ (on the lower or upper $\x$-axis, respectively) and vertical distribution of catalyst $\alpha h_0$ (color code). The solid lines refer to the rupture time expected from the linear solution by \eqref{eq:rupture_time}, while stars are obtained from a numerical solution of Eqs.~($\ref{eq:TFE-0} $-$\ref{eq:product-0}$) and \eqref{eq:fick-0} starting from a homogeneous initial condition with a small initial random perturbation. All relevant parameters are chosen as listed in the Material and Methods section.}
    \label{fig:rupture_time}
\end{figure}

It turns out that this is well described by the characteristic time of the system $1/\max(\eta_{m}(q))$ where $\eta_{m}$ is the largest eigenvalue of the dispersion relation $A(q)$ (see Sec.~\ref{sec:inhomogenous_solutes} of the supporting information).
We write the rupture time in a normalized form by referencing the rupture time $\tau_{r,p}$ for the passive case $\Gamma_P=\Gamma_R =0$, i.e. 
\begin{align}
\label{eq:rupture_time}
    \dfrac{\tau_r}{\tau_{r,p}}\approx \dfrac{\max\eta_{m,p}(q)}{\max\eta_{m}(q)},
\end{align}
where $\eta_{m,P}$ is the largest eigenvalue of the dispersion relation for the pure liquid. 
Fig.~\ref{fig:rupture_time} shows the estimates given by \eqref{eq:rupture_time} (solid lines) as well as the rupture time obtained from numerical simulations (symbols). The parameters are chosen as shown in the Material and Methods section. At $t=0$ all the fields are initialized homogeneously plus a small random perturbation in the liquid height ($h= h_0+\epsilon\mathcal{N}, \rho = \rho_0, \rho{P}=\rho_{0,P}, \rho_R= \rho_{0,R}$) for a random variable $\mathcal{N}$. 

We observe in Fig.~\ref{fig:rupture_time} that for  $\tilde{\Gamma}_P+\tilde{\Gamma}_R \lesssim 0.0225$ all data collapses onto a master curve. For $\Gamma_P=\Gamma_R=0$, we retrieve the passive case and thus $\tau_r/\tau_{r,p}=1$.  Considering $\tilde{\Gamma}_P+\tilde{\Gamma}_R \gtrsim 0.0225$, the situation becomes more complex. For $\alpha < 0$, in general the rupture time rises with $\tilde{\Gamma}_P+\tilde{\Gamma}_R$ until it reaches a threshold value $\Gamma_c(\rho_{0,P}+ \rho_{0,R}) l /(h_0 \gamma_0)$  where $\tau_r\to \infty$ which corresponds to the stabilization of the homogeneous film.  For $\alpha \ll 0$, that threshold value is as discussed before $\Gamma_c (\rho_{0,P}+\rho_{0,R}) l /(h_0 \gamma_0)=1-\cos\theta_0$. 
For $\alpha\approx 0$, the rupture time $\tau_r$ diverges for a smaller value of $\tilde{\Gamma}_P+\tilde{\Gamma}_R$. In the case of $\alpha > 0$, the rupture time initially increases with growing $\tilde{\Gamma}_P+\tilde{\Gamma}_R$ until it begins to decrease once more. This non-monotonous behavior corresponds to the transition from the unstable film (red stars in Fig.~\ref{fig:stability_portrait}) to the collapse (magenta stars  in Fig.~\ref{fig:stability_portrait}) regime, where the largest eigenvalue of the dispersion relation $\eta_{m}$ (see Materials and Methods) starts growing as $\tilde{\Gamma}_P+\tilde{\Gamma}_R$ increases. Despite the jump discontinuity in the transition of $q_{\max}$ to the collapse regime, $\max (\eta_{m})$ and $\tau_r$ exhibit continuity at the transition point for larger values of $\alpha$. This can be understood by considering the eigenvalue $\eta_{m}$ as shown in Fig.~S5. 
Concerning the reliability of the predictions of the LSA, Fig.~\ref{fig:rupture_time} shows good agreement of the rupture time predicted by the linearized solution \eqref{eq:rupture_time} (lines) with the simulations (symbols). We note that such an agreement is remarkable since the rupture occurs outside the linear regime where \eqref{eq:linear_soloution} used for the calculations in the LSA framework is not valid anymore. However, the LSA, via the ansatz in \eqref{eq:linear_soloution}, still qualitatively captures the timescale reasonably well. Quantitatively, the rupture time of the simulations does not match \eqref{eq:rupture_time} anymore for $\tilde{\Gamma}_P+\tilde{\Gamma}_R \gtrsim 0.045$ and $\alpha>0$. This is due to the wavenumber of the fastest growing mode becoming large here, i.e. the dominating wavelength is very small. The dominant wavelength builds up, but the thin film cannot break at such small wavelengths. Larger wavelengths have to grow for film rupture to occur. This process happens outside of the linear regime and is well illustrated by the data set represented by magenta stars in Fig.~\ref{fig:simulations} as well as in Fig.~S2 and the corresponding video \lstinline|sim_3.mp4| in the supplemental material. Film rupture in this case is not associated to the fastest growing wavenumber but to a smaller one.

\subsection*{Droplets}

\begin{figure*}
    \centering
    \includegraphics[width=1.0\linewidth]{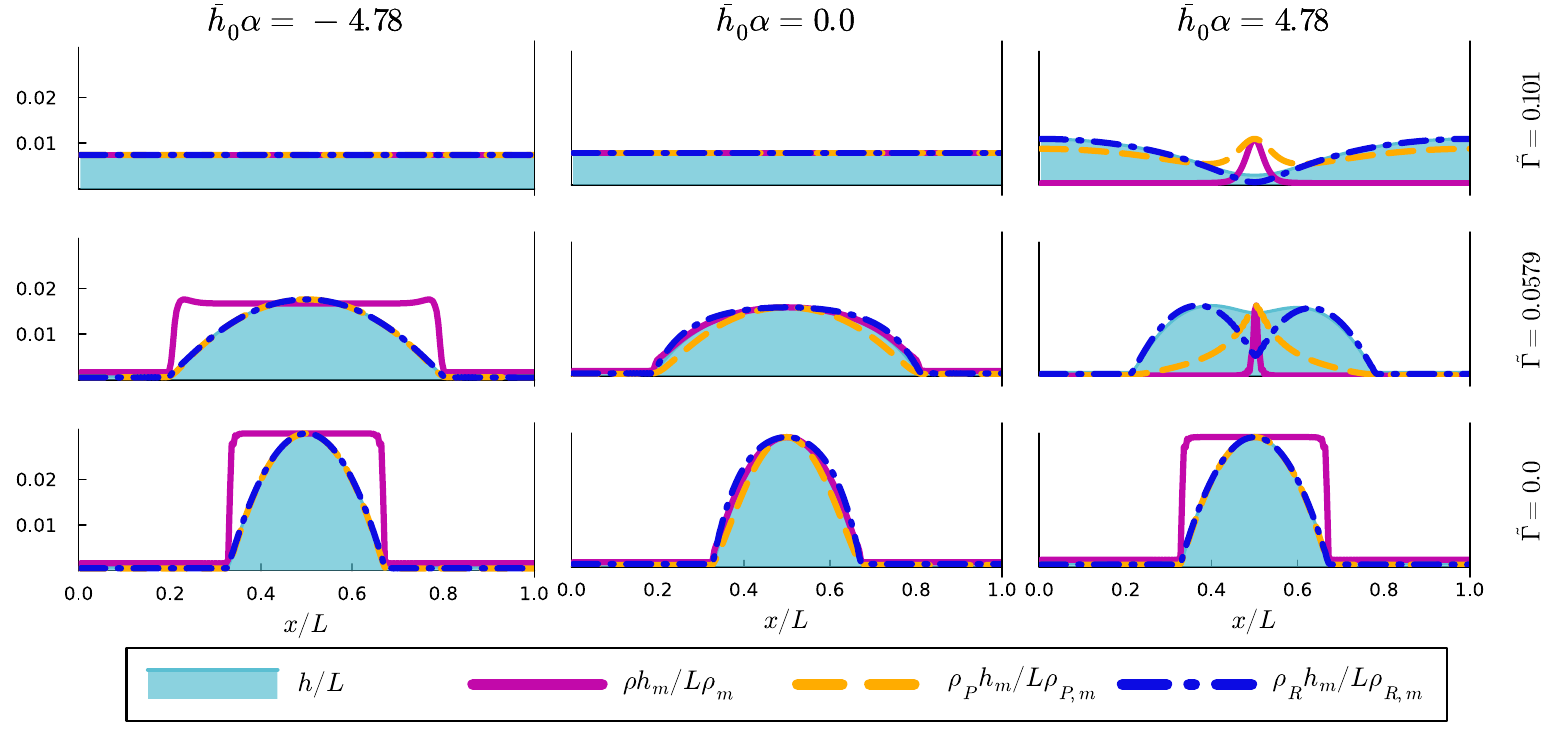}
    \caption{Steady state of a droplet with dissolved active catalyst particles for different parameters $\bar{h}\alpha$ and $\tilde{\Gamma}:=(\Gamma_P\bar{\rho}_{0,P}+\Gamma_R\bar{\rho}_{0,R})l/(\gamma_0 h_0)$, representing the assumed vertical distribution of catalyst over a length scale $1/\alpha$ and the percentage change of the surface tension according to \eqref{eq:surface_tension-0}.  As initial conditions we choose $h(0)$ a spherical cap with contact angle $\theta_0$, $\rho(0)\equiv$ for $h(0)>h^*$ and $\rho(0)\equiv 0$ for $h(0)\leq h^*$. The density of reactant is initially chosen as $\rho_{R}(0)=\rho_{R,Res}\sigma_R^\downarrow h(0)/(\omega \rho(0)+\sigma_B^\uparrow l)$ and the product density is $\rho_P=\omega \rho(0) \rho_R(0)/(l\sigma_P^\uparrow)$.  Negative $\alpha$ corresponds to the catalyst being enriched at the liquid-vapor interface, $\alpha=0$ to a vertically homogeneous distribution of catalyst, and positive $\alpha$ indicates that catalyst is enriched close to the solid substrate (see Fig.~\ref{fig:sketch}). The concentration fields $\rho$, $\rho_P$, and $\rho_R$ are divided by their respective maxima $\rho_m, \rho_{P,m}, \rho_{R,m}$ and multiplied by the maximum height of the system $h_m$ to visualize all fields in a single plot. The lengths are normalized by the system size $L$.  We run the simulations until no difference between two consecutive dumps until full convergence. The aspect ratio in all insets is fixed to $10$.}
    \label{fig:drop_sims}
\end{figure*}

Finally, once the film is broken 
droplets are formed. 
To reduce the required computing time when  investigating the steady state of droplets containing active catalytic particles, we start our simulations from a spherical cap, i.e. a portion of a sphere cut off by a plane\footnote{The spherical cap is the expected equilibrium shape of a liquid droplet with a radius smaller than the capillary length $r_c=\sqrt{\gamma_0/(\rho_l g)}$, where $g$ is the gravitational constant~\cite{doi_soft_2013}.}. The contact angle of this cap on the solid substrate is initially chosen as $\theta_0$, and the initial densities $\rho$, $\rho_P$, and $\rho_R$ are chosen constant inside the droplet i.e. $\rho(\x,0),\rho_P(\x,0),\rho_R(\x,0)=C$ where $h(\x,0)>h^*$ and $\rho(\x,0),\rho_P(\x,0),\rho_R(\x,0)=0$ for $h(\x,0)=0$. 
Fig.~\ref{fig:drop_sims} shows the droplet shape and the density of catalyst as well as of the chemicals for the three different scenarios, captured by  $\alpha$, and for the same parameter values used in Fig.~\ref{fig:stability_portrait}.

For $\alpha \ll 0$, catalyst particles accumulate at the liquid interface and hence they behave as an effective ($d-1$)-dimensional ideal gas, spread homogeneously across the droplet. In contrast, the products and reactants, due to their large diffusion coefficients as compared to the diffusion coefficient of the catalyst and vertically homogeneous concentration distribution, are homogeneously distributed across the film. 
For larger values of $\Gamma_R,\Gamma_P$ this leads to an overall decrease of the liquid-vapor surface tension and hence eventually to a perfectly wetting film (top row of the left column in Fig.~\ref{fig:drop_sims}). For smaller values of $\Gamma_R,\Gamma_P$ the decrease in $\gamma$ is not sufficient and the droplet persists (central and bottom rows of the left column in Fig.~\ref{fig:drop_sims}). 
Interestingly, in these cases, the liquid-vapor surface tension is not homogeneous anymore and it induces a weak Marangoni flow towards the contact lines that leads to a slight accumulation of catalyst at the three-phase contact line. 

For a homogeneous catalyst distribution, $\alpha=0$, at steady state all the densities are proportional to the droplet height $h$. Accordingly, the surface tension is homogeneous everywhere (center column of Fig.~\ref{fig:drop_sims}). As before, upon raising  values of the strength of the surface tension effects $\Gamma_P, \Gamma_R$ the surface tension lowers leading to spreading of the droplet (center) and eventually to a perfectly wetting liquid film (top). 

For $\alpha \gg 0$, the catalytic particles accumulate at the solid substrate. In this case, the eventual distribution of catalyst is very sensitive to $\Gamma_{R,P}$. For small values of $\Gamma_{R,P}$ an effective ($d-1$)-dimensional homogeneous layer is formed at the bottom of the film (bottom panel right column in Fig.~\ref{fig:drop_sims}). For larger values of $\Gamma_{R,P}$ (and symmetric initial conditions, see Supp. Mat.) the catalysts accumulate.  
Similarly to the collapse of catalyst observed in the initially homogeneous system, the Marangoni flows generated by a locally lower vapor-liquid surface tension at the location of high catalyst concentration deform the droplet, creating a pressure that pushes the catalyst toward the center. The product becomes enriched in the presence of more catalyst, and vice versa for the reactant. The concentration of catalyst at the droplet's center is similar to that observed in Ref.~\cite{singh_interface-mediated_2020} where catalytic \ce{TiO_2} heavy micro-particles (thus, $\alpha>0$) inside a liquid droplet of $3\%$ \ce{H2O2} solution, accumulate at the center of the droplet.
We mention that the relative position of the catalyst's maximum density and the droplet height depends on the initial conditions (see Supp. Mat. Fig.~S4). 
Finally, for even larger values of $\Gamma_{R,P}$ the liquid-vapor surface tension $\gamma$ gets according to our model so low that the liquid becomes perfectly wetting, and thus the droplets spread over the periodic boundary of the simulation box forming a closed film. However, for $\alpha>0$ the accumulation of catalyst persists leading to an inhomogeneous film thickness.


\subsection*{Sorption and strength of surface tension effects}

\begin{figure*}[hh]
    \centering
    \includegraphics[width=1.0\linewidth]{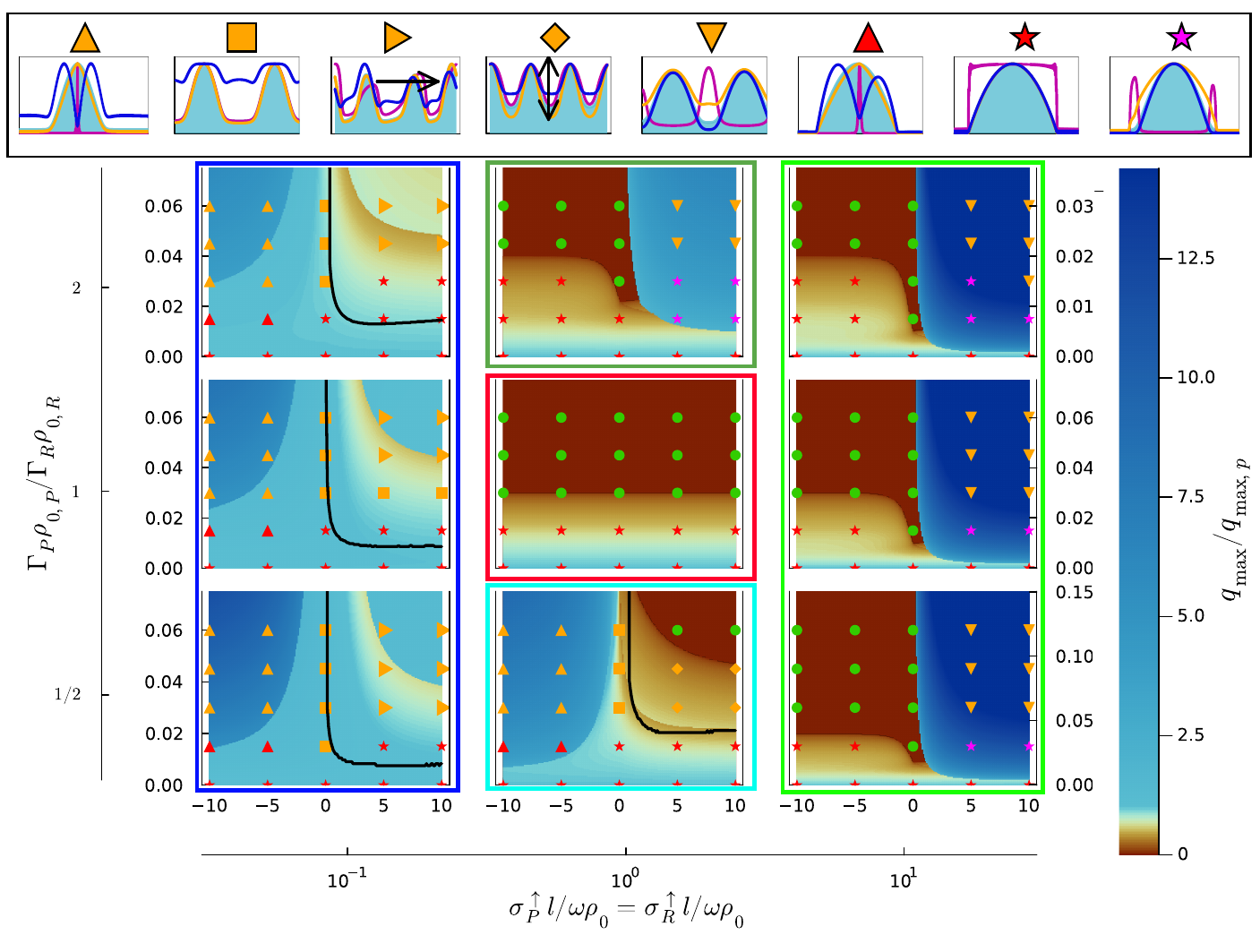}
    \caption{Stability diagrams depending on sorption and strength of surface tension effects. The nine plots at the bottom show the stability portraits for different ratios of the surface tension effects of reactant and product $\Gamma_P\rho_{0,P}/\Gamma_R\rho_{0,R}=1/2,1,2$ and sorption rates of product and reactant $\sigma_P^\uparrow l/\omega\rho_0=\sigma_R^\uparrow l/\omega\rho_0=0.1,1,10$. The color scheme of the background indicates the fastest growing mode wavenumber $q_{\max}$ divided by the passive case fastest growing wavenumber $q_{\max,p}$ ranging from stable films $q_{\max}=0$ in brown to the most excited states in dark blue. The symbols in the foreground correspond to simulations performed to validate and further asses the stability predicted by the LSA. Those simulations started from the initial condition $h=h_0 + \mathcal{N}, \rho=\rho_0, \rho_P=\rho_{P,0}, \rho_R=\rho_{R,0}$ and all relevant parameters but the sorption rates are chosen as reported in the Materials and Methods section. Red and magenta symbols indicate film rupture, while orange and pink symbols indicate excitement of certain wavelengths but no film rupture. In green simulations confirming a stable homogeneous state are reported. The marker shapes correspond to different observed steady state behavior, exemplary shown by simulation snapshots at the top. The diagrams highlighted in green correspond to the ones reported in Fig.~\ref{fig:stability_portrait}. The left $y$-axis of the diagrams is $\tilde{\Gamma}_P$, the right $y$-axis is $\tilde{\Gamma}_R$ and scales as $\Gamma_P\rho_{0,P}/\Gamma_R\rho_{0,R}$ times the left $y$-axis. The $x$-axis of the diagrams is $\alpha h_0$.}
    \label{fig:volatilities}
\end{figure*}

In the following, we analyze the dependence of the phenomena that we discussed so far on the ratio of the surface tension sensitivity to chemicals $\Gamma_P\rho_{0,P}/\Gamma_R\rho_{0,R}$ and on the rates of sorption $\sigma_P^\uparrow l/\omega\rho_0=\sigma_R^\uparrow l/\omega\rho_0$.
Indeed, as shown in Fig.~\ref{fig:volatilities}, the stability diagrams, like the one in Fig.~\ref{fig:stability_portrait}, are sensitive to these ratios.

There are three types of stability diagrams predicted by the LSA. 
First, some diagrams are shaped as the one presented in Fig.~\ref{fig:stability_portrait} (framed in dark or light green, where the dark green framed diagram corresponds to the one shown in Fig.~\ref{fig:stability_portrait}) as discussed in the previous sections. 
Second, 
for equally strong surface tension effects of product and reactant $\Gamma_P\rho_{0,P}/\Gamma_R\rho_{0,R}=1$ and sorption rates are equal to the reaction rate $\sigma_P^\uparrow/\omega\rho_0=\sigma_R^\uparrow/\omega\rho_0=1$ (central diagram framed in red). Here, reactants and products equally affect the surface tension and they have the same sorption. Accordingly, due to the symmetric roles of reactants and products, the fastest-growing mode is independent of the vertical distribution of catalyst $\alpha$. Stability is reached when \eqref{eq:cond-0} holds. Thus the catalyst can be effectively treated as a simple surfactant or solute lowering the surface tension of the liquid.
The third regime occurs for $\sigma_P^\uparrow/\omega\rho_0=\sigma_R^\uparrow/\omega\rho_0=0.1$ (framed in blue). Here the homogeneous state is never stable, at least not for the parameter range studied here. However, as we will discuss later using numerical simulations, this does not necessarily indicate film rupture. In the third type of stability diagram, we observe a second collapse mechanism in the left corner (high surface tension effects and catalyst enriched at the liquid surface) of the diagram.  Here the collapse is caused by catalyst enriched  at the liquid  surface ($\alpha<0$) effectively raising the local surface tension by consuming reactant and creating product such that the surface tension decreases locally. Therefore the Marangoni flow points from regions of low catalyst concentration to regions of higher catalyst concentration accumulating the catalyst. The catalyst behaves as if it were an "anti-surfactant". Furthermore, we obtain complex eigenvalues of the dissipation relation in the right corner of those diagrams, predicting a phase shift of excited wave modes with time in the linear regime. The black lines indicate the onset of such complex eigenvalues, where above the line significant imaginary parts appear in the eigenvalues of the dispersion relation. 
Finally, for $\Gamma_P\rho_{0,P}/\Gamma_R\rho_{0,R}=1/2$ and $\sigma_P^\uparrow/\omega\rho_0=\sigma_R^\uparrow/\omega\rho_0=1$ (framed in cyan) we observe a transition between a stability portrait of the first type and one of the third type. 

Numerical simulations show the onset of eventual regimes that are not present 
in Fig.~\ref{fig:stability_portrait}, 
including film-rupture, inhomogeneous film height with no film rupture, as well stable films with homogeneous height $h\equiv h_0$. For the films that break into droplets, we can distinguish between spherically shaped droplets obtained either via spinodal rupture (red stars) or via collapse (magenta stars), as well as droplets where the chemical concentration fields are shaped by the second kind of collapse, being catalyst enriched at the liquid interface and behaving as an effective ``anti-surfactant'' by consuming surface active reactant and replacing it with less surface active product (red up-triangles). 

For films that do not break, on top of films with homogeneous height (green circles), we observe films with inhomogeneous height. This is 
associated with local inhomogeneity in the chemical concentrations and the resulting flow dynamics. Of such steady states 
there are the ones resulting from the collapse mechanism discussed before (orange down-triangles) where the catalyst concentration is the highest where the film height is lowest and vice versa. Furthermore, patterns are emerging from the second collapse mechanism (orange up-triangles) that are characterized by a matching of the maxima of film height and catalyst concentration. On top of that, we observe not-ruptured perturbed states without any collapse for $\alpha=0$ (orange squares). 
Finally, for some parameters, we observe traveling waves~\footnote{In Fig.~\ref{fig:volatilities} a right moving pattern is shown but left moving patterns can be observed as well.} (orange right-triangles) and stationary waves (orange diamonds). These numerical results are supported by the LSA which shows non-zero imaginary components in the eigenvalues (above the black lines).


\section*{Discussion}

Our theoretical analysis pinpoints the transverse distribution of catalysts (captured by $\alpha$) and the net effect of the chemicals on the surface tension ($\Gamma_{R,P}\rho_{R,P}$) as the key parameters. Therefore, to derive conclusions that are relevant for real systems, it is crucial to bind the range of these parameters via available experimental values. 

For what concerns the tunability of the transverse distribution of catalysts, encoded by $\alpha$, experimental measurements have reported accumulation of catalysts at both liquid-gas~\cite{hemmeter_buoy_2023, hemmeter_structure_2023, steinruck_surface_2011} ($\alpha < 0$) and solid-liquid~\cite{frosch_wetting_2023, steinruck_surface_2011} ($\alpha > 0$) interfaces. 
On top of this, electric fields can be used to tune both the distribution of particles in a thin film as well as the range within which the disjoining pressure is operational~\cite{bhatt_electric_2023}\footnote{Ref.~\cite{bhatt_electric_2023} shows that applying a voltage of up to $60 V$ disjoining pressure becomes relevant for a film of thickness as large as $10 \mu m$. In such a scenario with a micrometer-thick film being prone to spinodal rupture it is possible to have catalytic particles that can be easily tracked by optical methods making it possible to observe the dynamics of the catalyst distribution experimentally while having a relevant effect from the disjoining pressure augmented by the electrical potential. 
In Ref.~\cite{john_ratchet-driven_2008} the authors show how to include such effects into the lubrication-approximation approach.}.

For what concerns the magnitude of $\Gamma_{R,P}\rho_{R,P}l/h_0$ as compared to the bare value $\gamma_0$, 
recent measurements~\cite{zhai_influence_2023} show that the surface tension of two selected ILs is significantly affected by the applied gas pressure and thus by the concentration of the dissolved gas in the liquid phase (also reported in the supporting information Fig.~\ref{fig:experimental_data}). In particular, in the case of carbon dioxide (\ce{CO2}) as an example of reactant gas, a gas pressure of $1\text{MPa}$, for example, reduces the vapor-liquid surface tension of the ILs \ce{[OMIM][PF6]} or \ce{[m(PEG2)2IM]I} by about $10\%$ for a temperature of $303.15\text{K}$. This relative change is smaller in the case of argon (\ce{Ar}) gas, where a decrease in the surface tension of about $2.5\%$ relative to the values for the pure ILs could be determined experimentally. 
These values are compatible with those used to derive Fig.~\ref{fig:stability_portrait} which shows that indeed even a few percent variation in the local surface tension is sufficient to observe the novel regimes predicted by our model.


The experimental data that we have commented on supports that,  upon varying the concentration of the suspended chemicals,  surface tension is influenced by magnitudes as large as the magnitudes studied above. However, they do not show that such changes in densities (and hence in surface tension) can be attained utilizing chemical reactions which is indeed crucial in our model.  On top of this, our predictions pinpoint the relevance of concentrations of chemicals. 
To estimate the magnitude of the local variations in chemical densities we recall that the mean concentration of product in the film is given by $\rho_{0,P}=\frac{\omega\rho_0}{\sigma_P^\uparrow l}\frac{\rho_{R,Res}\sigma_R^\downarrow h_0}{\omega \rho_0+\sigma_R^\uparrow l }$. Thus the amount of dissolved product in the liquid is proportional to the concentration of reactants in the atmosphere $\rho_{R,Res}$ which itself is proportional to the gas pressure of reactant. We calculate the mean concentration of reactant inside the film to be $\rho_{0,R}=\frac{\rho_{R,Res}\sigma_{R}^\downarrow h_0}{\sigma_R^\uparrow l+ \omega\rho_0}$. 
There are two regimes of interest here: First, if $\omega\rho_0 \gg \sigma_R^\uparrow l$ then $\rho_{0,P} \approx \rho_{R,Res} \frac{\sigma_R^\downarrow h_0}{\sigma_P^\uparrow l}$ and thus $\rho_{0,P}$ can be controlled by either increasing the gas pressure of reactant $R$ or by the ratio of the rates $\frac{\sigma_R^\downarrow}{\sigma_P^\uparrow}$. It is reasonable to assume that the sorption of reactant and product are comparable and thus the concentrations of reactant and product are alike. 
Second, if $\omega\rho_0 \ll \sigma_R^\uparrow l$  we obtain  $\rho_{0,P} \approx \frac{\omega\rho_0}{\sigma_P^\uparrow l}\rho_{R,Res}\frac{\sigma_R^\downarrow h_0}{\sigma_P^\uparrow l}$ resulting in an amplification factor $\frac{\omega\rho_0}{\sigma_P^\uparrow l}$. In fact, if $\sigma_P^\uparrow l \ll \omega \rho_0 \ll \sigma_R^\uparrow l$ we have that $\frac{\omega\rho_0}{\sigma_P^\uparrow l}\gg 1$ and thus $\rho_{0,P}\gg \rho_{0,R}$. Accordingly, if the product is much less volatile than the reactant significant changes of the vapor-liquid surface tension relative to that of the pure IL can be achieved also for lower gas pressures of reactant.

\section*{Conclusion}

We have developed a continuum model to describe the dynamics of thin liquid films accounting for the presence and varying distribution of catalyst particles as well as reactants and products based on the lubrication theory and the Fick-Jackobs approximation. We analyzed the model in terms of its linear stability and obtained predictions on the persistence of ultra-thin liquid layers as well as the time scales of the evolution of such films. That revealed a surprisingly rich phenomenology, reaching from the stabilization of an otherwise unstable film, due to the overall reduction of the vapor-liquid surface tension and Marangoni flows, over the tuning of its dominant wavelength to the accumulation of catalytic particles which is in qualitative agreement with experimental observations~\cite{singh_interface-mediated_2020}.
Not only the stability, but also the time scale needed to attain the steady state are strongly affected by the chemical reactions and, as shown in Fig.~\ref{fig:rupture_time}, the rupture time can be orders of magnitude larger as compared to the passive case. This is crucial for the design of measurement techniques studying film dynamics as well as for the design and optimization of catalytic materials. 

The predictions of the LSA have been confirmed 
by numerical simulations with a LBM scheme. In particular, the numerical data shed further light on the evolution of our model system. 
In fact, by numerical inspection, we found that in addition to stable films with homogeneous height, and spherical-cap droplets, other eventual regimes can be attained, such as films with inhomogeneous heights, non-spherical droplets as well as unsteady states such as traveling and stationary waves. 

All these numerical data have been obtained in the range of parameters that are compatible with current experimental measurements of both the transverse distribution of catalysts as well as the sensitivity of surface tension to added reactant chemicals such as dissolved gases. Hence we expect our prediction to be relevant in the design of novel catalysis concepts such as SLP and SILP technologies.



\subsection*{Linear Stability Analysis (LSA)}

\begin{figure}[hh]
    \centering
    \includegraphics[width=\linewidth]{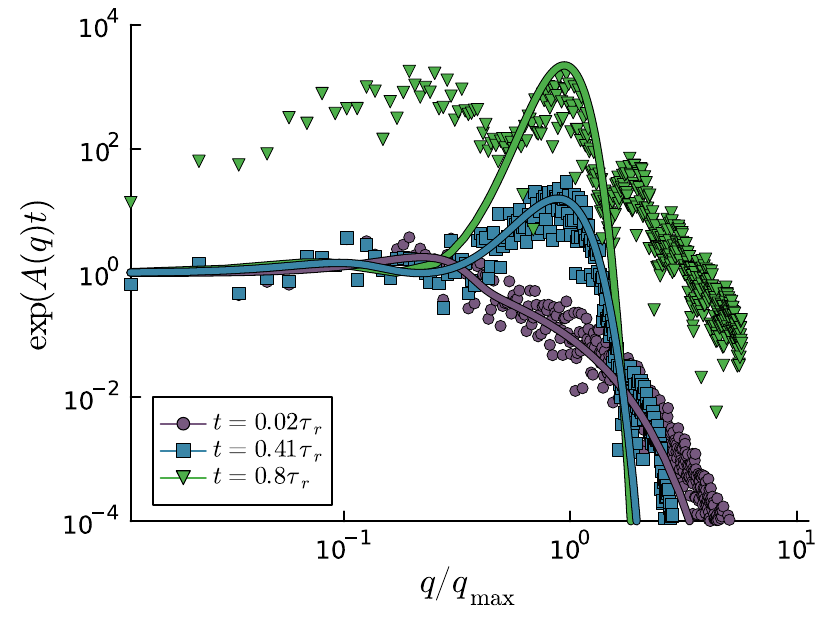}
    \caption{
    Comparison between the spectrum of the profile of a thin liquid film containing catalyst, reactant, and product species obtained from a simulation (points) with the linear solution \eqref{eq:linear_soloution} (lines) shows good agreement at various times. Different times are depicted by varying color and given in fractions of the rupture time $\tau_r$ defined in \eqref{eq:rupture_time_def}. $L$ is the simulation box size. } 
    \label{fig:fourier}
\end{figure}

 In order to study the system, we linearize \eqref{eq:TFE-0}, \eqref{eq:fick-0}, \eqref{eq:product-0} and \eqref{eq:reactant-0} by expanding them around the homogeneous state $h= h_0 + \delta h , \rho = \rho_0+ \delta \rho, \rho_P= \rho_{0,P}+\delta \rho_P , \rho_R 0 \rho_{0,R}+ \delta \rho_R$ and retaining only linear terms in the perturbations. By performing Fourier transformations according to $\widehat{\delta\rho}=\widehat{\rho}, \widehat{\delta h}= \widehat{h}, \widehat{\rho_P}=\widehat{\delta \rho_P}, \widehat{\rho_R}=\widehat{\delta \rho_R}$ the resulting expression is 
\begin{align}
\label{eq:Matrix_ODE}
 \partial_t \begin{pmatrix}
  \widehat{h} \\ \widehat{\rho}\\ \widehat{\rho_P} \\ \widehat{\rho_R}
 \end{pmatrix} =A(q)
 \cdot \begin{pmatrix}
 \widehat{h} \\ \widehat{\rho}\\ \widehat{\rho_P} \\ \widehat{\rho_R}
 \end{pmatrix}
\end{align}
The matrix $A(q)$ and the details of this computation can be found in the supplemental material. We solve this system to obtain the linear solution
\begin{align}
\label{eq:linear_soloution}
    \begin{pmatrix}
  \widehat{h} \\ \widehat{\rho}\\ \widehat{\rho_P} \\ \widehat{\rho_R}
 \end{pmatrix}(q,t)= \begin{pmatrix}
 \widehat{h} \\ \widehat{\rho}\\ \widehat{\rho_P} \\ \widehat{\rho_R} \end{pmatrix}(q,0)e^{A(q)t}. 
\end{align}
In Fig.~\ref{fig:fourier},  we validate the linear solution according to \eqref{eq:linear_soloution} against the spectrum of a numerical solution of the full model  Eqs.~($\ref{eq:TFE-0} $-$\ref{eq:product-0}$) and \eqref{eq:fick-0}. This solution is obtained by a lattice Boltzmann method for the TFE developed by \cite{zitz_lattice_2019, zitz_lattice_2021, zitz_controlling_2021} and is explained in detail in the supplemental material. The initial condition chosen in the simulation are homogeneous fields with a small perturbation $h(\x)=h_0+\epsilon \mathcal{N}, \rho(\x) = \rho_0, \rho_R=\rho_{0,R}:= \rho_{R,Res}\sigma_R^\downarrow h_0/(\omega \rho_0 +\sigma_R^\uparrow l ), \rho_P=\rho_{0,P}:= \omega \rho_{R,0} \rho_0 / (\sigma_P^\uparrow l) $. Here $\epsilon$ is a number such that $\epsilon\ll h_0$ and $\mathcal{N}$ is a normal distributed random variable. The linear solution performs well for short time scales, deteriorating as we approach film rupture. However, it still accurately captures the maxima of the film height spectrum, as shown in Fig.~\ref{fig:fourier}. We  utilize \eqref{eq:linear_soloution} to predict types of behaviors of the system and compare them against the numerical solution of the full model. Specifically, we focus on the eigenvalues $\eta_{m}(q)$ and $\eta_{2,3,4}$ of the matrix $A(q)$, where we choose the indices such that $\eta_{m} > \eta_{2,3,4}$. We define the wavenumber $q$ at which $\eta_{m}$ achieves its maximum positive value as the fastest growing mode, denoted as $q_{\max}$:
\begin{align}
\label{eq:q_max_def}
q_{\max}:= \argmax(\eta_{m}).
\end{align}
As we demonstrate, $q_{\max}$ and $\max (\eta_{m})$ effectively capture the system's stability and rupture time.

\subsection*{Parameters}
Parameters are chosen if not stated differently $R_c=1\text{nm}$, $R_P=0.5\text{\AA}$, $h_0=10\text{nm}$, $\nu=8.9\cdot 10^{-6}\text{m}^2/\text{s}$, $\rho_l=10^3\text{kg}/\text{m}^3$, $\gamma_0=0.073 \text{N}/\text{m}$, $\theta_0=\pi/9$, $\omega\rho_0/h_0=1.6\cdot10^{6} s^{-1}$, $\sigma_P l/h_0^\uparrow=\sigma_R^\uparrow l/h_0=1.6\cdot 10^{6} 1/\text{s}$, $\sigma_R^\downarrow=1.6\cdot 10^{7}1/\text{s}$, $h^*=0.1 h_0$, $h_0\alpha=5$, $\tilde{\Gamma}_P=2\tilde{\Gamma}_R=0.03$. Surface tension \cite{water_surface_tension}, density \cite{wagner_iapws_2002}, and viscosity \cite{huber2009new} are chosen to match the values of water at $T=303 \text{K}$. The reaction rate $\omega$ is derived from Ref.~\cite{brown_ionic_2014}, where spherical catalyst particles coated with platinum suspended in a $10\%$ aqueous \ce{H2O2} solution consume $0.0212\text{mol}/(\text{sm}^2)$ \ce{H2O2} molecules on their active sites. Here, we use these values and we assume that catalyst particles are $1\text{nm}$ in radius and suspended at a $4\%$ volume fraction. To reduce the number of free parameters, we set the transfer rates $\sigma_P^\uparrow l/h_0=\sigma_R^\uparrow l/h_0=\omega\rho_0/h_0=1.6 \cdot 10^6 s^{-1}$, and $\sigma_R^\downarrow=10\omega\rho_0=1.6 \cdot 10^7 s^{-1}$ for most parts if not stated otherwise.

\subsection*{Data Availability}
The data necessary to recreate this manuscript is available under \url{doi.org/10.5281/zenodo.10691669}. 

\subsection*{Competing Interests}
The authors declare no competing interest.

\subsection*{Acknowledgments}
We acknowledge funding by the Deutsche Forschungsgemeinschaft (DFG, German Research Foundation) Project-ID 431791331—SFB 1452, the Helmholtz Association of German Research Centers (HGF) and the Federal Ministry of Education and Research (BMBF), Germany for supporting the Innovation Pool project “Solar H2: Highly Pure and Compressed”, and Stefan Zitz for fruitful discussions.




\begin{thebibliography}{81}%
\makeatletter
\providecommand \@ifxundefined [1]{%
 \@ifx{#1\undefined}
}%
\providecommand \@ifnum [1]{%
 \ifnum #1\expandafter \@firstoftwo
 \else \expandafter \@secondoftwo
 \fi
}%
\providecommand \@ifx [1]{%
 \ifx #1\expandafter \@firstoftwo
 \else \expandafter \@secondoftwo
 \fi
}%
\providecommand \natexlab [1]{#1}%
\providecommand \enquote  [1]{``#1''}%
\providecommand \bibnamefont  [1]{#1}%
\providecommand \bibfnamefont [1]{#1}%
\providecommand \citenamefont [1]{#1}%
\providecommand \href@noop [0]{\@secondoftwo}%
\providecommand \href [0]{\begingroup \@sanitize@url \@href}%
\providecommand \@href[1]{\@@startlink{#1}\@@href}%
\providecommand \@@href[1]{\endgroup#1\@@endlink}%
\providecommand \@sanitize@url [0]{\catcode `\\12\catcode `\$12\catcode
  `\&12\catcode `\#12\catcode `\^12\catcode `\_12\catcode `\%12\relax}%
\providecommand \@@startlink[1]{}%
\providecommand \@@endlink[0]{}%
\providecommand \url  [0]{\begingroup\@sanitize@url \@url }%
\providecommand \@url [1]{\endgroup\@href {#1}{\urlprefix }}%
\providecommand \urlprefix  [0]{URL }%
\providecommand \Eprint [0]{\href }%
\providecommand \doibase [0]{http://dx.doi.org/}%
\providecommand \selectlanguage [0]{\@gobble}%
\providecommand \bibinfo  [0]{\@secondoftwo}%
\providecommand \bibfield  [0]{\@secondoftwo}%
\providecommand \translation [1]{[#1]}%
\providecommand \BibitemOpen [0]{}%
\providecommand \bibitemStop [0]{}%
\providecommand \bibitemNoStop [0]{.\EOS\space}%
\providecommand \EOS [0]{\spacefactor3000\relax}%
\providecommand \BibitemShut  [1]{\csname bibitem#1\endcsname}%
\let\auto@bib@innerbib\@empty
\bibitem [{\citenamefont {Wei}\ \emph {et~al.}(2020)\citenamefont {Wei},
  \citenamefont {Li}, \citenamefont {Cooks},\ and\ \citenamefont
  {Yan}}]{wei_accelerated_2020}%
  \BibitemOpen
  \bibfield  {author} {\bibinfo {author} {\bibfnamefont {Z.}~\bibnamefont
  {Wei}}, \bibinfo {author} {\bibfnamefont {Y.}~\bibnamefont {Li}}, \bibinfo
  {author} {\bibfnamefont {R.~G.}\ \bibnamefont {Cooks}}, \ and\ \bibinfo
  {author} {\bibfnamefont {X.}~\bibnamefont {Yan}},\ }\href {\doibase
  10.1146/annurev-physchem-121319-110654} {\bibfield  {journal} {\bibinfo
  {journal} {Annual Review of Physical Chemistry}\ }\textbf {\bibinfo {volume}
  {71}},\ \bibinfo {pages} {31} (\bibinfo {year} {2020})}\BibitemShut {NoStop}%
\bibitem [{\citenamefont {Fallah-Araghi}\ \emph {et~al.}(2014)\citenamefont
  {Fallah-Araghi}, \citenamefont {Meguellati}, \citenamefont {Baret},
  \citenamefont {Harrak}, \citenamefont {Mangeat}, \citenamefont {Karplus},
  \citenamefont {Ladame}, \citenamefont {Marques},\ and\ \citenamefont
  {Griffiths}}]{Fallah2014}%
  \BibitemOpen
  \bibfield  {author} {\bibinfo {author} {\bibfnamefont {A.}~\bibnamefont
  {Fallah-Araghi}}, \bibinfo {author} {\bibfnamefont {K.}~\bibnamefont
  {Meguellati}}, \bibinfo {author} {\bibfnamefont {J.-C.}\ \bibnamefont
  {Baret}}, \bibinfo {author} {\bibfnamefont {A.~E.}\ \bibnamefont {Harrak}},
  \bibinfo {author} {\bibfnamefont {T.}~\bibnamefont {Mangeat}}, \bibinfo
  {author} {\bibfnamefont {M.}~\bibnamefont {Karplus}}, \bibinfo {author}
  {\bibfnamefont {S.}~\bibnamefont {Ladame}}, \bibinfo {author} {\bibfnamefont
  {C.~M.}\ \bibnamefont {Marques}}, \ and\ \bibinfo {author} {\bibfnamefont
  {A.~D.}\ \bibnamefont {Griffiths}},\ }\href {\doibase
  10.1103/PhysRevLett.112.028301} {\bibfield  {journal} {\bibinfo  {journal}
  {Phys. Rev. Lett.}\ }\textbf {\bibinfo {volume} {112}},\ \bibinfo {pages}
  {028301} (\bibinfo {year} {2014})}\BibitemShut {NoStop}%
\bibitem [{\citenamefont {Bain}\ \emph {et~al.}(2016)\citenamefont {Bain},
  \citenamefont {Pulliam}, \citenamefont {Thery},\ and\ \citenamefont
  {Cooks}}]{Bain2016}%
  \BibitemOpen
  \bibfield  {author} {\bibinfo {author} {\bibfnamefont {R.~M.}\ \bibnamefont
  {Bain}}, \bibinfo {author} {\bibfnamefont {C.~J.}\ \bibnamefont {Pulliam}},
  \bibinfo {author} {\bibfnamefont {F.}~\bibnamefont {Thery}}, \ and\ \bibinfo
  {author} {\bibfnamefont {R.~G.}\ \bibnamefont {Cooks}},\ }\href {\doibase
  https://doi.org/10.1002/anie.201605899} {\bibfield  {journal} {\bibinfo
  {journal} {Angew. Chem. Int. Ed.}\ }\textbf {\bibinfo {volume} {55}},\
  \bibinfo {pages} {10478} (\bibinfo {year} {2016})}\BibitemShut {NoStop}%
\bibitem [{\citenamefont {Li}\ \emph {et~al.}(2018)\citenamefont {Li},
  \citenamefont {Liu}, \citenamefont {Gao}, \citenamefont {Helmy},
  \citenamefont {Wuelfing}, \citenamefont {Welch},\ and\ \citenamefont
  {Cooks}}]{Li2018}%
  \BibitemOpen
  \bibfield  {author} {\bibinfo {author} {\bibfnamefont {Y.}~\bibnamefont
  {Li}}, \bibinfo {author} {\bibfnamefont {Y.}~\bibnamefont {Liu}}, \bibinfo
  {author} {\bibfnamefont {H.}~\bibnamefont {Gao}}, \bibinfo {author}
  {\bibfnamefont {R.}~\bibnamefont {Helmy}}, \bibinfo {author} {\bibfnamefont
  {W.~P.}\ \bibnamefont {Wuelfing}}, \bibinfo {author} {\bibfnamefont {C.~J.}\
  \bibnamefont {Welch}}, \ and\ \bibinfo {author} {\bibfnamefont {R.~G.}\
  \bibnamefont {Cooks}},\ }\href {\doibase
  https://doi.org/10.1002/chem.201801176} {\bibfield  {journal} {\bibinfo
  {journal} {Chemistry - A European Journal}\ }\textbf {\bibinfo {volume}
  {24}},\ \bibinfo {pages} {7349} (\bibinfo {year} {2018})}\BibitemShut
  {NoStop}%
\bibitem [{\citenamefont {Shi}\ \emph {et~al.}(2022)\citenamefont {Shi},
  \citenamefont {Shang}, \citenamefont {Zhou}, \citenamefont {Zhao},\ and\
  \citenamefont {Zhang}}]{shi_interfacial_2022}%
  \BibitemOpen
  \bibfield  {author} {\bibinfo {author} {\bibfnamefont {R.}~\bibnamefont
  {Shi}}, \bibinfo {author} {\bibfnamefont {L.}~\bibnamefont {Shang}}, \bibinfo
  {author} {\bibfnamefont {C.}~\bibnamefont {Zhou}}, \bibinfo {author}
  {\bibfnamefont {Y.}~\bibnamefont {Zhao}}, \ and\ \bibinfo {author}
  {\bibfnamefont {T.}~\bibnamefont {Zhang}},\ }\href {\doibase
  10.1002/EXP.20210046} {\bibfield  {journal} {\bibinfo  {journal}
  {Exploration}\ }\textbf {\bibinfo {volume} {2}},\ \bibinfo {pages} {20210046}
  (\bibinfo {year} {2022})}\BibitemShut {NoStop}%
\bibitem [{\citenamefont {Buch}\ \emph {et~al.}(2007)\citenamefont {Buch},
  \citenamefont {Milet}, \citenamefont {Vácha}, \citenamefont {Jungwirth},\
  and\ \citenamefont {Devlin}}]{buch_water_2007}%
  \BibitemOpen
  \bibfield  {author} {\bibinfo {author} {\bibfnamefont {V.}~\bibnamefont
  {Buch}}, \bibinfo {author} {\bibfnamefont {A.}~\bibnamefont {Milet}},
  \bibinfo {author} {\bibfnamefont {R.}~\bibnamefont {Vácha}}, \bibinfo
  {author} {\bibfnamefont {P.}~\bibnamefont {Jungwirth}}, \ and\ \bibinfo
  {author} {\bibfnamefont {J.~P.}\ \bibnamefont {Devlin}},\ }\href {\doibase
  10.1073/pnas.0611285104} {\bibfield  {journal} {\bibinfo  {journal}
  {Proceedings of the National Academy of Sciences}\ }\textbf {\bibinfo
  {volume} {104}},\ \bibinfo {pages} {7342} (\bibinfo {year}
  {2007})}\BibitemShut {NoStop}%
\bibitem [{\citenamefont {Petersen}\ and\ \citenamefont
  {Saykally}(2008)}]{Petersen2008}%
  \BibitemOpen
  \bibfield  {author} {\bibinfo {author} {\bibfnamefont {P.~B.}\ \bibnamefont
  {Petersen}}\ and\ \bibinfo {author} {\bibfnamefont {R.~J.}\ \bibnamefont
  {Saykally}},\ }\href {\doibase https://doi.org/10.1016/j.cplett.2008.04.010}
  {\bibfield  {journal} {\bibinfo  {journal} {Chemical Physics Letters}\
  }\textbf {\bibinfo {volume} {458}},\ \bibinfo {pages} {255} (\bibinfo {year}
  {2008})}\BibitemShut {NoStop}%
\bibitem [{\citenamefont {Ruiz-Lopez}\ \emph {et~al.}(2020)\citenamefont
  {Ruiz-Lopez}, \citenamefont {Francisco}, \citenamefont {Martins-Costa},\ and\
  \citenamefont {Anglada}}]{ruiz-lopez_molecular_2020}%
  \BibitemOpen
  \bibfield  {author} {\bibinfo {author} {\bibfnamefont {M.~F.}\ \bibnamefont
  {Ruiz-Lopez}}, \bibinfo {author} {\bibfnamefont {J.~S.}\ \bibnamefont
  {Francisco}}, \bibinfo {author} {\bibfnamefont {M.~T.~C.}\ \bibnamefont
  {Martins-Costa}}, \ and\ \bibinfo {author} {\bibfnamefont {J.~M.}\
  \bibnamefont {Anglada}},\ }\href {\doibase 10.1038/s41570-020-0203-2}
  {\bibfield  {journal} {\bibinfo  {journal} {Nature Reviews Chemistry}\
  }\textbf {\bibinfo {volume} {4}},\ \bibinfo {pages} {459} (\bibinfo {year}
  {2020})}\BibitemShut {NoStop}%
\bibitem [{\citenamefont {Mondal}\ \emph {et~al.}(2018)\citenamefont {Mondal},
  \citenamefont {Acharya}, \citenamefont {Biswas}, \citenamefont {Bagchi},\
  and\ \citenamefont {Zare}}]{mondal_enhancement_2018}%
  \BibitemOpen
  \bibfield  {author} {\bibinfo {author} {\bibfnamefont {S.}~\bibnamefont
  {Mondal}}, \bibinfo {author} {\bibfnamefont {S.}~\bibnamefont {Acharya}},
  \bibinfo {author} {\bibfnamefont {R.}~\bibnamefont {Biswas}}, \bibinfo
  {author} {\bibfnamefont {B.}~\bibnamefont {Bagchi}}, \ and\ \bibinfo {author}
  {\bibfnamefont {R.~N.}\ \bibnamefont {Zare}},\ }\href {\doibase
  10.1063/1.5030114} {\bibfield  {journal} {\bibinfo  {journal} {J. Chem.
  Phys.}\ }\textbf {\bibinfo {volume} {148}},\ \bibinfo {pages} {244704}
  (\bibinfo {year} {2018})}\BibitemShut {NoStop}%
\bibitem [{\citenamefont {Singh}\ \emph {et~al.}(2020)\citenamefont {Singh},
  \citenamefont {Domínguez}, \citenamefont {Choudhury}, \citenamefont
  {Kottapalli}, \citenamefont {Popescu}, \citenamefont {Dietrich},\ and\
  \citenamefont {Fischer}}]{singh_interface-mediated_2020}%
  \BibitemOpen
  \bibfield  {author} {\bibinfo {author} {\bibfnamefont {D.~P.}\ \bibnamefont
  {Singh}}, \bibinfo {author} {\bibfnamefont {A.}~\bibnamefont {Domínguez}},
  \bibinfo {author} {\bibfnamefont {U.}~\bibnamefont {Choudhury}}, \bibinfo
  {author} {\bibfnamefont {S.~N.}\ \bibnamefont {Kottapalli}}, \bibinfo
  {author} {\bibfnamefont {M.~N.}\ \bibnamefont {Popescu}}, \bibinfo {author}
  {\bibfnamefont {S.}~\bibnamefont {Dietrich}}, \ and\ \bibinfo {author}
  {\bibfnamefont {P.}~\bibnamefont {Fischer}},\ }\href {\doibase
  10.1038/s41467-020-15713-y} {\bibfield  {journal} {\bibinfo  {journal}
  {Nature Communications}\ }\textbf {\bibinfo {volume} {11}},\ \bibinfo {pages}
  {2210} (\bibinfo {year} {2020})}\BibitemShut {NoStop}%
\bibitem [{\citenamefont {Wei}\ \emph {et~al.}(2018)\citenamefont {Wei},
  \citenamefont {Zhang}, \citenamefont {Wang}, \citenamefont {Zhang},
  \citenamefont {Zhang},\ and\ \citenamefont {Cooks}}]{Wei2018}%
  \BibitemOpen
  \bibfield  {author} {\bibinfo {author} {\bibfnamefont {Z.}~\bibnamefont
  {Wei}}, \bibinfo {author} {\bibfnamefont {X.}~\bibnamefont {Zhang}}, \bibinfo
  {author} {\bibfnamefont {J.}~\bibnamefont {Wang}}, \bibinfo {author}
  {\bibfnamefont {S.}~\bibnamefont {Zhang}}, \bibinfo {author} {\bibfnamefont
  {X.}~\bibnamefont {Zhang}}, \ and\ \bibinfo {author} {\bibfnamefont {R.~G.}\
  \bibnamefont {Cooks}},\ }\href {\doibase 10.1039/C8SC03382J} {\bibfield
  {journal} {\bibinfo  {journal} {Chem. Sci.}\ }\textbf {\bibinfo {volume}
  {9}},\ \bibinfo {pages} {7779} (\bibinfo {year} {2018})}\BibitemShut
  {NoStop}%
\bibitem [{\citenamefont {Zhao}\ \emph {et~al.}(2006)\citenamefont {Zhao},
  \citenamefont {Fujita},\ and\ \citenamefont {Arai}}]{zhao_developments_2006}%
  \BibitemOpen
  \bibfield  {author} {\bibinfo {author} {\bibfnamefont {F.}~\bibnamefont
  {Zhao}}, \bibinfo {author} {\bibfnamefont {S.-i.}\ \bibnamefont {Fujita}}, \
  and\ \bibinfo {author} {\bibfnamefont {M.}~\bibnamefont {Arai}},\ }\href
  {\doibase 10.2174/138527206778249586} {\bibfield  {journal} {\bibinfo
  {journal} {Current Organic Chemistry}\ }\textbf {\bibinfo {volume} {10}},\
  \bibinfo {pages} {1681} (\bibinfo {year} {2006})}\BibitemShut {NoStop}%
\bibitem [{\citenamefont {Zeng}\ \emph {et~al.}(2015)\citenamefont {Zeng},
  \citenamefont {Huang}, \citenamefont {Yang},\ and\ \citenamefont
  {Gu}}]{zeng_silica-supported_2015}%
  \BibitemOpen
  \bibfield  {author} {\bibinfo {author} {\bibfnamefont {K.}~\bibnamefont
  {Zeng}}, \bibinfo {author} {\bibfnamefont {Z.}~\bibnamefont {Huang}},
  \bibinfo {author} {\bibfnamefont {J.}~\bibnamefont {Yang}}, \ and\ \bibinfo
  {author} {\bibfnamefont {Y.}~\bibnamefont {Gu}},\ }\href {\doibase
  10.1016/S1872-2067(15)60910-X} {\bibfield  {journal} {\bibinfo  {journal}
  {Chinese Journal of Catalysis}\ }\textbf {\bibinfo {volume} {36}},\ \bibinfo
  {pages} {1606} (\bibinfo {year} {2015})}\BibitemShut {NoStop}%
\bibitem [{\citenamefont {Alsalahi}\ and\ \citenamefont
  {Trzeciak}(2021)}]{alsalahi_rhodium-catalyzed_2021}%
  \BibitemOpen
  \bibfield  {author} {\bibinfo {author} {\bibfnamefont {W.}~\bibnamefont
  {Alsalahi}}\ and\ \bibinfo {author} {\bibfnamefont {A.}~\bibnamefont
  {Trzeciak}},\ }\href {\doibase 10.1016/j.ccr.2020.213732} {\bibfield
  {journal} {\bibinfo  {journal} {Coordination Chemistry Reviews}\ }\textbf
  {\bibinfo {volume} {430}},\ \bibinfo {pages} {213732} (\bibinfo {year}
  {2021})}\BibitemShut {NoStop}%
\bibitem [{\citenamefont {Sch\"orner}\ \emph {et~al.}(2021)\citenamefont
  {Sch\"orner}, \citenamefont {Rothg\"ngel}, \citenamefont {Mitl\"ander},
  \citenamefont {Wisser}, \citenamefont {Thommes},\ and\ \citenamefont
  {Haumann}}]{schorner_gasphase_2021}%
  \BibitemOpen
  \bibfield  {author} {\bibinfo {author} {\bibfnamefont {M.}~\bibnamefont
  {Sch\"orner}}, \bibinfo {author} {\bibfnamefont {P.}~\bibnamefont
  {Rothg\"ngel}}, \bibinfo {author} {\bibfnamefont {K.}~\bibnamefont
  {Mitl\"ander}}, \bibinfo {author} {\bibfnamefont {D.}~\bibnamefont {Wisser}},
  \bibinfo {author} {\bibfnamefont {M.}~\bibnamefont {Thommes}}, \ and\
  \bibinfo {author} {\bibfnamefont {M.}~\bibnamefont {Haumann}},\ }\href
  {\doibase 10.1002/cctc.202100743} {\bibfield  {journal} {\bibinfo  {journal}
  {ChemCatChem}\ }\textbf {\bibinfo {volume} {13}},\ \bibinfo {pages} {4192}
  (\bibinfo {year} {2021})}\BibitemShut {NoStop}%
\bibitem [{\citenamefont {Scholten}\ \emph {et~al.}(2012)\citenamefont
  {Scholten}, \citenamefont {Leal},\ and\ \citenamefont
  {Dupont}}]{scholten_transition_2012}%
  \BibitemOpen
  \bibfield  {author} {\bibinfo {author} {\bibfnamefont {J.~D.}\ \bibnamefont
  {Scholten}}, \bibinfo {author} {\bibfnamefont {B.~C.}\ \bibnamefont {Leal}},
  \ and\ \bibinfo {author} {\bibfnamefont {J.}~\bibnamefont {Dupont}},\ }\href
  {\doibase 10.1021/cs200525e} {\bibfield  {journal} {\bibinfo  {journal} {ACS
  Catal.}\ }\textbf {\bibinfo {volume} {2}},\ \bibinfo {pages} {184} (\bibinfo
  {year} {2012})}\BibitemShut {NoStop}%
\bibitem [{\citenamefont {Masoud}\ and\ \citenamefont
  {Stone}(2014)}]{masoud_reciprocal_2014}%
  \BibitemOpen
  \bibfield  {author} {\bibinfo {author} {\bibfnamefont {H.}~\bibnamefont
  {Masoud}}\ and\ \bibinfo {author} {\bibfnamefont {H.~A.}\ \bibnamefont
  {Stone}},\ }\href {\doibase 10.1017/jfm.2014.8} {\bibfield  {journal}
  {\bibinfo  {journal} {Journal of Fluid Mechanics}\ }\textbf {\bibinfo
  {volume} {741}},\ \bibinfo {pages} {R4} (\bibinfo {year} {2014})}\BibitemShut
  {NoStop}%
\bibitem [{\citenamefont {Domínguez}\ \emph {et~al.}(2016)\citenamefont
  {Domínguez}, \citenamefont {Malgaretti}, \citenamefont {Popescu},\ and\
  \citenamefont {Dietrich}}]{dominguez_effective_2016}%
  \BibitemOpen
  \bibfield  {author} {\bibinfo {author} {\bibfnamefont {A.}~\bibnamefont
  {Domínguez}}, \bibinfo {author} {\bibfnamefont {P.}~\bibnamefont
  {Malgaretti}}, \bibinfo {author} {\bibfnamefont {M.}~\bibnamefont {Popescu}},
  \ and\ \bibinfo {author} {\bibfnamefont {S.}~\bibnamefont {Dietrich}},\
  }\href {\doibase 10.1103/PhysRevLett.116.078301} {\bibfield  {journal}
  {\bibinfo  {journal} {Phys. Rev. Lett.}\ }\textbf {\bibinfo {volume} {116}},\
  \bibinfo {pages} {078301} (\bibinfo {year} {2016})}\BibitemShut {NoStop}%
\bibitem [{\citenamefont {Jafari~Kang}\ \emph {et~al.}(2020)\citenamefont
  {Jafari~Kang}, \citenamefont {Sur}, \citenamefont {Rothstein},\ and\
  \citenamefont {Masoud}}]{jafari_kang_forward_2020}%
  \BibitemOpen
  \bibfield  {author} {\bibinfo {author} {\bibfnamefont {S.}~\bibnamefont
  {Jafari~Kang}}, \bibinfo {author} {\bibfnamefont {S.}~\bibnamefont {Sur}},
  \bibinfo {author} {\bibfnamefont {J.~P.}\ \bibnamefont {Rothstein}}, \ and\
  \bibinfo {author} {\bibfnamefont {H.}~\bibnamefont {Masoud}},\ }\href
  {\doibase 10.1103/PhysRevFluids.5.084004} {\bibfield  {journal} {\bibinfo
  {journal} {Phys. Rev. Fluids}\ }\textbf {\bibinfo {volume} {5}},\ \bibinfo
  {pages} {084004} (\bibinfo {year} {2020})}\BibitemShut {NoStop}%
\bibitem [{\citenamefont {Squarcini}\ and\ \citenamefont
  {Malgaretti}(2020)}]{squarcini_inhomogeneous_2020}%
  \BibitemOpen
  \bibfield  {author} {\bibinfo {author} {\bibfnamefont {A.}~\bibnamefont
  {Squarcini}}\ and\ \bibinfo {author} {\bibfnamefont {P.}~\bibnamefont
  {Malgaretti}},\ }\href {\doibase 10.1063/5.0025989} {\bibfield  {journal}
  {\bibinfo  {journal} {J. Chem. Phys.}\ }\textbf {\bibinfo {volume} {153}},\
  \bibinfo {pages} {234903} (\bibinfo {year} {2020})}\BibitemShut {NoStop}%
\bibitem [{\citenamefont {Imamura}\ and\ \citenamefont
  {Kawakatsu}(2021)}]{imamura_modeling_2021}%
  \BibitemOpen
  \bibfield  {author} {\bibinfo {author} {\bibfnamefont {S.}~\bibnamefont
  {Imamura}}\ and\ \bibinfo {author} {\bibfnamefont {T.}~\bibnamefont
  {Kawakatsu}},\ }\href {\doibase 10.1140/epje/s10189-021-00132-8} {\bibfield
  {journal} {\bibinfo  {journal} {The European Physical Journal E}\ }\textbf
  {\bibinfo {volume} {44}},\ \bibinfo {pages} {127} (\bibinfo {year} {2021})},\
  \bibinfo {note} {arXiv:2103.09483 [cond-mat]}\BibitemShut {NoStop}%
\bibitem [{\citenamefont {Steinrück}\ and\ \citenamefont
  {Wasserscheid}(2015)}]{steinruck_ionic_2015}%
  \BibitemOpen
  \bibfield  {author} {\bibinfo {author} {\bibfnamefont {H.-P.}\ \bibnamefont
  {Steinrück}}\ and\ \bibinfo {author} {\bibfnamefont {P.}~\bibnamefont
  {Wasserscheid}},\ }\href {\doibase 10.1007/s10562-014-1435-x} {\bibfield
  {journal} {\bibinfo  {journal} {Catalysis Letters}\ }\textbf {\bibinfo
  {volume} {145}},\ \bibinfo {pages} {380} (\bibinfo {year}
  {2015})}\BibitemShut {NoStop}%
\bibitem [{\citenamefont {Rasmus~Fehrmann}(2014)}]{Haumann_book}%
  \BibitemOpen
  \bibfield  {author} {\bibinfo {author} {\bibfnamefont {M.~H.}\ \bibnamefont
  {Rasmus~Fehrmann}, \bibfnamefont {Anders~Riisager}},\ }\href@noop {} {\emph
  {\bibinfo {title} {Supported Ionic Liquids}}}\ (\bibinfo  {publisher}
  {Wiley},\ \bibinfo {address} {New York},\ \bibinfo {year} {2014})\BibitemShut
  {NoStop}%
\bibitem [{\citenamefont {Hatanaka}\ \emph {et~al.}(2021)\citenamefont
  {Hatanaka}, \citenamefont {Yasuda}, \citenamefont {Uchiage}, \citenamefont
  {Nishida},\ and\ \citenamefont {Tominaga}}]{hatanaka_continuous_2021}%
  \BibitemOpen
  \bibfield  {author} {\bibinfo {author} {\bibfnamefont {M.}~\bibnamefont
  {Hatanaka}}, \bibinfo {author} {\bibfnamefont {T.}~\bibnamefont {Yasuda}},
  \bibinfo {author} {\bibfnamefont {E.}~\bibnamefont {Uchiage}}, \bibinfo
  {author} {\bibfnamefont {M.}~\bibnamefont {Nishida}}, \ and\ \bibinfo
  {author} {\bibfnamefont {K.-i.}\ \bibnamefont {Tominaga}},\ }\href {\doibase
  10.1021/acssuschemeng.1c02084} {\bibfield  {journal} {\bibinfo  {journal}
  {ACS Sustainable Chemistry \& Engineering}\ }\textbf {\bibinfo {volume}
  {9}},\ \bibinfo {pages} {11674} (\bibinfo {year} {2021})}\BibitemShut
  {NoStop}%
\bibitem [{\citenamefont {Marinkovic}\ \emph {et~al.}(2019)\citenamefont
  {Marinkovic}, \citenamefont {Riisager}, \citenamefont {Franke}, \citenamefont
  {Wasserscheid},\ and\ \citenamefont {Haumann}}]{marinkovic_fifteen_2019}%
  \BibitemOpen
  \bibfield  {author} {\bibinfo {author} {\bibfnamefont {J.~M.}\ \bibnamefont
  {Marinkovic}}, \bibinfo {author} {\bibfnamefont {A.}~\bibnamefont
  {Riisager}}, \bibinfo {author} {\bibfnamefont {R.}~\bibnamefont {Franke}},
  \bibinfo {author} {\bibfnamefont {P.}~\bibnamefont {Wasserscheid}}, \ and\
  \bibinfo {author} {\bibfnamefont {M.}~\bibnamefont {Haumann}},\ }\href
  {\doibase 10.1021/acs.iecr.8b04010} {\bibfield  {journal} {\bibinfo
  {journal} {Industrial \& Engineering Chemistry Research}\ }\textbf {\bibinfo
  {volume} {58}},\ \bibinfo {pages} {2409} (\bibinfo {year}
  {2019})}\BibitemShut {NoStop}%
\bibitem [{\citenamefont {Gu}\ and\ \citenamefont {Li}(2009)}]{gu_ionic_2009}%
  \BibitemOpen
  \bibfield  {author} {\bibinfo {author} {\bibfnamefont {Y.}~\bibnamefont
  {Gu}}\ and\ \bibinfo {author} {\bibfnamefont {G.}~\bibnamefont {Li}},\ }\href
  {\doibase 10.1002/adsc.200900043} {\bibfield  {journal} {\bibinfo  {journal}
  {Advanced Synthesis \& Catalysis}\ }\textbf {\bibinfo {volume} {351}},\
  \bibinfo {pages} {817} (\bibinfo {year} {2009})}\BibitemShut {NoStop}%
\bibitem [{\citenamefont {Riisager}\ \emph {et~al.}(2005)\citenamefont
  {Riisager}, \citenamefont {Fehrmann}, \citenamefont {Flicker}, \citenamefont
  {Van~Hal}, \citenamefont {Haumann},\ and\ \citenamefont
  {Wasserscheid}}]{riisager_very_2005}%
  \BibitemOpen
  \bibfield  {author} {\bibinfo {author} {\bibfnamefont {A.}~\bibnamefont
  {Riisager}}, \bibinfo {author} {\bibfnamefont {R.}~\bibnamefont {Fehrmann}},
  \bibinfo {author} {\bibfnamefont {S.}~\bibnamefont {Flicker}}, \bibinfo
  {author} {\bibfnamefont {R.}~\bibnamefont {Van~Hal}}, \bibinfo {author}
  {\bibfnamefont {M.}~\bibnamefont {Haumann}}, \ and\ \bibinfo {author}
  {\bibfnamefont {P.}~\bibnamefont {Wasserscheid}},\ }\href {\doibase
  10.1002/anie.200461534} {\bibfield  {journal} {\bibinfo  {journal} {Angew.
  Chem. Int. Ed.}\ }\textbf {\bibinfo {volume} {44}},\ \bibinfo {pages} {815}
  (\bibinfo {year} {2005})}\BibitemShut {NoStop}%
\bibitem [{\citenamefont {Knapp}\ \emph {et~al.}(2009)\citenamefont {Knapp},
  \citenamefont {Jentys},\ and\ \citenamefont {Lercher}}]{knapp_impact_2009}%
  \BibitemOpen
  \bibfield  {author} {\bibinfo {author} {\bibfnamefont {R.}~\bibnamefont
  {Knapp}}, \bibinfo {author} {\bibfnamefont {A.}~\bibnamefont {Jentys}}, \
  and\ \bibinfo {author} {\bibfnamefont {J.~A.}\ \bibnamefont {Lercher}},\
  }\href {\doibase 10.1039/b901318k} {\bibfield  {journal} {\bibinfo  {journal}
  {Green Chemistry}\ }\textbf {\bibinfo {volume} {11}},\ \bibinfo {pages} {656}
  (\bibinfo {year} {2009})}\BibitemShut {NoStop}%
\bibitem [{\citenamefont {Frosch}\ \emph {et~al.}(2023)\citenamefont {Frosch},
  \citenamefont {Tavera~Mendez}, \citenamefont {Koch}, \citenamefont
  {Sch\"orner}, \citenamefont {Haumann}, \citenamefont {Hartmann},\ and\
  \citenamefont {Wisser}}]{frosch_wetting_2023}%
  \BibitemOpen
  \bibfield  {author} {\bibinfo {author} {\bibfnamefont {M.}~\bibnamefont
  {Frosch}}, \bibinfo {author} {\bibfnamefont {C.~L.}\ \bibnamefont
  {Tavera~Mendez}}, \bibinfo {author} {\bibfnamefont {A.~B.}\ \bibnamefont
  {Koch}}, \bibinfo {author} {\bibfnamefont {M.}~\bibnamefont {Sch\"orner}},
  \bibinfo {author} {\bibfnamefont {M.}~\bibnamefont {Haumann}}, \bibinfo
  {author} {\bibfnamefont {M.}~\bibnamefont {Hartmann}}, \ and\ \bibinfo
  {author} {\bibfnamefont {D.}~\bibnamefont {Wisser}},\ }\href@noop {}
  {\bibfield  {journal} {\bibinfo  {journal} {J. Phys. Chem. C}\ }\textbf
  {\bibinfo {volume} {127}},\ \bibinfo {pages} {9196} (\bibinfo {year}
  {2023})}\BibitemShut {NoStop}%
\bibitem [{\citenamefont {Pereira}\ \emph
  {et~al.}(2007{\natexlab{a}})\citenamefont {Pereira}, \citenamefont
  {Trevelyan}, \citenamefont {Thiele},\ and\ \citenamefont
  {Kalliadasis}}]{pereira_interfacial_2007}%
  \BibitemOpen
  \bibfield  {author} {\bibinfo {author} {\bibfnamefont {A.}~\bibnamefont
  {Pereira}}, \bibinfo {author} {\bibfnamefont {P.~M.~J.}\ \bibnamefont
  {Trevelyan}}, \bibinfo {author} {\bibfnamefont {U.}~\bibnamefont {Thiele}}, \
  and\ \bibinfo {author} {\bibfnamefont {S.}~\bibnamefont {Kalliadasis}},\
  }\href {\doibase 10.1007/s10665-007-9143-9} {\bibfield  {journal} {\bibinfo
  {journal} {Journal of Engineering Mathematics}\ }\textbf {\bibinfo {volume}
  {59}},\ \bibinfo {pages} {207} (\bibinfo {year}
  {2007}{\natexlab{a}})}\BibitemShut {NoStop}%
\bibitem [{\citenamefont {Pereira}\ \emph
  {et~al.}(2007{\natexlab{b}})\citenamefont {Pereira}, \citenamefont
  {Trevelyan}, \citenamefont {Thiele},\ and\ \citenamefont
  {Kalliadasis}}]{pereira_dynamics_2007}%
  \BibitemOpen
  \bibfield  {author} {\bibinfo {author} {\bibfnamefont {A.}~\bibnamefont
  {Pereira}}, \bibinfo {author} {\bibfnamefont {P.~M.~J.}\ \bibnamefont
  {Trevelyan}}, \bibinfo {author} {\bibfnamefont {U.}~\bibnamefont {Thiele}}, \
  and\ \bibinfo {author} {\bibfnamefont {S.}~\bibnamefont {Kalliadasis}},\
  }\href {\doibase 10.1063/1.2775938} {\bibfield  {journal} {\bibinfo
  {journal} {Phys. Fluids}\ }\textbf {\bibinfo {volume} {19}},\ \bibinfo
  {pages} {112102} (\bibinfo {year} {2007}{\natexlab{b}})}\BibitemShut
  {NoStop}%
\bibitem [{\citenamefont {Bender}\ \emph {et~al.}(2017)\citenamefont {Bender},
  \citenamefont {Stephan},\ and\ \citenamefont
  {Gambaryan-Roisman}}]{bender_thin_2017}%
  \BibitemOpen
  \bibfield  {author} {\bibinfo {author} {\bibfnamefont {A.}~\bibnamefont
  {Bender}}, \bibinfo {author} {\bibfnamefont {P.}~\bibnamefont {Stephan}}, \
  and\ \bibinfo {author} {\bibfnamefont {T.}~\bibnamefont
  {Gambaryan-Roisman}},\ }\href {\doibase 10.1103/PhysRevFluids.2.084002}
  {\bibfield  {journal} {\bibinfo  {journal} {Phys. Rev. Fluids}\ }\textbf
  {\bibinfo {volume} {2}},\ \bibinfo {pages} {084002} (\bibinfo {year}
  {2017})}\BibitemShut {NoStop}%
\bibitem [{\citenamefont {Zhai}\ and\ \citenamefont
  {Koller}(2023)}]{zhai_influence_2023}%
  \BibitemOpen
  \bibfield  {author} {\bibinfo {author} {\bibfnamefont {Z.}~\bibnamefont
  {Zhai}}\ and\ \bibinfo {author} {\bibfnamefont {T.~M.}\ \bibnamefont
  {Koller}},\ }\href {\doibase 10.1016/j.molliq.2023.121491} {\bibfield
  {journal} {\bibinfo  {journal} {Journal of Molecular Liquids}\ }\textbf
  {\bibinfo {volume} {377}},\ \bibinfo {pages} {121491} (\bibinfo {year}
  {2023})}\BibitemShut {NoStop}%
\bibitem [{\citenamefont {Oron}\ \emph {et~al.}(1997)\citenamefont {Oron},
  \citenamefont {Davis},\ and\ \citenamefont {Bankoff}}]{oron_long-scale_1997}%
  \BibitemOpen
  \bibfield  {author} {\bibinfo {author} {\bibfnamefont {A.}~\bibnamefont
  {Oron}}, \bibinfo {author} {\bibfnamefont {S.~H.}\ \bibnamefont {Davis}}, \
  and\ \bibinfo {author} {\bibfnamefont {S.~G.}\ \bibnamefont {Bankoff}},\
  }\href {\doibase 10.1103/RevModPhys.69.931} {\bibfield  {journal} {\bibinfo
  {journal} {Reviews of Modern Physics}\ }\textbf {\bibinfo {volume} {69}},\
  \bibinfo {pages} {931} (\bibinfo {year} {1997})}\BibitemShut {NoStop}%
\bibitem [{\citenamefont {Craster}\ and\ \citenamefont
  {Matar}(2009)}]{craster_dynamics_2009}%
  \BibitemOpen
  \bibfield  {author} {\bibinfo {author} {\bibfnamefont {R.~V.}\ \bibnamefont
  {Craster}}\ and\ \bibinfo {author} {\bibfnamefont {O.~K.}\ \bibnamefont
  {Matar}},\ }\href {\doibase 10.1103/RevModPhys.81.1131} {\bibfield  {journal}
  {\bibinfo  {journal} {Reviews of Modern Physics}\ }\textbf {\bibinfo {volume}
  {81}},\ \bibinfo {pages} {1131} (\bibinfo {year} {2009})}\BibitemShut
  {NoStop}%
\bibitem [{\citenamefont {Zwanzig}(1992)}]{zwanzig_diffusion_1992}%
  \BibitemOpen
  \bibfield  {author} {\bibinfo {author} {\bibfnamefont {R.}~\bibnamefont
  {Zwanzig}},\ }\href {\doibase 10.1021/j100189a004} {\bibfield  {journal}
  {\bibinfo  {journal} {The Journal of Physical Chemistry}\ }\textbf {\bibinfo
  {volume} {96}},\ \bibinfo {pages} {3926} (\bibinfo {year}
  {1992})}\BibitemShut {NoStop}%
\bibitem [{\citenamefont {Reguera}\ and\ \citenamefont
  {Rubi}(2001)}]{Reguera2001}%
  \BibitemOpen
  \bibfield  {author} {\bibinfo {author} {\bibfnamefont {D.}~\bibnamefont
  {Reguera}}\ and\ \bibinfo {author} {\bibfnamefont {J.~M.}\ \bibnamefont
  {Rubi}},\ }\href@noop {} {\bibfield  {journal} {\bibinfo  {journal} {Phys.
  Rev. E}\ }\textbf {\bibinfo {volume} {64}},\ \bibinfo {pages} {061106}
  (\bibinfo {year} {2001})}\BibitemShut {NoStop}%
\bibitem [{\citenamefont {Malgaretti}\ \emph {et~al.}(2013)\citenamefont
  {Malgaretti}, \citenamefont {Pagonabarraga},\ and\ \citenamefont
  {Rubi}}]{Malgaretti2013}%
  \BibitemOpen
  \bibfield  {author} {\bibinfo {author} {\bibfnamefont {P.}~\bibnamefont
  {Malgaretti}}, \bibinfo {author} {\bibfnamefont {I.}~\bibnamefont
  {Pagonabarraga}}, \ and\ \bibinfo {author} {\bibfnamefont {J.}~\bibnamefont
  {Rubi}},\ }\href@noop {} {\bibfield  {journal} {\bibinfo  {journal} {Front.
  Phys.}\ }\textbf {\bibinfo {volume} {1}},\ \bibinfo {pages} {21} (\bibinfo
  {year} {2013})}\BibitemShut {NoStop}%
\bibitem [{\citenamefont {Malgaretti}\ and\ \citenamefont
  {Harting}(2023)}]{Malgaretti2023}%
  \BibitemOpen
  \bibfield  {author} {\bibinfo {author} {\bibfnamefont {P.}~\bibnamefont
  {Malgaretti}}\ and\ \bibinfo {author} {\bibfnamefont {J.}~\bibnamefont
  {Harting}},\ }\href {\doibase 10.3390/e25030470} {\bibfield  {journal}
  {\bibinfo  {journal} {Entropy}\ }\textbf {\bibinfo {volume} {25}},\ \bibinfo
  {pages} {470} (\bibinfo {year} {2023})}\BibitemShut {NoStop}%
\bibitem [{\citenamefont {Thiele}(2011)}]{thiele_note_nodate}%
  \BibitemOpen
  \bibfield  {author} {\bibinfo {author} {\bibfnamefont {U.}~\bibnamefont
  {Thiele}},\ }\href@noop {} {\bibfield  {journal} {\bibinfo  {journal} {Eur.
  Phys. J. Spec. Top.}\ }\textbf {\bibinfo {volume} {197}},\ \bibinfo {pages}
  {213} (\bibinfo {year} {2011})}\BibitemShut {NoStop}%
\bibitem [{\citenamefont {Thiele}\ \emph {et~al.}(2013)\citenamefont {Thiele},
  \citenamefont {Todorova},\ and\ \citenamefont
  {Lopez}}]{thiele_gradient_2013}%
  \BibitemOpen
  \bibfield  {author} {\bibinfo {author} {\bibfnamefont {U.}~\bibnamefont
  {Thiele}}, \bibinfo {author} {\bibfnamefont {D.~V.}\ \bibnamefont
  {Todorova}}, \ and\ \bibinfo {author} {\bibfnamefont {H.}~\bibnamefont
  {Lopez}},\ }\href {\doibase 10.1103/PhysRevLett.111.117801} {\bibfield
  {journal} {\bibinfo  {journal} {Phys. Rev. Lett.}\ }\textbf {\bibinfo
  {volume} {111}},\ \bibinfo {pages} {117801} (\bibinfo {year}
  {2013})}\BibitemShut {NoStop}%
\bibitem [{\citenamefont {Xu}\ \emph {et~al.}(2015)\citenamefont {Xu},
  \citenamefont {Thiele},\ and\ \citenamefont {Qian}}]{xu_variational_2015}%
  \BibitemOpen
  \bibfield  {author} {\bibinfo {author} {\bibfnamefont {X.}~\bibnamefont
  {Xu}}, \bibinfo {author} {\bibfnamefont {U.}~\bibnamefont {Thiele}}, \ and\
  \bibinfo {author} {\bibfnamefont {T.}~\bibnamefont {Qian}},\ }\href {\doibase
  10.1088/0953-8984/27/8/085005} {\bibfield  {journal} {\bibinfo  {journal}
  {Journal of Physics: Condensed Matter}\ }\textbf {\bibinfo {volume} {27}},\
  \bibinfo {pages} {085005} (\bibinfo {year} {2015})}\BibitemShut {NoStop}%
\bibitem [{\citenamefont {Steinrück}\ \emph {et~al.}(2011)\citenamefont
  {Steinrück}, \citenamefont {Libuda}, \citenamefont {Wasserscheid},
  \citenamefont {Cremer}, \citenamefont {Kolbeck}, \citenamefont {Laurin},
  \citenamefont {Maier}, \citenamefont {Sobota}, \citenamefont {Schulz},\ and\
  \citenamefont {Stark}}]{steinruck_surface_2011}%
  \BibitemOpen
  \bibfield  {author} {\bibinfo {author} {\bibfnamefont {H.-P.}\ \bibnamefont
  {Steinrück}}, \bibinfo {author} {\bibfnamefont {J.}~\bibnamefont {Libuda}},
  \bibinfo {author} {\bibfnamefont {P.}~\bibnamefont {Wasserscheid}}, \bibinfo
  {author} {\bibfnamefont {T.}~\bibnamefont {Cremer}}, \bibinfo {author}
  {\bibfnamefont {C.}~\bibnamefont {Kolbeck}}, \bibinfo {author} {\bibfnamefont
  {M.}~\bibnamefont {Laurin}}, \bibinfo {author} {\bibfnamefont
  {F.}~\bibnamefont {Maier}}, \bibinfo {author} {\bibfnamefont
  {M.}~\bibnamefont {Sobota}}, \bibinfo {author} {\bibfnamefont {P.~S.}\
  \bibnamefont {Schulz}}, \ and\ \bibinfo {author} {\bibfnamefont
  {M.}~\bibnamefont {Stark}},\ }\href {\doibase 10.1002/adma.201100211}
  {\bibfield  {journal} {\bibinfo  {journal} {Advanced Materials}\ }\textbf
  {\bibinfo {volume} {23}},\ \bibinfo {pages} {2571} (\bibinfo {year}
  {2011})}\BibitemShut {NoStop}%
\bibitem [{\citenamefont {Shylesh}\ \emph {et~al.}(2013)\citenamefont
  {Shylesh}, \citenamefont {Hanna}, \citenamefont {Mlinar}, \citenamefont
  {Kong}, \citenamefont {Reimer},\ and\ \citenamefont
  {Bell}}]{shylesh_situ_2013}%
  \BibitemOpen
  \bibfield  {author} {\bibinfo {author} {\bibfnamefont {S.}~\bibnamefont
  {Shylesh}}, \bibinfo {author} {\bibfnamefont {D.}~\bibnamefont {Hanna}},
  \bibinfo {author} {\bibfnamefont {A.}~\bibnamefont {Mlinar}}, \bibinfo
  {author} {\bibfnamefont {X.-Q.}\ \bibnamefont {Kong}}, \bibinfo {author}
  {\bibfnamefont {J.~A.}\ \bibnamefont {Reimer}}, \ and\ \bibinfo {author}
  {\bibfnamefont {A.~T.}\ \bibnamefont {Bell}},\ }\href {\doibase
  10.1021/cs3007445} {\bibfield  {journal} {\bibinfo  {journal} {ACS Catal.}\
  }\textbf {\bibinfo {volume} {3}},\ \bibinfo {pages} {348} (\bibinfo {year}
  {2013})}\BibitemShut {NoStop}%
\bibitem [{\citenamefont {Vrij}(1966)}]{vrij_possible_1966}%
  \BibitemOpen
  \bibfield  {author} {\bibinfo {author} {\bibfnamefont {A.}~\bibnamefont
  {Vrij}},\ }\href {\doibase 10.1039/df9664200023} {\bibfield  {journal}
  {\bibinfo  {journal} {Discuss. Faraday Soc.}\ }\textbf {\bibinfo {volume}
  {42}},\ \bibinfo {pages} {23} (\bibinfo {year} {1966})}\BibitemShut {NoStop}%
\bibitem [{\citenamefont {Doi}(2013)}]{doi_soft_2013}%
  \BibitemOpen
  \bibfield  {author} {\bibinfo {author} {\bibfnamefont {M.}~\bibnamefont
  {Doi}},\ }\href@noop {} {\emph {\bibinfo {title} {Soft matter physics}}}\
  (\bibinfo  {publisher} {Oxford University Press, USA},\ \bibinfo {year}
  {2013})\BibitemShut {NoStop}%
\bibitem [{\citenamefont {Rauscher}\ \emph {et~al.}(2008)\citenamefont
  {Rauscher}, \citenamefont {Blossey}, \citenamefont {Münch},\ and\
  \citenamefont {Wagner}}]{rauscher_spinodal_2008}%
  \BibitemOpen
  \bibfield  {author} {\bibinfo {author} {\bibfnamefont {M.}~\bibnamefont
  {Rauscher}}, \bibinfo {author} {\bibfnamefont {R.}~\bibnamefont {Blossey}},
  \bibinfo {author} {\bibfnamefont {A.}~\bibnamefont {Münch}}, \ and\ \bibinfo
  {author} {\bibfnamefont {B.}~\bibnamefont {Wagner}},\ }\href {\doibase
  10.1021/la802260b} {\bibfield  {journal} {\bibinfo  {journal} {Langmuir}\
  }\textbf {\bibinfo {volume} {24}},\ \bibinfo {pages} {12290} (\bibinfo {year}
  {2008})}\BibitemShut {NoStop}%
\bibitem [{\citenamefont {Fetzer}\ \emph {et~al.}(2007)\citenamefont {Fetzer},
  \citenamefont {Rauscher}, \citenamefont {Seemann}, \citenamefont {Jacobs},\
  and\ \citenamefont {Mecke}}]{fetzer_thermal_2007}%
  \BibitemOpen
  \bibfield  {author} {\bibinfo {author} {\bibfnamefont {R.}~\bibnamefont
  {Fetzer}}, \bibinfo {author} {\bibfnamefont {M.}~\bibnamefont {Rauscher}},
  \bibinfo {author} {\bibfnamefont {R.}~\bibnamefont {Seemann}}, \bibinfo
  {author} {\bibfnamefont {K.}~\bibnamefont {Jacobs}}, \ and\ \bibinfo {author}
  {\bibfnamefont {K.}~\bibnamefont {Mecke}},\ }\href {\doibase
  10.1103/PhysRevLett.99.114503} {\bibfield  {journal} {\bibinfo  {journal}
  {Phys. Rev. Lett.}\ }\textbf {\bibinfo {volume} {99}},\ \bibinfo {pages}
  {114503} (\bibinfo {year} {2007})}\BibitemShut {NoStop}%
\bibitem [{\citenamefont {Zitz}\ \emph {et~al.}(2021)\citenamefont {Zitz},
  \citenamefont {Scagliarini},\ and\ \citenamefont
  {Harting}}]{zitz_lattice_2021}%
  \BibitemOpen
  \bibfield  {author} {\bibinfo {author} {\bibfnamefont {S.}~\bibnamefont
  {Zitz}}, \bibinfo {author} {\bibfnamefont {A.}~\bibnamefont {Scagliarini}}, \
  and\ \bibinfo {author} {\bibfnamefont {J.}~\bibnamefont {Harting}},\ }\href
  {\doibase 10.1103/PhysRevE.104.034801} {\bibfield  {journal} {\bibinfo
  {journal} {Phys. Rev. E}\ }\textbf {\bibinfo {volume} {104}},\ \bibinfo
  {pages} {034801} (\bibinfo {year} {2021})}\BibitemShut {NoStop}%
\bibitem [{\citenamefont {Pelusi}\ \emph
  {et~al.}(2022{\natexlab{a}})\citenamefont {Pelusi}, \citenamefont {Sega},\
  and\ \citenamefont {Harting}}]{PSH22}%
  \BibitemOpen
  \bibfield  {author} {\bibinfo {author} {\bibfnamefont {F.}~\bibnamefont
  {Pelusi}}, \bibinfo {author} {\bibfnamefont {M.}~\bibnamefont {Sega}}, \ and\
  \bibinfo {author} {\bibfnamefont {J.}~\bibnamefont {Harting}},\ }\href
  {\doibase 10.1063/5.0093043} {\bibfield  {journal} {\bibinfo  {journal}
  {Phys. Fluids}\ }\textbf {\bibinfo {volume} {34}},\ \bibinfo {pages} {062109}
  (\bibinfo {year} {2022}{\natexlab{a}})}\BibitemShut {NoStop}%
\bibitem [{\citenamefont {Riisager}\ \emph {et~al.}(2006)\citenamefont
  {Riisager}, \citenamefont {Fehrmann}, \citenamefont {Haumann},\ and\
  \citenamefont {Wasserscheid}}]{riisagera_supported_2006}%
  \BibitemOpen
  \bibfield  {author} {\bibinfo {author} {\bibfnamefont {A.}~\bibnamefont
  {Riisager}}, \bibinfo {author} {\bibfnamefont {R.}~\bibnamefont {Fehrmann}},
  \bibinfo {author} {\bibfnamefont {M.}~\bibnamefont {Haumann}}, \ and\
  \bibinfo {author} {\bibfnamefont {P.}~\bibnamefont {Wasserscheid}},\ }\href
  {\doibase 10.1007/s11244-006-0111-9} {\bibfield  {journal} {\bibinfo
  {journal} {Topics in Catalysis}\ }\textbf {\bibinfo {volume} {40}},\ \bibinfo
  {pages} {91} (\bibinfo {year} {2006})}\BibitemShut {NoStop}%
\bibitem [{\citenamefont {De~Gennes}\ \emph {et~al.}(2004)\citenamefont
  {De~Gennes}, \citenamefont {Brochard-Wyart},\ and\ \citenamefont
  {Quéré}}]{de_gennes_capillarity_2004}%
  \BibitemOpen
  \bibfield  {author} {\bibinfo {author} {\bibfnamefont {P.-G.}\ \bibnamefont
  {De~Gennes}}, \bibinfo {author} {\bibfnamefont {F.}~\bibnamefont
  {Brochard-Wyart}}, \ and\ \bibinfo {author} {\bibfnamefont {D.}~\bibnamefont
  {Quéré}},\ }\href {\doibase 10.1007/978-0-387-21656-0} {{\selectlanguage
  {en}\emph {\bibinfo {title} {Capillarity and {Wetting} {Phenomena}}}}}\
  (\bibinfo  {publisher} {Springer New York},\ \bibinfo {address} {New York,
  NY},\ \bibinfo {year} {2004})\BibitemShut {NoStop}%
\bibitem [{\citenamefont {Herrero}\ and\ \citenamefont
  {Vel{\'a}zquez}(1996)}]{herrero_chemotactic_nodate}%
  \BibitemOpen
  \bibfield  {author} {\bibinfo {author} {\bibfnamefont {M.~A.}\ \bibnamefont
  {Herrero}}\ and\ \bibinfo {author} {\bibfnamefont {J.~J.}\ \bibnamefont
  {Vel{\'a}zquez}},\ }\href@noop {} {\bibfield  {journal} {\bibinfo  {journal}
  {JOMB}\ }\textbf {\bibinfo {volume} {35}},\ \bibinfo {pages} {177} (\bibinfo
  {year} {1996})}\BibitemShut {NoStop}%
\bibitem [{Note1()}]{Note1}%
  \BibitemOpen
  \bibinfo {note} {The spherical cap is the expected equilibrium shape of a
  liquid droplet with a radius smaller than the capillary length $r_c=\protect
  \sqrt {\gamma _0/(\rho _l g)}$, where $g$ is the gravitational constant~\cite
  {doi_soft_2013}.}\BibitemShut {Stop}%
\bibitem [{Note2()}]{Note2}%
  \BibitemOpen
  \bibinfo {note} {In Fig.~\ref {fig:volatilities} a right moving pattern is
  shown but left moving patterns can be observed as well.}\BibitemShut {Stop}%
\bibitem [{\citenamefont {Hemmeter}\ \emph
  {et~al.}(2023{\natexlab{a}})\citenamefont {Hemmeter}, \citenamefont
  {Kremitzl}, \citenamefont {Schulz}, \citenamefont {Wasserscheid},
  \citenamefont {Maier},\ and\ \citenamefont
  {Steinrück}}]{hemmeter_buoy_2023}%
  \BibitemOpen
  \bibfield  {author} {\bibinfo {author} {\bibfnamefont {D.}~\bibnamefont
  {Hemmeter}}, \bibinfo {author} {\bibfnamefont {D.}~\bibnamefont {Kremitzl}},
  \bibinfo {author} {\bibfnamefont {P.~S.}\ \bibnamefont {Schulz}}, \bibinfo
  {author} {\bibfnamefont {P.}~\bibnamefont {Wasserscheid}}, \bibinfo {author}
  {\bibfnamefont {F.}~\bibnamefont {Maier}}, \ and\ \bibinfo {author}
  {\bibfnamefont {H.}~\bibnamefont {Steinrück}},\ }\href {\doibase
  10.1002/chem.202203325} {\bibfield  {journal} {\bibinfo  {journal} {Chemistry
  - A European Journal}\ }\textbf {\bibinfo {volume} {29}},\ \bibinfo {pages}
  {e202203325} (\bibinfo {year} {2023}{\natexlab{a}})}\BibitemShut {NoStop}%
\bibitem [{\citenamefont {Hemmeter}\ \emph
  {et~al.}(2023{\natexlab{b}})\citenamefont {Hemmeter}, \citenamefont {Paap},
  \citenamefont {Maier},\ and\ \citenamefont
  {Steinrück}}]{hemmeter_structure_2023}%
  \BibitemOpen
  \bibfield  {author} {\bibinfo {author} {\bibfnamefont {D.}~\bibnamefont
  {Hemmeter}}, \bibinfo {author} {\bibfnamefont {U.}~\bibnamefont {Paap}},
  \bibinfo {author} {\bibfnamefont {F.}~\bibnamefont {Maier}}, \ and\ \bibinfo
  {author} {\bibfnamefont {H.-P.}\ \bibnamefont {Steinrück}},\ }\href
  {\doibase 10.3390/catal13050871} {\bibfield  {journal} {\bibinfo  {journal}
  {Catalysts}\ }\textbf {\bibinfo {volume} {13}},\ \bibinfo {pages} {871}
  (\bibinfo {year} {2023}{\natexlab{b}})}\BibitemShut {NoStop}%
\bibitem [{\citenamefont {Bhatt}\ \emph {et~al.}(2023)\citenamefont {Bhatt},
  \citenamefont {Gupta}, \citenamefont {Sumathi}, \citenamefont {Chandran},\
  and\ \citenamefont {Khare}}]{bhatt_electric_2023}%
  \BibitemOpen
  \bibfield  {author} {\bibinfo {author} {\bibfnamefont {B.}~\bibnamefont
  {Bhatt}}, \bibinfo {author} {\bibfnamefont {S.}~\bibnamefont {Gupta}},
  \bibinfo {author} {\bibfnamefont {V.}~\bibnamefont {Sumathi}}, \bibinfo
  {author} {\bibfnamefont {S.}~\bibnamefont {Chandran}}, \ and\ \bibinfo
  {author} {\bibfnamefont {K.}~\bibnamefont {Khare}},\ }\href {\doibase
  10.1002/admi.202202063} {\bibfield  {journal} {\bibinfo  {journal} {Advanced
  Materials Interfaces}\ }\textbf {\bibinfo {volume} {10}},\ \bibinfo {pages}
  {2202063} (\bibinfo {year} {2023})},\ \bibinfo {note} {\_eprint:
  https://onlinelibrary.wiley.com/doi/pdf/10.1002/admi.202202063}\BibitemShut
  {NoStop}%
\bibitem [{Note3()}]{Note3}%
  \BibitemOpen
  \bibinfo {note} {Ref.~\cite {bhatt_electric_2023} shows that applying a
  voltage of up to $60 V$ disjoining pressure becomes relevant for a film of
  thickness as large as $10 \mu m$. In such a scenario with a micrometer-thick
  film being prone to spinodal rupture it is possible to have catalytic
  particles that can be easily tracked by optical methods making it possible to
  observe the dynamics of the catalyst distribution experimentally while having
  a relevant effect from the disjoining pressure augmented by the electrical
  potential. In Ref.~\cite {john_ratchet-driven_2008} the authors show how to
  include such effects into the lubrication-approximation
  approach.}\BibitemShut {Stop}%
\bibitem [{\citenamefont {Zitz}\ \emph {et~al.}(2019)\citenamefont {Zitz},
  \citenamefont {Scagliarini}, \citenamefont {Maddu}, \citenamefont
  {Darhuber},\ and\ \citenamefont {Harting}}]{zitz_lattice_2019}%
  \BibitemOpen
  \bibfield  {author} {\bibinfo {author} {\bibfnamefont {S.}~\bibnamefont
  {Zitz}}, \bibinfo {author} {\bibfnamefont {A.}~\bibnamefont {Scagliarini}},
  \bibinfo {author} {\bibfnamefont {S.}~\bibnamefont {Maddu}}, \bibinfo
  {author} {\bibfnamefont {A.~A.}\ \bibnamefont {Darhuber}}, \ and\ \bibinfo
  {author} {\bibfnamefont {J.}~\bibnamefont {Harting}},\ }\href {\doibase
  10.1103/PhysRevE.100.033313} {\bibfield  {journal} {\bibinfo  {journal}
  {Phys. Rev. E}\ }\textbf {\bibinfo {volume} {100}},\ \bibinfo {pages}
  {033313} (\bibinfo {year} {2019})}\BibitemShut {NoStop}%
\bibitem [{\citenamefont {Zitz}\ \emph {et~al.}(2023)\citenamefont {Zitz},
  \citenamefont {Scagliarini},\ and\ \citenamefont
  {Harting}}]{zitz_controlling_2021}%
  \BibitemOpen
  \bibfield  {author} {\bibinfo {author} {\bibfnamefont {S.}~\bibnamefont
  {Zitz}}, \bibinfo {author} {\bibfnamefont {A.}~\bibnamefont {Scagliarini}}, \
  and\ \bibinfo {author} {\bibfnamefont {J.}~\bibnamefont {Harting}},\ }\href
  {\doibase 10.1103/PhysRevFluids.8.L122001} {\bibfield  {journal} {\bibinfo
  {journal} {Phys. Rev. Fluids}\ }\textbf {\bibinfo {volume} {8}},\ \bibinfo
  {pages} {L122001} (\bibinfo {year} {2023})}\BibitemShut {NoStop}%
\bibitem [{\citenamefont {T.}(2014)}]{water_surface_tension}%
  \BibitemOpen
  \bibfield  {author} {\bibinfo {author} {\bibfnamefont {P.}~\bibnamefont
  {T.}},\ }\href {http://www.iapws.org/relguide/Surf-H2O.html} {\bibfield
  {journal} {\bibinfo  {journal} {The International Association for the
  Properties of Water and Steam}\ } (\bibinfo {year} {2014})}\BibitemShut
  {NoStop}%
\bibitem [{\citenamefont {Wagner}\ and\ \citenamefont
  {Pruß}(2002)}]{wagner_iapws_2002}%
  \BibitemOpen
  \bibfield  {author} {\bibinfo {author} {\bibfnamefont {W.}~\bibnamefont
  {Wagner}}\ and\ \bibinfo {author} {\bibfnamefont {A.}~\bibnamefont {Pruß}},\
  }\href {\doibase 10.1063/1.1461829} {\bibfield  {journal} {\bibinfo
  {journal} {Journal of Physical and Chemical Reference Data}\ }\textbf
  {\bibinfo {volume} {31}},\ \bibinfo {pages} {387} (\bibinfo {year}
  {2002})}\BibitemShut {NoStop}%
\bibitem [{\citenamefont {Huber}\ \emph {et~al.}(2009)\citenamefont {Huber},
  \citenamefont {Perkins}, \citenamefont {Laesecke}, \citenamefont {Friend},
  \citenamefont {Sengers}, \citenamefont {Assael}, \citenamefont {Metaxa},
  \citenamefont {Vogel}, \citenamefont {Mare{\v{s}}},\ and\ \citenamefont
  {Miyagawa}}]{huber2009new}%
  \BibitemOpen
  \bibfield  {author} {\bibinfo {author} {\bibfnamefont {M.~L.}\ \bibnamefont
  {Huber}}, \bibinfo {author} {\bibfnamefont {R.~A.}\ \bibnamefont {Perkins}},
  \bibinfo {author} {\bibfnamefont {A.}~\bibnamefont {Laesecke}}, \bibinfo
  {author} {\bibfnamefont {D.~G.}\ \bibnamefont {Friend}}, \bibinfo {author}
  {\bibfnamefont {J.~V.}\ \bibnamefont {Sengers}}, \bibinfo {author}
  {\bibfnamefont {M.~J.}\ \bibnamefont {Assael}}, \bibinfo {author}
  {\bibfnamefont {I.~N.}\ \bibnamefont {Metaxa}}, \bibinfo {author}
  {\bibfnamefont {E.}~\bibnamefont {Vogel}}, \bibinfo {author} {\bibfnamefont
  {R.}~\bibnamefont {Mare{\v{s}}}}, \ and\ \bibinfo {author} {\bibfnamefont
  {K.}~\bibnamefont {Miyagawa}},\ }\href@noop {} {\bibfield  {journal}
  {\bibinfo  {journal} {Journal of Physical and Chemical Reference Data}\
  }\textbf {\bibinfo {volume} {38}},\ \bibinfo {pages} {101} (\bibinfo {year}
  {2009})}\BibitemShut {NoStop}%
\bibitem [{\citenamefont {Brown}\ and\ \citenamefont
  {Poon}(2014)}]{brown_ionic_2014}%
  \BibitemOpen
  \bibfield  {author} {\bibinfo {author} {\bibfnamefont {A.}~\bibnamefont
  {Brown}}\ and\ \bibinfo {author} {\bibfnamefont {W.}~\bibnamefont {Poon}},\
  }\href {\doibase 10.1039/C4SM00340C} {\bibfield  {journal} {\bibinfo
  {journal} {Soft Matter}\ }\textbf {\bibinfo {volume} {10}},\ \bibinfo {pages}
  {4016} (\bibinfo {year} {2014})}\BibitemShut {NoStop}%
\bibitem [{\citenamefont {John}\ \emph {et~al.}(2008)\citenamefont {John},
  \citenamefont {H\"anggi},\ and\ \citenamefont
  {Thiele}}]{john_ratchet-driven_2008}%
  \BibitemOpen
  \bibfield  {author} {\bibinfo {author} {\bibfnamefont {K.}~\bibnamefont
  {John}}, \bibinfo {author} {\bibfnamefont {P.}~\bibnamefont {H\"anggi}}, \
  and\ \bibinfo {author} {\bibfnamefont {U.}~\bibnamefont {Thiele}},\ }\href
  {\doibase 10.1039/b718850a} {\bibfield  {journal} {\bibinfo  {journal} {Soft
  Matter}\ }\textbf {\bibinfo {volume} {4}},\ \bibinfo {pages} {1183} (\bibinfo
  {year} {2008})}\BibitemShut {NoStop}%
\bibitem [{\citenamefont {Popescu}\ \emph {et~al.}(2012)\citenamefont
  {Popescu}, \citenamefont {Oshanin}, \citenamefont {Dietrich},\ and\
  \citenamefont {Cazabat}}]{popescu_precursor_2012}%
  \BibitemOpen
  \bibfield  {author} {\bibinfo {author} {\bibfnamefont {M.~N.}\ \bibnamefont
  {Popescu}}, \bibinfo {author} {\bibfnamefont {G.}~\bibnamefont {Oshanin}},
  \bibinfo {author} {\bibfnamefont {S.}~\bibnamefont {Dietrich}}, \ and\
  \bibinfo {author} {\bibfnamefont {A.-M.}\ \bibnamefont {Cazabat}},\ }\href
  {\doibase 10.1088/0953-8984/24/24/243102} {\bibfield  {journal} {\bibinfo
  {journal} {Journal of Physics: Condensed Matter}\ }\textbf {\bibinfo {volume}
  {24}},\ \bibinfo {pages} {243102} (\bibinfo {year} {2012})}\BibitemShut
  {NoStop}%
\bibitem [{\citenamefont {Pelusi}\ \emph
  {et~al.}(2022{\natexlab{b}})\citenamefont {Pelusi}, \citenamefont {Sega},\
  and\ \citenamefont {Harting}}]{pelusi_liquid_2022}%
  \BibitemOpen
  \bibfield  {author} {\bibinfo {author} {\bibfnamefont {F.}~\bibnamefont
  {Pelusi}}, \bibinfo {author} {\bibfnamefont {M.}~\bibnamefont {Sega}}, \ and\
  \bibinfo {author} {\bibfnamefont {J.}~\bibnamefont {Harting}},\ }\href
  {\doibase 10.1063/5.0093043} {\bibfield  {journal} {\bibinfo  {journal}
  {Phys. Fluids}\ }\textbf {\bibinfo {volume} {34}},\ \bibinfo {pages} {062109}
  (\bibinfo {year} {2022}{\natexlab{b}})}\BibitemShut {NoStop}%
\bibitem [{\citenamefont {Peschka}\ \emph {et~al.}(2019)\citenamefont
  {Peschka}, \citenamefont {Haefner}, \citenamefont {Marquant}, \citenamefont
  {Jacobs}, \citenamefont {Münch},\ and\ \citenamefont
  {Wagner}}]{peschka_signatures_2019}%
  \BibitemOpen
  \bibfield  {author} {\bibinfo {author} {\bibfnamefont {D.}~\bibnamefont
  {Peschka}}, \bibinfo {author} {\bibfnamefont {S.}~\bibnamefont {Haefner}},
  \bibinfo {author} {\bibfnamefont {L.}~\bibnamefont {Marquant}}, \bibinfo
  {author} {\bibfnamefont {K.}~\bibnamefont {Jacobs}}, \bibinfo {author}
  {\bibfnamefont {A.}~\bibnamefont {Münch}}, \ and\ \bibinfo {author}
  {\bibfnamefont {B.}~\bibnamefont {Wagner}},\ }\href {\doibase
  10.1073/pnas.1820487116} {\bibfield  {journal} {\bibinfo  {journal}
  {Proceedings of the National Academy of Sciences}\ }\textbf {\bibinfo
  {volume} {116}},\ \bibinfo {pages} {9275} (\bibinfo {year}
  {2019})}\BibitemShut {NoStop}%
\bibitem [{\citenamefont {Fetzer}\ and\ \citenamefont
  {Jacobs}(2007)}]{fetzer_slippage_2007}%
  \BibitemOpen
  \bibfield  {author} {\bibinfo {author} {\bibfnamefont {R.}~\bibnamefont
  {Fetzer}}\ and\ \bibinfo {author} {\bibfnamefont {K.}~\bibnamefont
  {Jacobs}},\ }\href {\doibase 10.1021/la701746r} {\bibfield  {journal}
  {\bibinfo  {journal} {Langmuir}\ }\textbf {\bibinfo {volume} {23}},\ \bibinfo
  {pages} {11617} (\bibinfo {year} {2007})}\BibitemShut {NoStop}%
\bibitem [{\citenamefont {Reiter}(2001)}]{reiter_dewetting_2001}%
  \BibitemOpen
  \bibfield  {author} {\bibinfo {author} {\bibfnamefont {G.}~\bibnamefont
  {Reiter}},\ }\href {\doibase 10.1103/PhysRevLett.87.186101} {\bibfield
  {journal} {\bibinfo  {journal} {Phys. Rev. Lett.}\ }\textbf {\bibinfo
  {volume} {87}},\ \bibinfo {pages} {186101} (\bibinfo {year}
  {2001})}\BibitemShut {NoStop}%
\bibitem [{\citenamefont {Pototsky}\ \emph {et~al.}(2005)\citenamefont
  {Pototsky}, \citenamefont {Bestehorn}, \citenamefont {Merkt},\ and\
  \citenamefont {Thiele}}]{pototsky_morphology_2005}%
  \BibitemOpen
  \bibfield  {author} {\bibinfo {author} {\bibfnamefont {A.}~\bibnamefont
  {Pototsky}}, \bibinfo {author} {\bibfnamefont {M.}~\bibnamefont {Bestehorn}},
  \bibinfo {author} {\bibfnamefont {D.}~\bibnamefont {Merkt}}, \ and\ \bibinfo
  {author} {\bibfnamefont {U.}~\bibnamefont {Thiele}},\ }\href {\doibase
  10.1063/1.1927512} {\bibfield  {journal} {\bibinfo  {journal} {J. Chem.
  Phys.}\ }\textbf {\bibinfo {volume} {122}},\ \bibinfo {pages} {224711}
  (\bibinfo {year} {2005})}\BibitemShut {NoStop}%
\bibitem [{\citenamefont {Landau}\ and\ \citenamefont
  {Lifshitz}(2013)}]{landau2013fluid}%
  \BibitemOpen
  \bibfield  {author} {\bibinfo {author} {\bibfnamefont {L.~D.}\ \bibnamefont
  {Landau}}\ and\ \bibinfo {author} {\bibfnamefont {E.~M.}\ \bibnamefont
  {Lifshitz}},\ }\href@noop {} {\emph {\bibinfo {title} {Fluid Mechanics:
  Landau and Lifshitz: Course of Theoretical Physics, Volume 6}}},\
  Vol.~\bibinfo {volume} {6}\ (\bibinfo  {publisher} {Elsevier},\ \bibinfo
  {year} {2013})\BibitemShut {NoStop}%
\bibitem [{\citenamefont {Hack}\ \emph {et~al.}(2020)\citenamefont {Hack},
  \citenamefont {Tewes}, \citenamefont {Xie}, \citenamefont {Datt},
  \citenamefont {Harth}, \citenamefont {Harting},\ and\ \citenamefont
  {Snoeijer}}]{hack_self-similar_2020}%
  \BibitemOpen
  \bibfield  {author} {\bibinfo {author} {\bibfnamefont {M.~A.}\ \bibnamefont
  {Hack}}, \bibinfo {author} {\bibfnamefont {W.}~\bibnamefont {Tewes}},
  \bibinfo {author} {\bibfnamefont {Q.}~\bibnamefont {Xie}}, \bibinfo {author}
  {\bibfnamefont {C.}~\bibnamefont {Datt}}, \bibinfo {author} {\bibfnamefont
  {K.}~\bibnamefont {Harth}}, \bibinfo {author} {\bibfnamefont
  {J.}~\bibnamefont {Harting}}, \ and\ \bibinfo {author} {\bibfnamefont
  {J.~H.}\ \bibnamefont {Snoeijer}},\ }\href {\doibase
  10.1103/PhysRevLett.124.194502} {\bibfield  {journal} {\bibinfo  {journal}
  {Phys. Rev. Lett.}\ }\textbf {\bibinfo {volume} {124}},\ \bibinfo {pages}
  {194502} (\bibinfo {year} {2020})}\BibitemShut {NoStop}%
\bibitem [{\citenamefont {Thiele}\ \emph {et~al.}(2012)\citenamefont {Thiele},
  \citenamefont {Archer},\ and\ \citenamefont
  {Plapp}}]{thiele_thermodynamically_2012}%
  \BibitemOpen
  \bibfield  {author} {\bibinfo {author} {\bibfnamefont {U.}~\bibnamefont
  {Thiele}}, \bibinfo {author} {\bibfnamefont {A.~J.}\ \bibnamefont {Archer}},
  \ and\ \bibinfo {author} {\bibfnamefont {M.}~\bibnamefont {Plapp}},\ }\href
  {\doibase 10.1063/1.4758476} {\bibfield  {journal} {\bibinfo  {journal}
  {Phys. Fluids}\ }\textbf {\bibinfo {volume} {24}},\ \bibinfo {pages} {102107}
  (\bibinfo {year} {2012})}\BibitemShut {NoStop}%
\bibitem [{\citenamefont {Thiele}\ \emph {et~al.}(2016)\citenamefont {Thiele},
  \citenamefont {Archer},\ and\ \citenamefont {Pismen}}]{thiele_gradient_2016}%
  \BibitemOpen
  \bibfield  {author} {\bibinfo {author} {\bibfnamefont {U.}~\bibnamefont
  {Thiele}}, \bibinfo {author} {\bibfnamefont {A.~J.}\ \bibnamefont {Archer}},
  \ and\ \bibinfo {author} {\bibfnamefont {L.~M.}\ \bibnamefont {Pismen}},\
  }\href {\doibase 10.1103/PhysRevFluids.1.083903} {\bibfield  {journal}
  {\bibinfo  {journal} {Phys. Rev. Fluids}\ }\textbf {\bibinfo {volume} {1}},\
  \bibinfo {pages} {083903} (\bibinfo {year} {2016})}\BibitemShut {NoStop}%
\bibitem [{\citenamefont {Trinschek}\ \emph {et~al.}(2018)\citenamefont
  {Trinschek}, \citenamefont {John},\ and\ \citenamefont
  {Thiele}}]{trinschek_modelling_2018}%
  \BibitemOpen
  \bibfield  {author} {\bibinfo {author} {\bibfnamefont {S.}~\bibnamefont
  {Trinschek}}, \bibinfo {author} {\bibfnamefont {K.}~\bibnamefont {John}}, \
  and\ \bibinfo {author} {\bibfnamefont {U.}~\bibnamefont {Thiele}},\ }\href
  {\doibase 10.1039/c8sm00422f} {\bibfield  {journal} {\bibinfo  {journal}
  {Soft Matter}\ }\textbf {\bibinfo {volume} {14}},\ \bibinfo {pages} {4464}
  (\bibinfo {year} {2018})},\ \bibinfo {note} {arXiv:1803.00522 [cond-mat,
  physics:physics]}\BibitemShut {NoStop}%
\bibitem [{\citenamefont {Krüger}\ \emph {et~al.}(2017)\citenamefont
  {Krüger}, \citenamefont {Kusumaatmaja}, \citenamefont {Kuzmin},
  \citenamefont {Shardt}, \citenamefont {Silva},\ and\ \citenamefont
  {Viggen}}]{kruger_lattice_2017}%
  \BibitemOpen
  \bibfield  {author} {\bibinfo {author} {\bibfnamefont {T.}~\bibnamefont
  {Krüger}}, \bibinfo {author} {\bibfnamefont {H.}~\bibnamefont
  {Kusumaatmaja}}, \bibinfo {author} {\bibfnamefont {A.}~\bibnamefont
  {Kuzmin}}, \bibinfo {author} {\bibfnamefont {O.}~\bibnamefont {Shardt}},
  \bibinfo {author} {\bibfnamefont {G.}~\bibnamefont {Silva}}, \ and\ \bibinfo
  {author} {\bibfnamefont {E.~M.}\ \bibnamefont {Viggen}},\ }\href {\doibase
  10.1007/978-3-319-44649-3} {{\selectlanguage {en}\emph {\bibinfo {title} {The
  {Lattice} {Boltzmann} {Method}: {Principles} and {Practice}}}}},\ Graduate
  {Texts} in {Physics}\ (\bibinfo  {publisher} {Springer International
  Publishing},\ \bibinfo {address} {Cham},\ \bibinfo {year} {2017})\BibitemShut
  {NoStop}%
\bibitem [{\citenamefont {Baumgartner}\ \emph {et~al.}(2022)\citenamefont
  {Baumgartner}, \citenamefont {Shiri}, \citenamefont {Sinha}, \citenamefont
  {Karpitschka},\ and\ \citenamefont {Cira}}]{baumgartner_marangoni_2022}%
  \BibitemOpen
  \bibfield  {author} {\bibinfo {author} {\bibfnamefont {D.~A.}\ \bibnamefont
  {Baumgartner}}, \bibinfo {author} {\bibfnamefont {S.}~\bibnamefont {Shiri}},
  \bibinfo {author} {\bibfnamefont {S.}~\bibnamefont {Sinha}}, \bibinfo
  {author} {\bibfnamefont {S.}~\bibnamefont {Karpitschka}}, \ and\ \bibinfo
  {author} {\bibfnamefont {N.~J.}\ \bibnamefont {Cira}},\ }\href {\doibase
  10.1073/pnas.2120432119} {\bibfield  {journal} {\bibinfo  {journal}
  {Proceedings of the National Academy of Sciences}\ }\textbf {\bibinfo
  {volume} {119}},\ \bibinfo {pages} {e2120432119} (\bibinfo {year}
  {2022})}\BibitemShut {NoStop}%
\bibitem [{\citenamefont {Karpitschka}\ and\ \citenamefont
  {Riegler}(2014)}]{karpitschka_sharp_2014}%
  \BibitemOpen
  \bibfield  {author} {\bibinfo {author} {\bibfnamefont {S.}~\bibnamefont
  {Karpitschka}}\ and\ \bibinfo {author} {\bibfnamefont {H.}~\bibnamefont
  {Riegler}},\ }\href {\doibase 10.1017/jfm.2014.73} {\bibfield  {journal}
  {\bibinfo  {journal} {Journal of Fluid Mechanics}\ }\textbf {\bibinfo
  {volume} {743}},\ \bibinfo {pages} {R1} (\bibinfo {year} {2014})}\BibitemShut
  {NoStop}%
\bibitem [{\citenamefont {Seidl}\ \emph {et~al.}(2022)\citenamefont {Seidl},
  \citenamefont {Bosch}, \citenamefont {Paap}, \citenamefont {Livraghi},
  \citenamefont {Zhai}, \citenamefont {Wick}, \citenamefont {Koller},
  \citenamefont {Wasserscheid}, \citenamefont {Maier}, \citenamefont {Smith},
  \citenamefont {Bachmann}, \citenamefont {Steinrück},\ and\ \citenamefont
  {Meyer}}]{seidl_bis-polyethylene_2022}%
  \BibitemOpen
  \bibfield  {author} {\bibinfo {author} {\bibfnamefont {V.}~\bibnamefont
  {Seidl}}, \bibinfo {author} {\bibfnamefont {M.}~\bibnamefont {Bosch}},
  \bibinfo {author} {\bibfnamefont {U.}~\bibnamefont {Paap}}, \bibinfo {author}
  {\bibfnamefont {M.}~\bibnamefont {Livraghi}}, \bibinfo {author}
  {\bibfnamefont {Z.}~\bibnamefont {Zhai}}, \bibinfo {author} {\bibfnamefont
  {C.~R.}\ \bibnamefont {Wick}}, \bibinfo {author} {\bibfnamefont {T.~M.}\
  \bibnamefont {Koller}}, \bibinfo {author} {\bibfnamefont {P.}~\bibnamefont
  {Wasserscheid}}, \bibinfo {author} {\bibfnamefont {F.}~\bibnamefont {Maier}},
  \bibinfo {author} {\bibfnamefont {A.-S.}\ \bibnamefont {Smith}}, \bibinfo
  {author} {\bibfnamefont {J.}~\bibnamefont {Bachmann}}, \bibinfo {author}
  {\bibfnamefont {H.-P.}\ \bibnamefont {Steinrück}}, \ and\ \bibinfo {author}
  {\bibfnamefont {K.}~\bibnamefont {Meyer}},\ }\href {\doibase
  https://doi.org/10.1016/j.jil.2022.100041} {\bibfield  {journal} {\bibinfo
  {journal} {Journal of Ionic Liquids}\ }\textbf {\bibinfo {volume} {2}},\
  \bibinfo {pages} {100041} (\bibinfo {year} {2022})}\BibitemShut {NoStop}%
\end{thebibliography}
%

\section*{Supporting Information}
\subsection{Model}
\label{sec:model}

In this section, we introduce the mathematical and physical model used to describe the dynamics of a thin liquid film containing dissolved catalysts, reactants, and products. Our model is based on fundamental equations including corresponding assumptions, which we outline below.

Within a $d$-dimensional coordinate system in the $\x,z$-space, where z = 0 is in the plane of the solid-liquid interface and $\bf{x}\in \mathbb{R}^{d-1}$, let $h(\x)$ be the height of the liquid film with mass density $\rho_l$ and horizontal velocity $u(\x,z)\in \mathbb{R}^{d-1}$ and the vertical velocity $v(\x,z)\in \mathbb{R}$.  We assume homogeneous and constant density of the film, implying incompressibility. Additionally, we introduce the number densities of the catalyst $\tilde{\rho}(\x,z)$, the products $P$ $\tilde{\rho}_p(\x,z)$, and the reactants $R$ $\tilde{\rho}_R(\x,z)$ which are all dissolved in the liquid film.

In many technologically relevant situations, the height of the film is much smaller than its extension in the horizontal or longitudinal direction. Via this so-called lubrication approximation, we make the ansatz that the pressure $p$ is constant in the direction normal to the substrate, i.e. $\partial_z p(\x,z) =0$. In the absence of net charges at the interfaces, the total pressure is the sum of the Laplace pressure and the disjoining pressure, as described by the equation~\cite{oron_long-scale_1997}:
\begin{subequations}
\label{eq:pressure}
    \begin{align}
    \label{eq:pressure_0}
      p(\x)=&-\gamma \nabla^2 h(\x) - \gamma(\x) (1-\cos \theta(\x) )f(h(\x))\\
      \label{eq:f_pressure}
      f(h(\x))=& \dfrac{(n-1)(m-1)}{(n-m)h^*}\left(\left(\dfrac{h^*}{h(\x)}\right)^n-\left(\dfrac{h^*}{h(\x)}\right)^m\right).
\end{align}
\end{subequations}
In these equations, $\gamma $ denotes the surface or interfacial tension between the liquid and gas/vapor phase and $h^*$ is the height of zero disjoining pressure that we consider as a dry film. \eqref{eq:f_pressure} introduces a negligible small layer of minimal liquid height $h^*$, called the precursor layer. The main purpose of this layer is to avoid division by zero when the liquid film dewets even though there is evidence that indeed, also in a real system, a layer of molecular thickness may be formed on dry substrates preceding liquid films or droplets \cite{popescu_precursor_2012}. $n$ and $m$ are both integers such that $n> m $, representing long-range attractive and short-range repulsive  interactions between the substrate and the atmosphere. The functional form of $f(h)$ can be derived by integrating the Lennard-Jones potential between gas molecules and solid-substrate molecules \cite{doi_soft_2013, oron_long-scale_1997, pelusi_liquid_2022}. Choosing the usual $(12,6)$ Lennard-Johnes potential one obtains $(n,m)=(9,3)$~\cite{peschka_signatures_2019, doi_soft_2013}.
The three-phase contact angle, $\theta$, is obtained from Young's law for wetting on solid smooth surfaces in thermodynamic equilibrium, $\cos \theta= \frac{\gamma_{sv}-\gamma_{l s}}{\gamma}$  \cite{doi_soft_2013}. Here, the indices "sv" and "ls" reflect the interfacial tensions between solid-vapor and liquid-solid, respectively. 

In principle, the height of the film is determined by the solution of the Navier-Stokes equation under the pressure drop provided by \eqref{eq:pressure}.
However, within the current length scale separation we can integrate the Navier-Stokes equation along the transverse direction and obtain an equation for the time evolution of the height of the film. To do so proper boundary conditions should be imposed at the solid-liquid and liquid-gas interfaces. 
It has been shown experimentally that, to match the experimental data, a finite slip length at the liquid-solid interface has to be introduced~\cite{peschka_signatures_2019, fetzer_slippage_2007, reiter_dewetting_2001, oron_long-scale_1997} 
Theoretically, this is necessary to allow the three-phase contact line of the film to move after rupture due to pressure gradients \cite{peschka_signatures_2019}. 
At the liquid-vapor interface we impose a free boundary condition $\partial_t h(\x,t) + u(\x,h(\x,t),t)\cdot \nabla h(\x,t) = v(\x,h(\x,t),t)$ \cite{oron_long-scale_1997, pototsky_morphology_2005}
Additionally, we apply the no penetration boundary condition at the solid-liquid contact plate located at $z=0$. That means that no liquid is flowing in or out of the solid. The no penetration boundary condition reads $(u(\x,0,t),v(\x,0,t))\cdot \n =0$, where $\n$ is the normal of the fluid surface. 
Finally, we model the atmosphere to be inviscous and assume shear stress due to surface tension gradients at the liquid-vapor interface. This gives the boundary condition $\partial_z u(\x,h(\x,t),t)=\nabla \gamma / \mu$ \cite{landau2013fluid}. 
With these assumptions, we integrate the Navier-Stokes equations along the $z$-direction and get the following evolution equations for the liquid film height and flux~\cite{hack_self-similar_2020,peschka_signatures_2019,oron_long-scale_1997}:
\begin{subequations}
\label{eq:thin_sheet}
    \begin{align}
\label{eq:thin_sheet_0}
    \partial_t h + \nabla\cdot hU  &=0 \\
\label{eq:thin_sheet_1}
    \partial_ t hU + \nabla\cdot (hU\otimes U)&=- \dfrac{h}{\rho_l}\nabla p + \dfrac{h^2+ bh }{2\rho_l M(h)}\nabla \gamma-\dfrac{\mu}{\rho_l} \dfrac{hU }{M(h)}.
\end{align}
\end{subequations}
with 
\begin{align}
\label{eq:flux}
    U:=\frac{1}{h}\int u(\x,z) dz.
\end{align}
In \eqref{eq:thin_sheet} $\mu$ is the dynamic viscosity, and $M(h)$ is the mobility function defined as $M(h)=\frac{2h^2 + 6h b + 3b^2}{6}$. The mobility function gives the coupling between the height $h$ and the pressure $p$ as we will see in \eqref{eq:TFE}. By assuming the Stokes regime, i.e. $Re \ll 1 $ we can further simplify \eqref{eq:thin_sheet} to arrive at the well-established thin film equation (TFE) \cite{oron_long-scale_1997}
\begin{align}
\label{eq:TFE}
    \partial_t h + \nabla\cdot \left(-\dfrac{M(h)h}{\mu}\nabla p + \dfrac{h^2 + 2 bh}{2\mu}\nabla \gamma \right)=0. 
\end{align}

Concerning the catalyst density, we focus on its height-integrated number density $\rho(\x) = \int_0^{h(\x)} \tilde{\rho}(\x,z)dz$,
which represents the number of catalytic particles to be found in the fluid column at horizontal location $\x$. Considering again the much smaller film height compared to the longitudinal length scale, we can assume the catalyst density quickly equilibrates with respect to time along the vertical direction and can therefore be considered at equilibrium. Thus, we assume that the density attains the equilibrium Boltzmann weight along the transverse direction and that it factorizes  in 
\begin{align}
\tilde{\rho}(\x,z)= \rho(\x) \tilde{\rho}_z(z,h(\x))\,.     
\end{align}
with 
\begin{align}
    \tilde{\rho}_z (z,h(\x)) =\dfrac{e^{-\beta \xi(z,h(\x))}}{\int_0^{h(\x)} e^{-\beta\xi(z,h(\x))}}.
\end{align}
Here 
$\xi$ is the effective potential that accounts for all effective interactions between the catalysts and the fluid and solid interfaces. Clearly, $\xi$ is sensitive to the molecular details of the catalysts as well as the interfaces. Accordingly, in order to account for repulsion ($\alpha>0$) attraction ($\alpha<0$) and neutral ($\alpha=0$) interactions with the fluid interface we model $\xi$ as
\begin{align}
\label{eq:potential}
    \beta\xi(z,h(\x)) = \begin{cases}
    \alpha z & \text{for }\alpha> 0\\
    0 & \text{for }\alpha = 0\\
    \alpha (z-h(\x))& \text{for }\alpha< 0.
    \end{cases}
\end{align}
This results in
\begin{align}
\label{eq:sinking_distribution}
   \tilde{\rho}_z(z,h(\x)) = 
  \begin{cases}
   \dfrac{\alpha e^{-\alpha z}}{1-e^{-\alpha h(\x)}}& \text{for }\alpha\geq0\\
    \dfrac{\alpha e^{-\alpha (z-h(\x))}}{-1+e^{\alpha h(\x)}}& \text{for }\alpha\leq0.
    \end{cases}
\end{align}
This modeling reflects different scenarios for the distribution of the catalyst within the liquid film along the vertical z-direction in the form of an exponential behavior with a characteristic length scale of $1/\alpha$. For $\alpha>0$ the catalyst is distributed towards the solid substrate while for $\alpha<0$ it is attracted to the liquid-vapor interface. For $\alpha = 0$, the catalyst is assumed to be homogeneously distributed in the vertical direction. 
See Figure 1
for an illustration of the three cases.

To project the $3D$ dynamics of the catalyst onto the plane of the solid substrate we follow the Fick-Jacobs approximation~\cite{zwanzig_diffusion_1992,Reguera2001,Malgaretti2013} and we integrate the $3D$ advection-diffusion equation governing the time evolution of the density along the transverse direction  \cite{thiele_note_nodate, thiele_gradient_2013, xu_variational_2015} yielding
\begin{align}
    \label{eq:fick}
    \partial_t \rho(\x) &= \nabla \cdot\left( D\nabla \rho(\x) - v(\x)\rho(\x) + D\beta \rho(\x) \nabla \mathcal{F}(\x)\right).
\end{align}
with effective potential
\begin{align}
    \mathcal{F}(\x)= - \dfrac{1}{\beta} \ln \left( \int_0^{h(\x)} e^{-\beta \xi(z,h(\x))} dz \right).  
\end{align}
The effective velocity $v(\x)$ in \eqref{eq:fick} is the integrated flow of catalyst across the transverse direction
\begin{align}
\label{eq:colloid_advection}
\begin{split}
     v(\x)=& \int\limits_0^{h(\x)} \tilde{\rho}_z(z,h(\x)) u(z,\x) dz\\ =&   M_\gamma(h(\x))\dfrac{\nabla \gamma(\x)}{\mu}-M_p(h(\x))\dfrac{\nabla p(\x)}{\mu}.
     \end{split}
\end{align}
where the local velocity profile, $u(z,\x)$ reads
\begin{align}
u =\left(\dfrac{z^2}{2}-zh-\dfrac{b(b+2h)}{2}\right)\dfrac{\nabla p}{\mu}+(z+b)\dfrac{\nabla \gamma }{\mu}
\end{align}
and we have introduced the mobility functions:
\begin{subequations}\label{eq:mob-cat}
\begin{align}
M_\gamma(h) = &\dfrac{1}{\int_0^{h} e^{-\beta \xi(z,h)}d\x}\int_0^{h} e^{-\beta \xi(z,h)}(z+b) dz \label{eq:mob-cat1}\\
M_p(h) =& \dfrac{1}{2}\dfrac{\int_0^{h} e^{-\beta \xi(z,h) }\left[z^2-2hz-2hb -b^2\right] dz}{\int_0^{h} e^{-\beta \xi(z,h)}d\x}\,. \label{eq:mob-cat2}
\end{align}
\end{subequations}
They couple the dynamics of the integrated number density of catalyst $\rho$ to the gradients of pressure and surface tension $\nabla p, \nabla \gamma$. The evaluation of the above integrals is carried out in section \ref{sec:inhomogenous_solutes}. One should note that both $\nabla \mathcal{F}(\x)$ and $v(\x)$ depend on the parameter $\alpha$. 

Finally, for what concerns the dynamics of the reactant and products we assume their linear dimensions to be much smaller than that of the catalysts and we do not account for any effective interactions with the interfaces (this can be easily implemented in the model when needed). Accordingly, both reactants and products are assumed to be homogeneously distributed along the transverse direction irrespective of the distribution of catalysts.
Additionally, we assume that the product concentration in the liquid film is also very diluted. In this context, we account for both advection and diffusion while assuming that reactants are transformed into the products at the catalytic particles at a rate of $\omega \rho(\x,t)/h(\x,t)$ where $\omega$ is the reaction rate per catalytic particle. Furthermore, we incorporate the transport of the reactant to the surrounding gas at a rate of $\sigma_R^\uparrow$ as well as the transport from the surrounding gas atmosphere, as depicted in Figure 1.
Additionally, we assume a constant supply of reactant from the surrounding gas atmosphere. Therefore, the evolution equation for the reactant is written \cite{thiele_note_nodate, thiele_gradient_2013, xu_variational_2015}:
\begin{align}
\label{eq:reactant}
\begin{split}
    \partial_t \rho_R =& \nabla \cdot \left( D_R \nabla \rho_R - v_{R} \rho_R + D_R \beta \rho_R \nabla\mathcal{F}_{0}\right) \\
    &- \frac{\omega \rho\rho_R}{h}  - \sigma_R^\uparrow \dfrac{l}{h}\rho_R + \rho_{R,Res}\sigma_R^\downarrow.
    \end{split}
\end{align}

Similarly, the density of products within the liquid film evolves according to \cite{thiele_note_nodate, thiele_gradient_2013, xu_variational_2015}:
\begin{align}
\label{eq:product}
\begin{split}
    \partial_t \rho_P =& \nabla \cdot \left( D_P \nabla \rho_P - v_{P} \rho_P + D_P \beta \rho_P \nabla\mathcal{F}_{0}\right) \\
    &+ \omega \rho\rho_R  - \sigma_P^\uparrow \dfrac{l}{h}\rho_P.
    \end{split}
\end{align}
We remark that $v_{R}=v_{P}=U$ and $\nabla \mathcal{F}_{\alpha_P=0}=-k_BT\dfrac{\nabla h}{h}$ as calculated in section \ref{sec:inhomogenous_solutes}. 


The last necessary ingredient in our model is the dependency of the vapor-liquid surface tension on the concentration $\rho_P$ and $\rho_R$. The vapor-liquid surface tension $\gamma(\x)$ varies as a function of the $\x$-coordinate and depends on the concentration of product at the surface $\frac{l}{h(\x)}\rho_P(\x)$ as well as that of the reactant $\frac{l}{h(\x)}\rho_R(\x)$. In the regime where the concentrations of both reactants and products are very small we can use a linear dependency of the surface tension with respect to their concentrations \cite{thiele_thermodynamically_2012, thiele_gradient_2016,squarcini_inhomogeneous_2020}, i.e. 
\begin{align}
\label{eq:surface_tension}
\begin{split}
    \gamma(\x)=& \gamma_0 -\Gamma_P\dfrac{l}{h(\x)}\rho_P(\x)-\Gamma_R\dfrac{l}{h(\x)}\rho_R(\x). \\
\end{split}
\end{align}
In \eqref{eq:surface_tension} $\gamma_0 $ represents the vapor-liquid surface tension in the absence of any dissolved chemicals in the form of reactants and products. $\Gamma_P$ and $\Gamma_R$ are control parameters describing the influences of products, and reactants respectively, on the vapor-liquid surface tension. If $\Gamma_P>\Gamma_R$ the surface tension is effectively reduced, as compared to $\gamma_0 -\Gamma_R\rho_{R,Res}\sigma_R^\downarrow h_0 / (\sigma_R^\uparrow l)$ in the presence of catalytic particles. Here $\gamma_0 -\Gamma_R\rho_{R,Res}\sigma_R^\downarrow h_0 / (\sigma_R^\uparrow l)$ is the surface tension of the liquid in the absence of catalytic particles and consequently also the absence of product, at an equilibrium amount of reactants. This effective reduction of surface tension happens because less surface active reactant is replaced by more surface active product. If $\Gamma_P<\Gamma_R$ vice versa. The TFE \eqref{eq:TFE} is oblivious to the absolute values of the solid-liquid and solid-vapor surface tensions. The only thing that matters is their difference $\gamma_{sv}-\gamma_{sl}$ setting the contact angle by Young's law of wetting. We define the passive contact angle $\theta_0$ by 
\begin{align}
    \label{eq:theta_0}
    \cos \theta_0:=\dfrac{\gamma_{sv}-\gamma_{sl}}{\gamma_0}.
\end{align}
It is the contact angle attained when products and reactants are modeled as not affecting the surface tension, i.e. $\Gamma_P=\Gamma_R=0$ and thus $\gamma(\x)=\gamma_0$. For simplicity, we assume the difference $\gamma_{sv}-\gamma_{sl}$ to be independent of the concentrations of products and reactants, thus constant in space and time. With this definition, we can rewrite the prefactor of the disjoining pressure in \eqref{eq:pressure}  as
\begin{align}
    -\gamma(1-\cos\theta)=-\gamma+\gamma_0\cos \theta_0
\end{align}
where we use $\cos\theta=(\gamma_{sv}-\gamma_{sl})/\gamma$. 

In the limit $ \alpha\to -\infty$,  $\omega, \sigma_P^\uparrow \to \infty$ one retrieves exactly the equations usually used to describe thin liquid films covered with a dilute layer of surfactant \cite{thiele_thermodynamically_2012, trinschek_modelling_2018}. That is so far expected as this limit represents colloids bound to the liquid surface. The product in this limit is produced and immediately removed from the system leading to $\rho_P\propto \rho$ and thus the surface tension reads $\gamma= \gamma_0 - \Gamma \rho$. Hence the colloids effectively behave like a surfactant. Choosing $\Gamma_P=\Gamma_R =0$, $\alpha=0$ we get the equations for thin liquid films of chemical solutions \cite{thiele_note_nodate, xu_variational_2015}. That is also to be expected as in this limit we model vertically homogeneous distributed particles with no surface tension effect. The limits are calculated and discussed in more detail in section \ref{sec:inhomogenous_solutes}.

\subsection{Numerical Method}

We present a numerical integrator for Eqs. \eqref{eq:thin_sheet_0}, \eqref{eq:thin_sheet_1}, \eqref{eq:fick}, \eqref{eq:product}, \eqref{eq:reactant} and \eqref{eq:surface_tension}. 

To solve \eqref{eq:thin_sheet_0}, \eqref{eq:thin_sheet_1} we use a Lattice Boltzmann Method (LBM) scheme developed by \cite{zitz_controlling_2021, zitz_lattice_2019, zitz_lattice_2021}. We perform simulations in one dimension only, as due to the varying behavior of the system, for certain parameters we need to run very long simulations that are more easily feasible in one dimension. Also our main focus is the stability of a liquid layer of homogeneous height where the dimension of the system is not expected to have any influence. As meshing we use a so-called $D1Q3$ lattice, which is a discretization of the phase space consisting of a set of equidistant points in space, with a set of three velocity directions $c_i=-1,0,1$ attached to every such point. On this lattice the discrete velocity distribution functions $(f_i)_{i=1,2,3}$ are considered \cite{kruger_lattice_2017}. The height and the flux can be obtained from the discrete-velocity distribution functions via
\begin{align}
    h=\sum f_i,\quad  hU = \sum c_i h_i.
\end{align}
The heart of the algorithm is the streaming and collision step, which consist of updating the discrete-velocity distribution functions at each time step according to
 \begin{align}
     \label{eq:collision_step}
         f_i&(x+c_i, t+1)=f_i(x,t)-\frac{1}{\tau}\left(f_i(x,t)-f_i^{eq}(\rho(x,t), u(x,t))\right) + \frac{c_i}{2}F.
 \end{align}
\eqref{eq:collision_step} contains the relaxation time $\tau$ that we will choose as $\tau=1$ but that is not the only valid choice. Further \eqref{eq:collision_step} uses the equilibrium distribution $f_i^{eq}$ defined as
\begin{align}
\label{eq:equilibrium_0}
    f_0^{eq}&=h\left(1-U^2\right)\\
    \label{eq:equilibrium}
    f_i^{eq}&= h\left( \frac{1}{2} c_i\cdot U +\frac{1}{2}U^2\right).
\end{align}
Finally, we directly calculate the force term $F$ by centred differences
\begin{align}
    F=- h\nabla p + \frac{h^2+ b }{2M(h)}\nabla \gamma-\mu \frac{hU }{M(h)}
\end{align}
which is the right hand side of \eqref{eq:thin_sheet_1}. With a Chapman-Ensk\o g analysis, one can show that by the scheme described above we indeed solve \eqref{eq:thin_sheet_0},\eqref{eq:thin_sheet_1} second-order accurate.

To solve \eqref{eq:fick} we introduce a second set $(g_i)_{i=1,2,3}$ of discrete-velocity distribution functions on the same lattice. We obtain $\rho$ by summing
\begin{align}
    \rho = \sum g_i
\end{align}
but unlike before we calculate 
\begin{align}
\label{eq:vel_fick}
    V=-D\frac{\nabla \rho}{\rho}+ v(x)+D\nabla\mathcal{F}.
\end{align}
The collision and streaming step is the same as \eqref{eq:collision_step}, and the equilibrium distribution functions also stay the same as \eqref{eq:equilibrium_0}, \eqref{eq:equilibrium}  where in both cases we substitute $f_i \mapsto g_i$, $U \mapsto V$, and $F\mapsto F_g$, where the forcing $F_g$ we still have to specify. Together with $F_g=0$ that already solves \eqref{eq:fick}. Remark that this scheme is quite similar to the standard LBM model for advection-diffusion processes \cite{kruger_lattice_2017} but \eqref{eq:equilibrium_0} and \eqref{eq:equilibrium} are lacking the diffusive term. That gives us more precise control over the diffusion coefficient $D$ via \eqref{eq:vel_fick} at the cost of numerically having to deal with the term $\nabla \rho / \rho$ that might introduce numerical difficulties. Upon dewetting it becomes non-trivial to control the concentration of catalyst in the precursor layer. We aim at a constant small density $\rho^*-\rho_{crit}$ on the dry dewetted areas that are in the spirit of the precursor film $h=h^*$ the minimum value of catalyst concentration. To numerically achieve that we set the forcing for the catalyst concentration to:
\begin{align}
    F_g= - \rho \nabla \kappa \left( -\left(\frac{\rho^*}{\rho+\rho_{crit}}\right)^k+ \left(\frac{h^*}{h}\right)^l\right)
\end{align}
where $k,l$ are constant integers, $\kappa$ is a constant with dimensions of a pressure,  and $h^*,\rho^*$ are small as compared to $h_0, \rho_0$. This force is essentially $0$ as long as $h\gg h^*$, $\rho\gg\rho^*$ and therefore does not affect the behavior in the bulk. When $\rho$ becomes small the first term becomes larger and ensures that always $\rho>0$ prohibiting the advection of more catalyst than is actually present. The second term becomes active when $h$ becomes small and forces the catalyst to evacuate the precursor film. Due to this pressure, we will get a small precursor film of $\rho=\rho^*-\rho_{crit}$ whenever the film is dewetted i.e. $h=h^*$.  This precursor film is to be considered as no catalysts. 

To handle \eqref{eq:product} we introduce a third set $(g^P_i)_{i=1,2,3}$ of discrete velocity distribution functions.  We calculate 
\begin{align}
    \rho_P(t+\Delta t) = \sum g_i^P + \frac{\omega \rho(t) \rho_R(t)}{h}- \sigma_P^\uparrow \frac{l}{h(t)} \rho_P(t).
\end{align}
Further, we take $V_P=U$ at every time-step, collision, and streaming we perform as in \eqref{eq:collision_step}. The equilibrium distribution we calculate by the equivalent of \eqref{eq:equilibrium_0} \eqref{eq:equilibrium}. We include an additional force 
\begin{align}
    F_g^P= -\rho_P \nabla \kappa \left( - \left(\frac{\rho_P^*}{\rho_P +\rho_{P,crit}}\right)^k+\left(\frac{h^*}{h}\right)^l\right).
\end{align}
This again prevents the appearance of negative densities and creates a precursor layer of $\rho_P=\rho_P^*-\rho_{P,crit}$ whenever $h=h^*$.

For the reactant $\rho_R$ we proceed completely analogue to $\rho_P$. We introduce $(g^R_i)_{i=1,2,3}$. The density of reactant is calculated by 
\begin{align}
    \rho_{R}(t+\Delta t)= \sum g_i^R - \frac{\omega \rho(t) \rho_R(t)}{h}-\sigma_R^\uparrow \rho_R(t) \frac{l}{h(t)}+ \rho_{R,Res} \sigma_R^\downarrow.
\end{align}
Furthermore, we set $V_R=U$. The collision and streaming steps we perform as in \eqref{eq:collision_step}. The equilibrium distribution we calculate once again by the equivalent of \eqref{eq:equilibrium_0} \eqref{eq:equilibrium}. Finally, the additional force we chose to be 
\begin{align}
    F_g^R= -\rho_R \nabla \kappa \left( - \left(\frac{\rho_R^*}{\rho_R +\rho_{R,crit}}\right)^k+\left(\frac{h^*}{h}\right)^l\right).
\end{align}

The parameters $\rho_{P,crit}$ and $\rho_{R,crit}$ are chosen
\begin{align}
    \rho_{R,crit}=\rho_R^*- \frac{\rho_{R,Res}\sigma_R^\downarrow h^*}{\omega (\rho^*-\rho_{crit}) + \sigma_R^\uparrow l}
\end{align}
and 
\begin{align}
    \rho_{P,crit}=\rho_{P}^* - \frac{\omega(\rho^*- \rho_{crit})(\rho_R^*- \rho_{R,crit})}{\sigma_P^\uparrow l}
\end{align}
The choices of $\rho_{P,crit}$ and $\rho_{R,crit}$ ensure that sink and source terms cancel out in the precursor film such that no reaction takes place in the dewetted areas.

In the collapse regime it at times becomes necessary to apply a moving average on the densities with a center weight of $1-10^{-5}$ to preserve numerical stability.  

\subsection{Examples of numerical simulations}

\begin{figure}[hh]
    \centering
    \includegraphics[width=\linewidth]{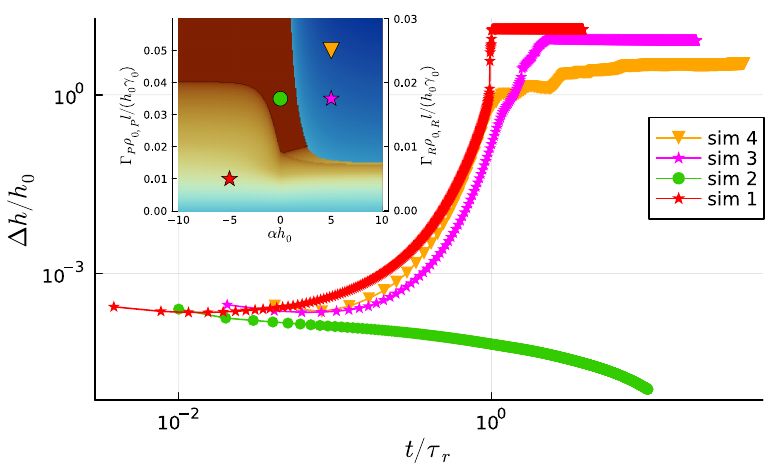}
    \caption{The maximum height difference in time plotted for four exemplary simulations, showing the different possible behaviors of the system. All simulations are started in the perturbed homogeneous state $h=h_0+ \epsilon \mathcal{N}_0^1, \rho=\rho_0, \rho_P=\rho_{0,P}, \rho_R=\rho_{0,R}$. All parameters are chosen as given in the materials and methods section. The values of $\Gamma_P,\Gamma_R$ and $\alpha$ are chosen as follows: sim 1: $\Gamma_P\rho_{0,P}l /(h_0\gamma_0)=2\Gamma_P\rho_{0,R}l /(h_0\gamma_0)=0.02$, $\alpha h_0=-5.0$; sim 2: $\Gamma_P\rho_{0,P}l /(h_0\gamma_0)=2\Gamma_P\rho_{0,R}l /(h_0\gamma_0)=0.035$, $\alpha h_0=-0.0$; sim 3: $\Gamma_P\rho_{0,P}l /(h_0\gamma_0)=2\Gamma_P\rho_{0,R}l /(h_0\gamma_0)=0.035$, $\alpha h_0=5.0$ and sim 4: $\Gamma_P\rho_{0,P}l /(h_0\gamma_0)=2\Gamma_P\rho_{0,R}l /(h_0\gamma_0)=0.05$, $\alpha h_0=5.0$ . The zeroth time-step is omitted due to the logarithmic scale of the axes. In the inset, we show the position of those simulations in the stability diagram as colored symbols. For every simulation, a video is available numbered in the same order as the legend here. In simulation 2 it holds $\tau_r=\infty$ thus we chose an arbitrary value instead, to show its evolution in the same plot as the other three. This figure contains the same information as Figure 3 of the main text.}
    \label{fig:simulations_si}
\end{figure}

\begin{figure}
    \centering
    \includegraphics[width=\linewidth]{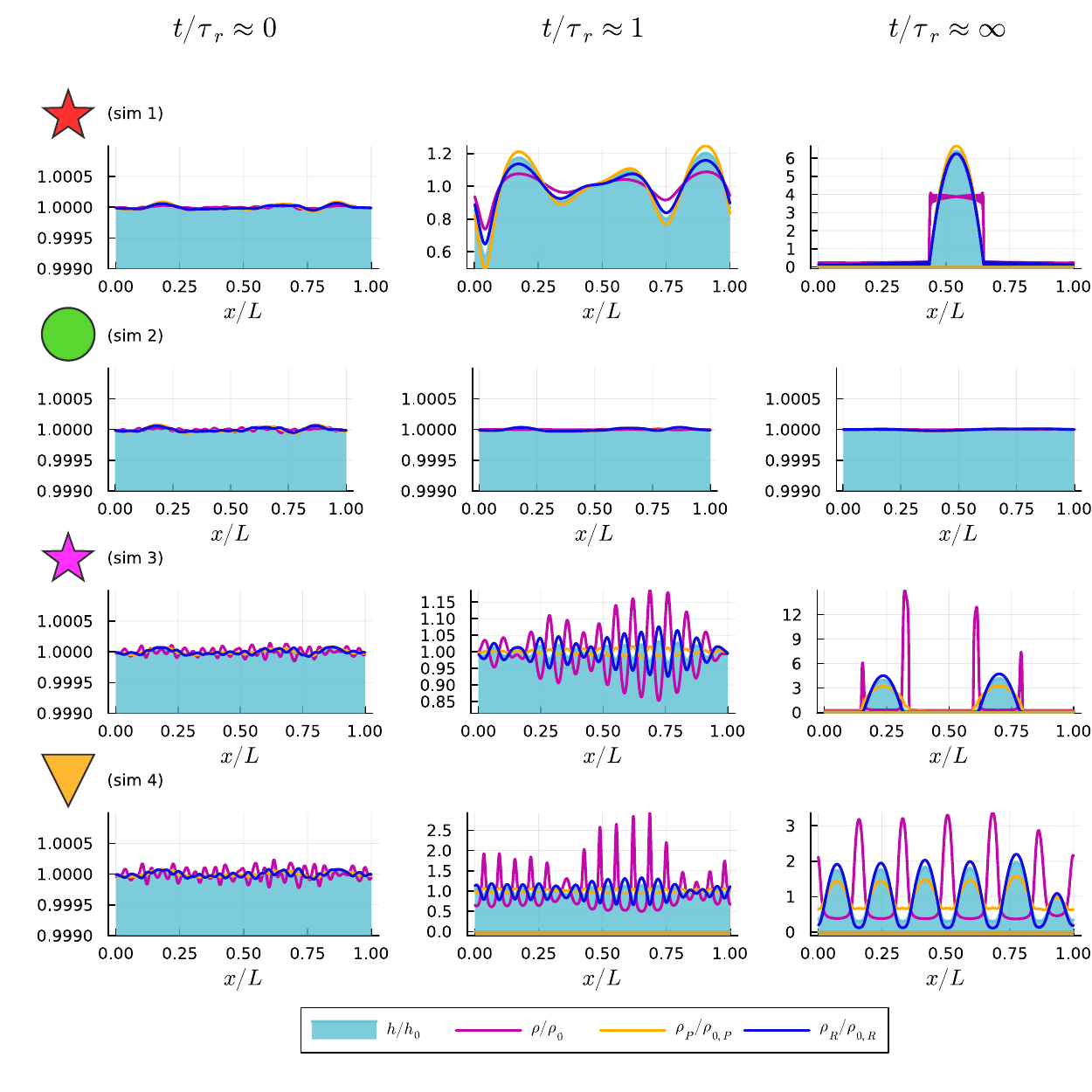}
    \caption{Snapshots of the four exemplary simulations described in the caption of Figure \ref{fig:simulations_si} showing height $h$, density of catalyst $\rho$, reactant $\rho_{R}$ and product $\rho_{P}$ normalized by their respective initial value. Shown are the dumps at the following time steps: the smallest time step that is not the initial condition ($t/\tau_r\approx 0$), the biggest time step that is smaller than the rupture time predicted by the LSA ($t/\tau_r\approx1$) and the last time step ($t/\tau_r\approx \infty$). Videos of those simulations can be found under the same name. }
    \label{fig:snapshots}
\end{figure}

Starting a simulation from a perturbed homogeneous state, i.e. with initial data $h(t=0)=h_0+ \epsilon \mathcal{N}, \rho(t=0)=\rho_0, \rho_P(t=0)=\rho_{0,P}$ and $\rho_R(t=0)=\rho_{0,R}$ where $\epsilon\ll 1 $ and $\mathcal{N}$ is the normal random distributed random variable we obtain the results as shown exemplarily in the four supporting videos as well as in Figures \ref{fig:simulations_si} and \ref{fig:snapshots}. For the shown simulations we used the same parameters as reported in the Materials and Methods section, but with a better spatial resolution to study the dynamics more resolved. In the videos, we see on the left-hand side the four fields of interest in real space. The y-axis is not fixed. On the right-hand side we see the Fourier transform of film height $\widehat{h}$ as blue dots compared with the linear solution \eqref{eq:linear_soloution}, showing good agreement at initial time steps and disagreeing more and more as second-order terms step in. There are four different behaviors shown:
\begin{enumerate}
    \item     \emph{Simulation 1, Spinoidal dewetting:} The liquid film breaks due to disjoining pressure, while the concentration fields follow the dynamics of the fluid. Reduced surface tension slows down this process. Heterogeneity in concentrations has minor effects. Rupture time and dominant wavelength at film rupture are well predicted by the linear solution \eqref{eq:linear_soloution}.
    \item  \emph{Simulation 2, Stabilisation:} Due to the reduced surface tension the homogeneous state becomes stable. Every perturbation decays over time. The linear solution \eqref{eq:linear_soloution} can almost perfectly describe the whole time evolution of the system.
    \item \emph{Simulation 3, Collapse leading to rupture:} In the collapse regime catalyst is accumulated and thus a small wavelength is excited. Initially the linear solution \eqref{eq:linear_soloution} predicts the exponential growth of this wavelength well. But then Laplace pressure prevents the small excited wavelength from further growth. The growth of $\Delta h/h_0$ is halted, and \eqref{eq:linear_soloution}   disagrees with the actual numerical solution. Now the system coarsens unifying single peaks and thereby increasing the dominant wavelength in the system. Finally, film rupture happens at a wavelength a lot larger than the $q_{\max}$ predicted by the LSA. This coarsening may happen rather quickly such that no significant delay in the rupture time can be seen or it might delay rupture by up to one order of magnitude as compared to the rupture time predicted by the LSA (see main text).
    \item \emph{Simulation 4, Collapse not leading to rupture:} At large $(\Gamma_P \rho_{0,P}+\Gamma_R\rho_{0,R})l/(h_0\gamma_0)$ we also observe the growth of a small wavelength coming from the accumulation of catalyst, resulting an appreciable corrugation of the liquid film. Next, the system coarsens but does not break. That is due to the low average surface tension $\langle \gamma\rangle = \gamma_0- \Gamma_P \rho_{0,P} - \Gamma_R \rho_{0,R}$ that makes the liquid perfectly wetting. The expected equilibrium state is a single cluster of catalyst and a closed liquid film. 
\end{enumerate}

\subsection{Interpolating the flow-field}

The full flow field $(u,v)$ can be obtained just  from just the horizontal component $u(x,z)$ via the incompressibility condition $
\nabla \cdot (u,v)=0 $
to be 
\begin{align}
    v(x,z)=\int_0^z \frac{dv}{dy}dy + v(x,0)=-\int_0^ z\frac{du}{dx}dy + 0
\end{align}
The local flow field in horizontal direction $u(x,z)$ can be interpolated from the pressure and surface tension gradients via
\begin{align}
u(x,z) =\left(\dfrac{z^2}{2}-zh-\dfrac{b(b+2h)}{2}\right)\dfrac{\nabla p(x)}{\mu}+(z+b)\dfrac{\nabla \gamma(x)}{\mu}. 
\end{align}
From the considerations above we obtain
\begin{align}
v(x,z)=-\left(\dfrac{z^3}{6}-\dfrac{z^2h}{2}-\dfrac{b(b+2h)z}{2}\right)\dfrac{\nabla^2 p(x)}{\mu}-\dfrac{z^2 + 2bz}{2}\dfrac{\nabla^2 \gamma(x)}{\mu}.
\end{align}
This is used to interpolate the flow-fields shown in Figure 3 
in the main text.

\subsection{Droplets revisited}

\begin{figure}[hh]
    \centering
    \includegraphics[width=\linewidth]{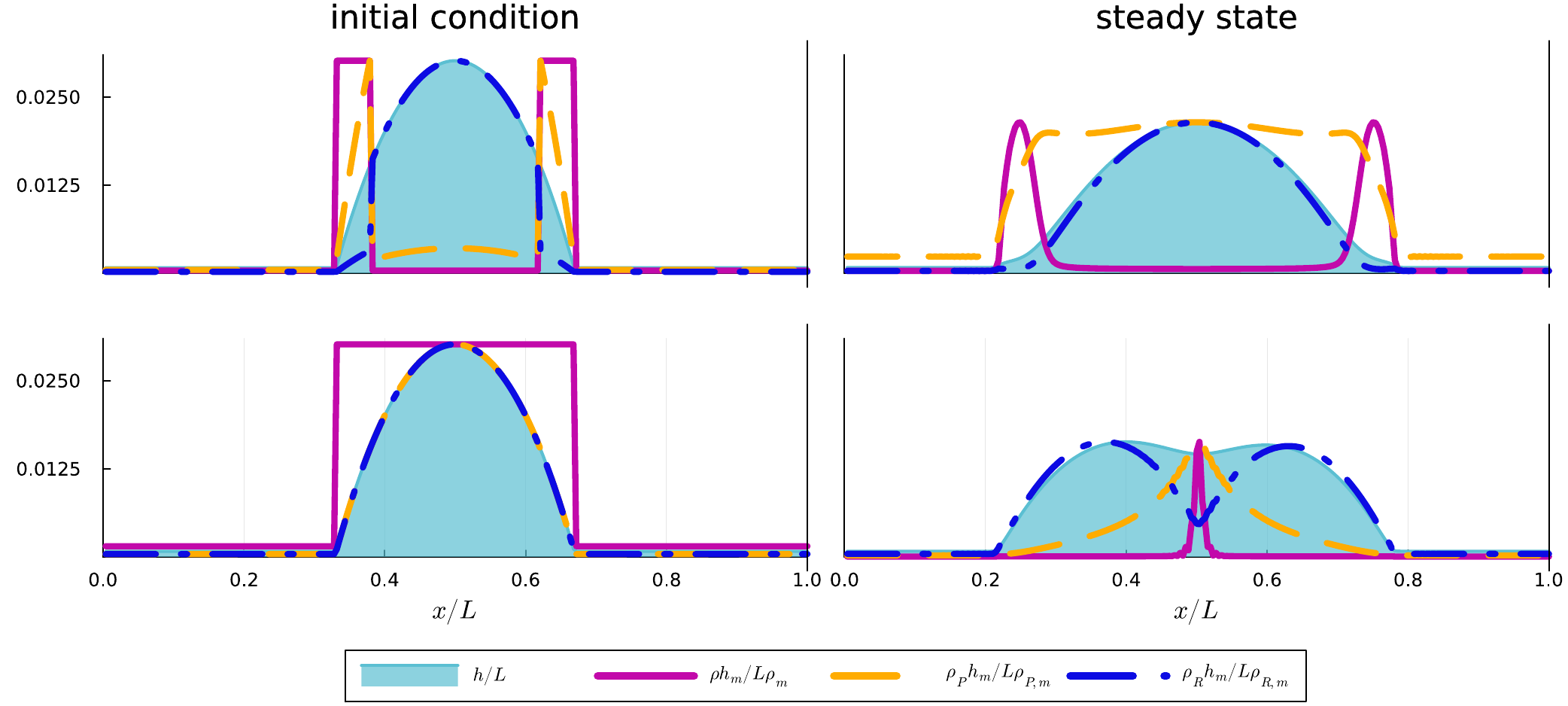}
    \caption{Steady state of a droplet with dissolved active catalyst particles for different initial conditions. Parameters are chosen the same as for the center-right subplot of Figure 6 in the main text. The labeling is also the same as reported there. The seemingly higher value in the top right sub-figure is due to the normalization and the low product concentration within the droplet. The aspect ratio is again $10$ to better highlight the deformation of the droplets.}
    \label{fig:drops_initial_condition}
\end{figure}

As already mentioned in the main text the equilibrium state of a droplet in the collapse regime may depend on the initial condition. In Figure \ref{fig:drops_initial_condition} we compare two simulations started at different initial conditions for $\alpha>5$. Depending on the initial condition the position of the formed catalyst clusters is altered. When coming from a ruptured film the state of the system immediately after rupture resembles closer to the initial condition of the simulation shown at the top of Figure \ref{fig:drops_initial_condition}. What will be the equilibrium state of a three-dimensional droplet is not easily deduced from two-dimensional simulations and remains a question for further research. This is because the two-dimensional droplet has two distinct contact points while a three-dimensional drop has an actual contact line. Outside the collapse regime, the eventual steady state does not depend on the initial condition.

\subsection{Vertically Heterogeneous Solutes and their Mobility's}
\label{sec:inhomogenous_solutes}

The building block of the model presented in the main text is a solvent with a vertical distribution inside a liquid film. This model has to the best of our knowledge not been proposed in literature so far so we are going to briefly discuss it. This will include some repetition of what has already been presented in section \ref{sec:model}. While doing so we explicitly calculate the integral in the definitions of the mobility functions of the catalyst \eqref{eq:mob-cat}. Further, we calculate three limits of interest. Vertically homogeneous distributed catalyst ($\alpha=0$), catalyst that forms a single layer at the vapor-liquid interface ($\alpha\to -\infty$), and catalyst that forms a single layer at the solid substrate ($\alpha\to \infty$). Especially vertically homogeneous catalyst reproduces the model used for a solute in a liquid film  used by \cite{baumgartner_marangoni_2022, karpitschka_sharp_2014} while a single layer of catalyst at the liquid surface reproduces the thin film equation including soluted surfactants \cite{thiele_thermodynamically_2012}.

The height of the liquid film $h$ evolves, as already discussed, according to the thin film equation 
\begin{align}
    \partial_t h &= \nabla \cdot \left(\frac{M(h) h }{\mu}\nabla p - \frac{h(h+2b)}{2\mu}\nabla \gamma \right),
\end{align}
while the height-integrated density of a solute $\varrho$ follows an advection-diffusion equation with a Fick-Jackobs correction term for the geometry of the film 
\begin{align}
    \partial_t \varrho(x,t)=&\nabla\cdot\left(\underbrace{D\nabla\varrho(x,t)}_{\text{diffusion}}-\underbrace{v(x,t)\varrho(x,t)}_{\text{advection}}+\underbrace{D\beta\varrho\nabla F}_{\text{Fick-Jacobs}}\right).
\end{align}

We assume that the actual three-dimensional density of the solute $\tilde{\varrho}(\mathbf{x},z)$ can be decomposed into $\tilde{\varrho}(\mathbf{x},z)=\varrho(\mathbf{x})\varrho_z(z,h(x))$, $\varrho_z$ is the vertical distribution of solute that due to the length scale separation be horizontal and vertical length scales is assumed to be at steady state. Further we assume that $\varrho(z,h(x))$ is given by an external potential $\xi=\xi(z,h(x))$
\begin{align}
\label{eq:distribution_2}
    \tilde{\varrho}_z (z,h(x))= \frac{e^{-\beta \xi(z,h(x))}}{\int_0^{h(x)} e^{-\beta\xi(z,h(x))}}.
\end{align}
To better illustrate the model we choose the potential $\xi$ to be given by \eqref{eq:potential} i.e.
\begin{align}
    \beta\xi(x,z) = \begin{cases}
    \alpha z & \text{for }\alpha\geq 0\\
    \alpha (z-h(x))& \text{for }\alpha\leq 0.
    \end{cases}
\end{align}
This gives an exponential distribution as we will see shortly and covers a lot of relevant scenarios. One can repeat the ongoing derivation also for a different potential $\xi$ if needed. Accordingly the partition function $Z=\int_0^h \exp(-\beta \xi) dz$ calculates as 
\begin{align}
    Z(x) =
    \begin{cases}
    \int_0^{h(x)}  e^{-\alpha z} dz = \frac{1-e^{-\alpha h(x)}}{\alpha}& \text{for } \alpha\geq 0\\
    \int_0^{h(x)}  e^{-\alpha (z-h(x))} dz = \frac{-1+e^{\alpha h(x)}}{\alpha}& \text{for } \alpha\leq 0.
    \end{cases}
\end{align}
The distribution of catalyst corresponding to this is given by \eqref{eq:sinking_distribution} / \eqref{eq:distribution_2} i.e. the exponential distribution
\begin{align}
    \tilde{\varrho}(x,z) =
    \begin{cases}
    \frac{\alpha e^{-\alpha z}}{1-e^{-\alpha h(x)}}& \text{for }\alpha\geq0\\
     \frac{\alpha e^{-\alpha (z-h(x))}}{-1+e^{\alpha h(x)}}& \text{for }\alpha\leq0.
     \end{cases}
\end{align}
For $\alpha<0$ a high concentration of catalyst at the liquid-vapor interface with exponential decay into the bulk  is modeled while for $\alpha>0$ we obtain a high concentration of catalyst close to the solid substrate with exponential decay into the bulk. In both cases, the exponential distribution has  a segmentation length $\lambda=\frac{1}{|\alpha|}$. Compare to \cite{doi_soft_2013}. 

The free energy gradient then is for $\alpha\geq 0$
\begin{align}
\begin{split}
    \nabla \mathcal{F}(h)&=-k_BT \nabla \ln \int_0^{h}  e^{-\alpha z} dz\\
    &= -k_BT \frac{\alpha\partial_x h}{e^{\alpha h}-1}
    \end{split}
\end{align}
For $\alpha\leq0$ we obtain 
\begin{align}
\begin{split}
    \nabla \mathcal{F}(h)&=-k_BT \nabla \ln \int_0^{h} \alpha e^{-\alpha (z-h))} dz\\
    &= -k_BT \frac{\alpha e^{\alpha h}\partial_x h}{e^{\alpha h}-1}.
    \end{split}
\end{align}
For the advective velocities, we have 
\begin{align}
\begin{split}
     v(x)=& \int\limits_0^{h(x)}\!\!\! \tilde{\varrho}_z(z,h(x)) u(z,x) dz\\ =&   \frac{\nabla \gamma(x)}{\mu}M_\gamma(h(x))-\frac{\nabla p(x)}{\mu}M_p(h(x)).
     \end{split}
\end{align}
where we used the mobility functions 
\begin{subequations}
    \begin{align}
M_\gamma(h) = &\frac{1}{\int_0^{h} e^{-\beta \xi(z,h}dx}\int_0^{h} e^{-\beta \xi(z,h)}(z+b) dz\\
     M_p(h) =& \frac{1}{2}\dfrac{\int_0^{h} e^{-\beta \xi(z,h) }\left[z^2-2hz-2hb -b^2\right] dz}{\int_0^{h} e^{-\beta \xi(z,h)}dx}. 
\end{align}
\end{subequations}
Those integrals we  calculate for $\alpha\geq 0$
\begin{subequations}
    \begin{align}
    M_\gamma(h) &=\frac{1}{1-e^{-\alpha h}}\left( -be^{-\alpha h} + \frac{1-e^{-\alpha h}(\alpha h +1)}{\alpha}+ b \right)\\
    M_p(h&)= - \frac{2 - 2 \alpha h - \alpha^2 b (b + 2 h) + e^{-\alpha h} ( \alpha^2 (b + h)^2-2)}{2 \alpha^2(1-e^{-\alpha h})}
\end{align}
\end{subequations}
and for $\alpha\leq 0$
\begin{subequations}
    \begin{align}
    M_\gamma(h) &=\frac{1}{e^{\alpha h}-1}\left( b(e^{\alpha h}-1)+ \frac{e^{\alpha h}-1}{\alpha}-h \right)\\
    M_p(h)&= - \frac{-2 + \alpha^2 (b + h)^2 - e^{\alpha h}\left(-2 + 2 \alpha h + \alpha^2 b (b + 2 h)\right)}{2 a^2(e^{\alpha h}-1)}.
\end{align}
\end{subequations}

The surface liquid-gas surface tension depends only on the solvent that is fairly close to the liquid-gas interface. Thus we introduce a prefactor 
\begin{align}
\label{eq:surface_tension_prefactor}
    o(x)=\int_{h-l}^h\tilde{\varrho}(z,h(x))dz
\end{align}
where in \eqref{eq:surface_tension_prefactor} $l$ is a molecular length scale. In most cases, especially for a dilute solution a linear dependency of the surface tension on the solute concentration is justified \cite{thiele_thermodynamically_2012,zhai_influence_2023}, thus we write the surface tension to be 
\begin{align}
    \gamma=\gamma_0 - \Gamma o(x) \varrho(x). 
\end{align}
In a similar way the solid-liquid surface tension reads 
\begin{align}
    \gamma_{sl}=\gamma_{0,sl}-\Gamma_{sl}o_{sl}\varrho
\end{align}
where the prefactor $o_{sl}$ is given by
\begin{align}
    o_{sl}=\int_0^l\tilde{\varrho}(z,h(x)) dz.
\end{align}

The model used in the main text can be derived from this model by choosing $\varrho=\rho,\rho_P,\rho_R$ and keeping $\alpha$ for the catalyst density $\rho$ while choosing $\alpha=0$ for the densities of product $\rho_P$ and reactant $\rho_R$. Further, we have chosen $\Gamma=\Gamma_{sl}=0$ for the catalyst, neglecting any direct effect of the catalyst on the surface tension. It only acts indirectly by consuming surface-active reactant and producing surface-active product. Further, we have  used $\Gamma_{sl}=0$ for both reactant and product neglecting the effects of the chemicals on the solid surface energies. That has been done to simplify the model and to reduce the number of free variables. Of course, one could also include the non-homogeneous distribution of chemicals in the vertical direction ($\alpha\neq 0$) and the effects of the catalyst and reactant on both, the liquid-gas and liquid-solid surface energy.

Let us stress the three limiting cases of this distributions 
\begin{enumerate}

    \item $\underset{\alpha \to 0}{\lim}$: This limit indicates the segmentation length $\frac{1}{\alpha}$ becoming infinite meaning the solute vertically homogeneously distributed. We get 
    \begin{align}
        \tilde{\varrho}(x,z)\sim \lim_{\alpha\to 0}  \begin{cases}
    \frac{\alpha e^{-\alpha z}}{1-e^{-\alpha h(x)}}& \text{for }\alpha\geq0\\
     \frac{\alpha e^{-\alpha (z-h(x))}}{-1+e^{\alpha h(x)}}& \text{for }\alpha\leq0
     \end{cases} = \frac{1}{h(x)}
    \end{align}
    and
    \begin{align}
    \begin{split}
        \nabla \mathcal{F}(h)&= -k_BT \underset{\alpha \to 0}{\lim} \nabla \ln Z(h)= -k_BT \underset{\alpha \to 0}{\lim}\frac{\alpha\nabla h}{e^{\alpha h}-1}\\
        &= -k_BT\frac{\nabla h}{h}= -k_BT \nabla \ln h.  
        \end{split}
    \end{align}
    The upper and lower limits agree. Furthermore, the mobilities calculate to be
    \begin{align}
    \begin{split}
        M_\gamma(h) &= \underset{\alpha \to 0}{\lim} \frac{1}{1-e^{-\alpha h}}\left( -be^{-\alpha h} + \frac{1-e^{-\alpha h}(\alpha h +1)}{\alpha}+ b \right)\\
        &= b + \frac{h}{2}
        \end{split}
    \end{align}
    and 
    \begin{align}
    \begin{split}
        M_p(h) &= -\underset{\alpha \to 0}{\lim}  \frac{2 - 2 \alpha h - \alpha^2 b (b + 2 h) + e^{-\alpha h} ( \alpha^2 (b + h)^2-2)}{2 a^2(1-e^{-\alpha h})}\\
        &= \frac{1}{6} (3 b^2 + 6 b h + 2 h^2)=M(h)
        \end{split}
    \end{align}
    giving 
    \begin{align}
        v(x)=\frac{\nabla \gamma(x)}{\mu} M_\gamma(h(x)) - \frac{\nabla p(h(x))}{\mu}M_p(h(x))= U. 
    \end{align}
        Finally, the surface tension prefactor evaluates to be 
    \begin{align}
       o=\lim_{\alpha\to 0}\int_{h-l}^{h}  \frac{\alpha e^{-\alpha z}}{1-e^{-\alpha h(x)}} dz = \frac{l}{h(x)}.
    \end{align}
    By symmetry, we have the solid-liquid surface tension prefactor to be 
    \begin{align}
        o_{sl}=o.
    \end{align}
    
    We would have gotten the same result when choosing $\xi=0$ and then calculating $\nabla \mathcal{F}$ and $v$ from there on. 
    Also, the left limit yields the same result. This limit retrieves the model used by \cite{baumgartner_marangoni_2022, karpitschka_sharp_2014}.

    \item $\underset{\alpha \to \infty}{\lim}$: This limit indicates the segmentation length $\frac{1}{\alpha}$ becoming $0$, meaning the catalysts are confined to the substrate. We obtain the vertical distribution
    \begin{align}
        \tilde{\varrho}(x,z)\sim \lim_{\alpha\to \infty}  \frac{\alpha e^{-\alpha z}}{1-e^{-\alpha h(x)}} = \delta_0
    \end{align}
    and the free energy gradient
    \begin{align}
        \nabla \mathcal{F}(h)&= -k_BT \underset{\alpha \to \infty}{\lim} \nabla \ln Z(h)= -k_BT \underset{\alpha \to \infty}{\lim}\frac{\alpha\nabla h}{e^{\alpha h}-1}=0.  
    \end{align}
    Here it is important not to interchange limit and derivative as $\lim_{\alpha \to \infty} Z=-\infty$. Furthermore, the mobilities are given by
    \begin{align}
        M_\gamma(h) &= \underset{\alpha \to \infty}{\lim}\frac{1}{1-e^{-\alpha h}}\left( -be^{-\alpha h} + \frac{1-e^{-\alpha h}(\alpha h +1)}{\alpha}+ b \right)=b
    \end{align}
    and 
    \begin{align}
    \begin{split}
        M_p(h) &= -\underset{\alpha \to \infty}{\lim}  \frac{2 - 2 \alpha h - \alpha^2 b (b + 2 h) + e^{-\alpha h} ( \alpha^2 (b + h)^2-2)}{2 a^2(1-e^{-\alpha h})}\\
        &= \frac{b^2}{2}+bh
        \end{split}
    \end{align}
    which implies
    \begin{align}
        v(x)=\frac{\nabla \gamma(x)}{\mu} M_\gamma(h(x)) - \frac{\nabla p(h(x))}{\mu}M_p(h(x))= \frac{\nabla \gamma(x)}{\mu}b -\frac{\nabla p(h(x))}{\mu}(\frac{b^2}{2}+bh(x))=u(x,z=0). 
    \end{align}
     We finally derive the surface tension prefactors
    \begin{align}
        o=\lim_{\alpha\to \infty}\int_{h-l}^{h}  \frac{\alpha e^{-\alpha z}}{1-e^{-\alpha h(x)}} dz = 0
    \end{align}
    and 
    \begin{align}
        o_{sl}=\lim_{\alpha\to \infty}\int_{0}^{l}  \frac{\alpha e^{-\alpha z}}{1-e^{-\alpha h(x)}} dz = 1
    \end{align}
    Thus the liquid-vapor surface tension is not influenced by the solute, while the solid-liquid surface tension is dependent on the full amount of soluted particles, as is evident from the fact that the solute in this limit is distributed in a single layer at the solid interface.
    
    We would have gotten the same result when choosing $e^{-\beta \xi}=\delta_0$ the Dirac-delta and then calculating $\nabla \mathcal{F}$ and $v$ from there on.

    \item $\underset{\alpha \to -\infty}{\lim}$: In this limit the segmentation length $\frac{1}{-\alpha}$ becomes $0$ and thus  the catalysts are confined to the fluid vapor interface. We calculate 
    \begin{align}
        \tilde{\varrho}(x,z)\sim \lim_{\alpha\to \infty}   \frac{\alpha e^{-\alpha (z-h(x))}}{-1+e^{\alpha h(x)}} = \delta_h
    \end{align}
    while the free energy gradient is given by
    \begin{align}
        \nabla \mathcal{F}(h)&= -k_BT \underset{\alpha \to -\infty}{\lim} \nabla \ln Z(h)= -k_BT \underset{\alpha \to -\infty}{\lim}\frac{\alpha e^{-\alpha z}\nabla h}{e^{\alpha h}-1}=0.  
    \end{align}
    Furthermore the mobilities read
    \begin{align}
        M_\gamma(h) &= \underset{\alpha \to -\infty}{\lim}\frac{1}{e^{\alpha h}-1}\left( b(e^{\alpha h}-1)+ \frac{e^{\alpha h}-1}{\alpha}-h \right)=b+h
    \end{align}
    and 
    \begin{align}
    \begin{split}
        M_p (h)&= -\underset{\alpha \to -\infty}{\lim}   \frac{-2 + \alpha^2 (b + h)^2 - e^{\alpha h}\left(-2 + 2 \alpha h + \alpha^2 b (b + 2 h)\right)}{2 a^2(e^{\alpha h}-1)}\\
        &= \frac{1}{2}(b^2 + 2bh + h^2)
        \end{split}
    \end{align}
    resulting in 
    \begin{align}
        v(x)=\frac{\nabla \gamma(x)}{\mu} M_\gamma(h(x)) - \frac{\nabla p(h(x))}{\mu}M_p(h(x))= \frac{\nabla \gamma(x)}{\mu}(b+h(x)) -\frac{\nabla p(h(x))}{\mu}\frac{1}{2}(b^2 + 2bh(x) + h(x)^2)=u(x,z=h(x)). 
    \end{align}
    The surface tension prefactor  we calculate as 
    \begin{align}
        o=\lim_{\alpha\to -\infty}\int_{h-l}^{h} \frac{\alpha e^{-\alpha (z-h(x))}}{-1+e^{\alpha h(x)}} dz = 1.
    \end{align}
    Thus all solute contributes to the surface tension effect, as to be expected for solute arranged in a single layer at the liquid-vapor interface. The solid-liquid surface tension prefactor in this limit is given by 
    \begin{align}
        o_{sl}=\lim_{\alpha\to -\infty}\int_{0}^{l} \frac{\alpha e^{-\alpha (z-h(x))}}{-1+e^{\alpha h(x)}} dz = 0.
    \end{align}
    Thus the solid-liquid surface tension is not affected by the solutes as to be expected for all solutes immersed at the liquid interface.

    We would have gotten the same result when choosing $e^{-\beta \xi}=\delta_h$ the Dirac-delta in $h$ and then calculating $\nabla \mathcal{F}$ and $v$ from there on. We have retrieved the thin-film model containing a mono-layer of surfactants absorbed at the liquid interface described by e.g. \cite{thiele_thermodynamically_2012}. 
\end{enumerate}

\subsection{Linear Stability Analysis}

We want to linearise the system \eqref{eq:TFE}, \eqref{eq:fick}, \eqref{eq:product}, \eqref{eq:reactant} and \eqref{eq:surface_tension}. This procedure will reduce the model to an ODE
\begin{align}
\label{eq:Matrix_ODE_si}
 \partial_t \begin{pmatrix}
  \widehat{h} \\ \widehat{\rho}\\ \widehat{\rho_P} \\ \widehat{\rho_R}
 \end{pmatrix} =A(q)
 \cdot \begin{pmatrix}
 \widehat{h} \\ \widehat{\rho}\\ \widehat{\rho_P} \\ \widehat{\rho_R}
 \end{pmatrix}
\end{align}
that solve to be 
\begin{align}
\label{eq:linear_soloution_si}
    \begin{pmatrix}
  \widehat{h} \\ \widehat{\rho}\\ \widehat{\rho_P} \\ \widehat{\rho_R}
 \end{pmatrix}(q,t)= \begin{pmatrix}
 \widehat{h} \\ \widehat{\rho}\\ \widehat{\rho_P} \\ \widehat{\rho_R} \end{pmatrix}(q,0)e^{A(q)t}. 
\end{align}
In order to obtain the explicit form of the matrix $A$ let all fields be approximated by their first-order representation 
\begin{subequations}
\label{eq:first_order}
    \begin{align}
 h(x,t)\simeq& h_0+\delta h(x,t)\\
 p(x,t)\simeq& p_0+\delta p(x,t)\\
 \rho(x,t)\simeq& \rho_0+\delta \rho(x,t)\\
 \rho_P(x,t) \simeq& \rho_{0,P}+ \delta \rho_P(x,t)\\
  \rho_R(x,t) \simeq& \rho_{0,R}+ \delta \rho_R(x,t)\\
 v(x,t) \simeq& \underbrace{v_0}_{=0}+\delta v(x,t)\\
  v_P(x,t) =v_R(x,t) \simeq& \underbrace{v_0}_{=0}+\delta v_{P,R}(x,t)\\
 \gamma(x,t) \simeq& \gamma_0 -\Gamma_P\frac{l\rho_{0,P}}{h_0}-\Gamma_P \frac{l \delta \rho_{P}(x,t)}{h_0}+\Gamma_P \frac{l \delta h(x,t) \rho_{0,P}}{h_0^2} \\
 &-\Gamma_R\frac{l\rho_{0,R}}{h_0}-\Gamma_R \frac{l \delta \rho_{R}(x,t)}{h_0}+\Gamma_R \frac{l \delta h(x,t) \rho_{0,R}}{h_0^2}. 
\end{align}
\end{subequations}

Where we chose the concentration constants of product and reactant such that they are in local equilibrium, meaning the non-conserved terms of \eqref{eq:product} and \eqref{eq:reactant} cancel each other out:
\begin{subequations}
\label{eq:equilibrium_intial_condition}
    \begin{align}
    0=&\frac{\omega \rho_0\rho_{0,R}}{h_0}-\sigma_P^\uparrow\frac{l}{h_0}\rho_{0,P}\\
    0=&-\frac{\omega\rho_0\rho_{0,R}}{h_0}-\sigma_R^\uparrow\frac{l}{h_0}\rho_{0,R}+ \sigma_R^\downarrow  \rho_{R,Res}.
\end{align}
\end{subequations}
We can solve \eqref{eq:equilibrium_intial_condition} for $\rho_{0,P}$ and $\rho_{0,R}$ obtaining:
\begin{subequations}
    \begin{align}
    \rho_{0,R}=& \frac{ h_0 \rho_{R,Res} \sigma_R^\downarrow}{ \omega \rho_0 + \sigma_R^\uparrow l}\\
        \rho_{0,P}=&\frac{ \omega \rho_0 \rho_{R,Res}}{\sigma_P^\uparrow l }
\end{align}
\end{subequations}
According to \eqref{eq:first_order}  we get up to linear order terms (dropping the dependencies)
\begin{subequations}
    \begin{align}
    \delta p=&\Gamma_P \left(  \frac{l \delta \rho_P}{h_0}- \frac{l \rho_{0,P} \delta h }{h_0^2}\right) f(h_0)  + \Gamma_R \left(  \frac{l \delta \rho_R}{h_0}- \frac{l \rho_{0,R} \delta h }{h_0^2}\right) f(h_0) \nonumber \\
    &- \left(\gamma_0- \cos\theta_0 \gamma_0 - \Gamma_P\frac{l\rho_{0,P}}{h_0} - \Gamma_R\frac{l\rho_{0,R}}{h_0}\right) f'(h_0)\delta h \nonumber\\
    &-\left(\gamma_0-\Gamma_P \frac{l \rho_{0,P}}{h_0}-\Gamma_R \frac{l \rho_{0,R}}{h_0}\right)\nabla^2\delta h
    \end{align}
    \begin{align}
 \partial_t \delta h=&\nabla\cdot \left(\frac{M(h_0)h_0}{\mu}\nabla\delta p+ \Gamma_P\frac{h_0^2+2bh_0}{2\mu}\nabla \left(\frac{l \delta \rho_P}{h_0}- \frac{l\rho_{0,P}\delta h}{h_0^2}\right)+\Gamma_R\frac{h_0^2+2bh_0}{2\mu}\nabla \left(\frac{l \delta \rho_R}{h_0}- \frac{l\rho_{0,R}\delta h}{h_0^2}\right)\right)
    \end{align}
    \begin{align}
 \partial_t \delta \rho=&\nabla\cdot\left(D\nabla\delta\rho-\delta v \rho_0+D\beta\rho_0\partial_h \mathcal{F}(h_0) \nabla \delta h\right)
    \end{align}
    \begin{align}
  \partial_t \delta \rho_P=&\nabla\cdot\left(D_P\nabla\delta\rho_P-\delta v_P \rho_{0,P}+D_P\beta\rho_{0,P}\partial_h \mathcal{F}(h_0) \nabla \delta h\right)\nonumber\\
  &+\omega\left(\frac{\rho_0\delta \rho_R+\rho_{0,R}\delta\rho}{h_0}-\frac{\rho_{0,R}\rho_0}{h_0^2}\delta h\right)  - \sigma_P^\uparrow \frac{l \delta \rho_P}{h_0 }+ \sigma_P^\uparrow \frac{l \rho_{0,P} \delta h }{h_0^2} 
     \end{align}
    \begin{align}
  \partial_t \delta \rho_R=&\nabla\cdot\left(D_R\nabla\delta\rho_R-\delta v_R \rho_{0,R}+D_R\beta\rho_{0,R}\partial_h \mathcal{F}(h_0) \nabla \delta h\right)\nonumber\\
    &-\omega\left(\frac{\rho_0\delta \rho_R+\rho_{0,R}\delta\rho}{h_0}-\frac{\rho_{0,R}\rho_0}{h_0^2}\delta h\right)- \sigma_R^\uparrow \frac{l \delta \rho_R}{h_0 }+ \sigma_R^\uparrow \frac{l\rho_{R,0}\delta h  }{h_0^2}
       \end{align}
    \begin{align}
  \delta v=& -\frac{\Gamma_P }{\mu}M_\gamma(h_0) \nabla \left( \frac{l \delta \rho_P }{h_0}- \frac{l \rho_{0,P}\delta h}{h_0^2}\right) -\frac{\Gamma_R }{\mu}M_\gamma(h_0) \nabla \left( \frac{l \delta \rho_R }{h_0}- \frac{l \rho_{0,R}\delta h}{h_0^2}\right) - M_p(h_0)\frac{\nabla \delta p}{\mu}  
  \end{align}
    \begin{align}
    \delta v_{P,R}=& -\frac{\Gamma_P }{\mu}M_{\gamma,\alpha=0}(h_0) \nabla \left( \frac{l \delta \rho_P }{h_0}- \frac{l \rho_{0,P}\delta h}{h_0^2}\right)  -\frac{\Gamma_R }{\mu}M_{\gamma,\alpha=0}(h_0) \nabla \left( \frac{l \delta \rho_R }{h_0}- \frac{l \rho_{0,R}\delta h}{h_0^2}\right) - M_{p,\alpha=0}(h_0)\frac{\nabla \delta p}{\mu}
\end{align}
\end{subequations}
where $V_h$ and $V_\rho$ have to be determined from the vertical potential $\xi$.

Fourier transforming to $\hat{\delta h }= \hat{\delta h }(q)$,  $\hat{\delta \rho}=\hat{\delta \rho}(q)$,  $\hat{\delta \rho_P}=\hat{\delta \rho_P}(q)$ and  $\hat{\delta \rho_R}=\hat{\delta \rho_R}(q)$ we obtain the evolution equations \eqref{eq:TFE}, \eqref{eq:fick}, \eqref{eq:product} and \eqref{eq:reactant} become the following:
\begin{subequations}
\label{eq:linearized}
    \begin{align}
\begin{split}
     \partial_t \widehat{\delta h} =& -q^2 \frac{M(h_0)h_0}{\mu} \left( \Gamma_P \left(\frac{l\widehat{\delta \rho_P}}{h_0}- \frac{l\rho_{0,P}\widehat{\delta h}}{h_0^2} \right)f(h_0) \right)  -q^2 \frac{M(h_0)h_0}{\mu} \left( \Gamma_R \left(\frac{l\widehat{\delta \rho_R}}{h_0}- \frac{l\rho_{0,R}\widehat{\delta h}}{h_0^2} \right)f(h_0) \right)\\
        &q^2 \frac{M(h_0)h_0}{\mu} \left( \left(\gamma_0 -\cos\theta_0 \gamma_0  -\Gamma_P  \frac{l\rho_{0,P}}{h_0}  -\Gamma_R  \frac{l\rho_{0,R}}{h_0} \right) f'(h_0) \widehat{\delta h} \right)\\
         &-q^4 \frac{M(h_0)h_0}{\mu} \left(\left(\gamma_0 - \Gamma_P \frac{l\rho_{0,P}}{h_0} - \Gamma_R \frac{l\rho_{0,R}}{h_0}\right) \widehat{\delta h} \right)\\
         &- q^2\Gamma_P\frac{h_0^2 + 2bh_0}{2\mu}\left(\frac{l\widehat{\delta \rho_P}}{h_0}- \frac{l\rho_{0,P} \widehat{\delta h}}{h_0^2}\right) - q^2\Gamma_R\frac{h_0^2 + 2bh_0}{2\mu}\left(\frac{l\widehat{\delta \rho_R}}{h_0}- \frac{l\rho_{0,R} \widehat{\delta h}}{h_0^2}\right)
\end{split}
   \end{align}
    \begin{align}
\begin{split}
    \partial_t \widehat{\delta \rho} =& -D q^2 \widehat{\delta \rho}- q^2 D \beta \rho_0 \partial_h \mathcal{F}\widehat{\delta h}- q^2 \rho_0 \frac{\Gamma_P}{\mu}M_\gamma(h_0)\left( \frac{l\widehat{\delta \rho_P}}{h_0}- \frac{l\rho_{0,P}\widehat{\delta h}}{h_0^2}\right)\\
    &- q^2 \rho_0 \frac{\Gamma_R}{\mu}M_\gamma(h_0)\left( \frac{l\widehat{\delta \rho_R}}{h_0}- \frac{l\rho_{0,R}\widehat{\delta h}}{h_0^2}\right)\\
        &-q^2\rho_0\frac{M_p(h_0)}{\mu}\left(\Gamma_P \left(\frac{l\widehat{\delta \rho_P}}{h_0}- \frac{l\rho_{0,P}\widehat{\delta h}}{h_0^2} \right)f(h_0) + \Gamma_R \left(\frac{l\widehat{\delta \rho_R}}{h_0}- \frac{l\rho_{0,R}\widehat{\delta h}}{h_0^2} \right)f(h_0)\right)\\
        &+ q^2\rho_0\frac{M_p(h_0)}{\mu}\left( \left(\gamma_0 -\cos\theta_0 \gamma_0  -\Gamma_P  \frac{l\rho_{0,P}}{h_0} -\Gamma_R  \frac{l\rho_{0,R}}{h_0}  \right) f'(h_0) \widehat{\delta h}     \right)\\
         &-q^4\rho_0\frac{M_p(h_0)}{\mu}\left( \left(\gamma_0 - \Gamma_P \frac{l\rho_{0,P}}{h_0}- \Gamma_R \frac{l\rho_{0,R}}{h_0}\right) \widehat{\delta h}  \right)
\end{split}   
\end{align}
    \begin{align}
\begin{split}
    \partial_t \widehat{\delta \rho_P} =& -D_P q^2 \widehat{\delta \rho_P}- q^2 D_P \beta \rho_{0,P} \partial_h \mathcal{F}_{\alpha=0}\widehat{\delta h}- q^2 \rho_{0,P} \frac{\Gamma_P}{\mu}M_{\gamma,\alpha=0}(h_0)\left( \frac{l\widehat{\delta \rho_P}}{h_0}- \frac{l\rho_{0,P}\widehat{\delta h}}{h_0^2}\right)\\
    &- q^2 \rho_{0,P} \frac{\Gamma_R}{\mu}M_{\gamma,\alpha=0}(h_0)\left( \frac{l\widehat{\delta \rho_R}}{h_0}- \frac{l\rho_{0,R}\widehat{\delta h}}{h_0^2}\right)\\
        &-q^2\rho_{0,P}\frac{M_{p,\alpha=0}(h_0)}{\mu}\left(\Gamma_P \left(\frac{l\widehat{\delta \rho_P}}{h_0}- \frac{l\rho_{0,P}\widehat{\delta h}}{h_0^2}\right)f(h_0) + \Gamma_R \left(\frac{l\widehat{\delta \rho_R}}{h_0}- \frac{l\rho_{0,R}\widehat{\delta h}}{h_0^2} \right)f(h_0)\right)\\
        &+q^2\rho_{0,P}\frac{M_{p,\alpha=0}(h_0)}{\mu}\left(\left(\gamma_0 -\cos\theta_0 \gamma_0  -\Gamma_P \frac{l\rho_{0,P}}{h_0}  -\Gamma_R  \frac{l\rho_{0,R}}{h_0} \right) f'(h_0) \widehat{\delta h} \right)\\
        &-q^4\rho_{0,P}\frac{M_{p,\alpha=0}(h_0)}{\mu}\left( \left(\gamma_0 - \Gamma_P \frac{l\rho_{0,P}}{h_0}- \Gamma_R \frac{l\rho_{0,R}}{h_0}\right) \widehat{\delta h}  \right)\\
        &+\frac{\omega \widehat{\delta \rho}\rho_{0,R}}{h_0}+\frac{\omega\rho_0 \widehat{\delta\rho_R}}{h_0} - \frac{\omega \rho_{0,R} \rho_0}{h_0^2}\widehat{\delta h}- \sigma_P^\uparrow \frac{l \widehat{\delta \rho_P}}{h_0 }+ \sigma_P^\uparrow \frac{l \rho_{0,P} \widehat{\delta h }}{h_0^2}
\end{split}
\end{align}
\begin{align}
\begin{split}
    \partial_t \widehat{\delta \rho_R} =& -D_R q^2 \widehat{\delta \rho_R}- q^2 D_R \beta \rho_{0,P} \partial_h \mathcal{F}_{\alpha=0}\widehat{\delta h}- q^2 \rho_{0,R} \frac{\Gamma_P}{\mu}M_{\gamma,\alpha=0}(h_0)\left( \frac{l\widehat{\delta \rho_P}}{h_0}- \frac{l\rho_{0,P}\widehat{\delta h}}{h_0^2}\right)\\
    &- q^2 \rho_{0,R} \frac{\Gamma_R}{\mu}M_{\gamma,\alpha=0}(h_0)\left( \frac{l\widehat{\delta \rho_R}}{h_0}- \frac{l\rho_{0,R}\widehat{\delta h}}{h_0^2}\right)\\
        &-q^2\rho_{0,R}\frac{M_{p,\alpha=0}(h_0)}{\mu}\left(\Gamma_P \left(\frac{l\widehat{\delta \rho_P}}{h_0}- \frac{l\rho_{0,P}\widehat{\delta h}}{h_0^2}\right)f(h_0) + \Gamma_R \left(\frac{l\widehat{\delta \rho_R}}{h_0}- \frac{l\rho_{0,R}\widehat{\delta h}}{h_0^2} \right)f(h_0)\right)\\
        &+q^2\rho_{0,R}\frac{M_{p,\alpha=0}(h_0)}{\mu}\left( \left(\gamma_0 -\cos\theta_0 \gamma_0  -\Gamma_P  \frac{l\rho_{0,P}}{h_0} -\Gamma_R  \frac{l\rho_{0,R}}{h_0} \right) f'(h_0) \widehat{\delta h} \right)\\
         &-q^4\rho_{0,R}\frac{M_{p,\alpha=0}(h_0)}{\mu}\left(\left(\gamma_0 - \Gamma_P \frac{l\rho_{0,P}}{h_0} - \Gamma_R \frac{l\rho_{0,R}}{h_0}\right) \widehat{\delta h}  \right)\\
        &-\frac{\omega \widehat{\delta \rho}\rho_{0,R}}{h_0}-\frac{\omega\rho_0 \widehat{\delta\rho_R}}{h_0} + \frac{\omega \rho_{0,R} \rho_0}{h_0^2}\widehat{\delta h}  - \sigma_R^\uparrow \frac{l\widehat{\delta \rho_R}}{h_0 }+ \sigma_R^\uparrow \frac{l\rho_{0,R}\widehat{\delta h}}{h_0^2} 
\end{split}
\end{align}
\end{subequations}

Lets collect the term
\begin{align}
   \Sigma(q) := q^2 (\gamma_0 -\cos \theta_0 \gamma_0 - \Gamma_P \frac{l\rho_{0,P}}{h_0}  - \Gamma_R \frac{l\rho_{0,R}}{h_0}) f'(h_0) - q^4 (\gamma_0 - \Gamma_P \frac{l\rho_{0,P}}{h_0}  - \Gamma_R \frac{l\rho_{0,R}}{h_0}) .
\end{align}
We can rewrite the linearized equations in Fourier space \eqref{eq:linearized} as  
\begin{align}
    \partial_t \begin{pmatrix}
    \widehat{\delta h}\\
    \widehat{\delta \rho}\\
    \widehat{\delta \rho_P}\\
    \widehat{\delta \rho_R}
    \end{pmatrix}= \begin{pmatrix}
    a_1^1 & a_1^2 & a_1^3 & a_1^4 \\
    a_2^1 & a_2^2 & a_2^3 & a_2^4\\
    a_3^1 & a_3^2 & a_3^3 & a_3^4\\
    a_4^1 & a_4^2 & a_4^3 & a_4^4
    \end{pmatrix}  \begin{pmatrix}
    \widehat{\delta h}\\
    \widehat{\delta \rho}\\
    \widehat{\delta \rho_P}\\
    \widehat{\delta\rho_R}
    \end{pmatrix}
\end{align}
where
\begin{subequations}
    \begin{align}
    a_1^1 =& q^2 \frac{M(h_0)h_0}{\mu} \left(\Gamma_P \frac{l\rho_{0,P}}{h_0^2}+ \Gamma_R \frac{l\rho_{0,R}}{h_0^2}\right)f(h_0)+\frac{M(h_0)h_0}{\mu}\Sigma(q)\\
    &+q^2 \frac{h_0^2+2bh_0}{2\mu}\left(\Gamma_P\frac{l\rho_{0,P}}{h_0^2}+\Gamma_R\frac{l\rho_{0,R}}{h_0^2}\right)\\
    a_1^2 =& 0\\
    a_1^3 =& -q^2 \frac{M(h_0)h_0 }{\mu }\Gamma_P \frac{l}{h_0}f(h_0)-q^2 \Gamma_P \frac{h_0^2+2bh_0}{2\mu} \frac{l}{h_0}\\
    a_1^4 =& -q^2 \frac{M(h_0)h_0 }{\mu }\Gamma_R \frac{l}{h_0}f(h_0)-q^2 \Gamma_R \frac{h_0^2+2bh_0}{2\mu} \frac{l}{h_0}
\end{align}
\begin{align}
    a_2^1=&- q^2 D\beta \rho_0 \partial_h \mathcal{F}+ q^2 \frac{M_\gamma \rho_0}{\mu}\left(\Gamma_P \frac{l \rho_{0,P}}{h_0^2}+\Gamma_R \frac{l \rho_{0,R}}{h_0^2}\right)+\frac{M_p\rho_0}{\mu}\Sigma(q)\\
    &+ q^2 \frac{M_p\rho_0 }{\mu}\left(\Gamma_P\frac{l\rho_{0,P}}{h_0^2}+ \Gamma_R \frac{l \rho_{0,R}}{h_0^2}\right)f(h_0)\\
    a_2^2=& -Dq^2\\
    a_2^3 =& -q^2 \frac{M_\gamma\rho_0}{\mu}\Gamma_P\frac{l}{h_0}- q^2\frac{M_p\rho_0}{\mu }\Gamma_P\frac{l}{h_0}f(h_0)\\
    a_2^4 =& -q^2 \frac{M_\gamma\rho_0}{\mu}\Gamma_R\frac{l}{h_0}- q^2\frac{M_p\rho_0}{\mu }\Gamma_R\frac{l}{h_0}f(h_0)
\end{align}
\begin{align}
    a_3^1=& - q^2 D_P\beta \rho_{0,A }\partial_h \mathcal{F}_{0}+ q^2\frac{M_{\gamma,0}\rho_{0,P}}{\mu}\left(\Gamma_P \frac{l\rho_{0,P}}{h_0^2}+\Gamma_R \frac{l\rho_{0,R}}{h_0^2}\right)-\frac{\omega\rho_{0,R}\rho_0}{h_0^2}\\
    &+q^2 \frac{M_{p,0}\rho_{0,P}}{\mu}\left(\Gamma_P \frac{l\rho_{0,P}}{h_0^2}+\Gamma_R \frac{l\rho_{0,R}}{h_0^2}\right)f(h_0)+\frac{M_{p,0}\rho_{0,P}}{\mu} \Sigma(q)+ \sigma_P^\uparrow \frac{l \rho_{0,P} }{h_0^2} \\
    a_3^2 =& \frac{\omega\rho_{0,R}}{h_0}\\
    a_3^3 =& -q^2 D_P -q^2 \frac{M_{\gamma,0}\rho_{0,P}}{\mu}\Gamma_P \frac{l}{h_0}- q^2\frac{M_{p,0}\rho_{0,P}}{\mu}\Gamma_P\frac{l}{h_0}f(h_0) -\sigma_P^\uparrow\frac{l}{h_0}\\
    a_3^4=& -q^2 \frac{M_{\gamma,0}\rho_{0,P}}{\mu}\Gamma_R \frac{l}{h_0}- q^2\frac{M_{p,0}\rho_{0,P}}{\mu}\Gamma_R\frac{l}{h_0}f(h_0) +\frac{\omega\rho_0}{h_0}
\end{align}
\begin{align}
    a_4^1=& - q^2 D_R\beta \rho_{0,B }\partial_h \mathcal{F}_{0}+ q^2\frac{M_{\gamma,0}\rho_{0,R}}{\mu}\left(\Gamma_P \frac{l\rho_{0,P}}{h_0^2}+\Gamma_R \frac{l\rho_{0,R}}{h_0^2}\right) + \frac{\omega\rho_{0,R}\rho_0}{h_0^2}\\
    &+q^2 \frac{M_{p,0}\rho_{0,R}}{\mu}\left(\Gamma_P \frac{l\rho_{0,P}}{h_0^2}+\Gamma_R \frac{l\rho_{0,R}}{h_0^2}\right)f(h_0)+\frac{M_{p,0}\rho_{0,R}}{\mu} \Sigma(q)+\sigma_R^\uparrow \frac{l \rho_{0,R}}{h_0^2}\\
    a_4^2 =& -\frac{\omega\rho_{0,R}}{h_0}\\
    a_4^3 =&-q^2 \frac{M_{\gamma,0}\rho_{0,R}}{\mu}\Gamma_P \frac{l}{h_0}- q^2\frac{M_{p,0}\rho_{0,R}}{\mu}\Gamma_P\frac{l}{h_0}f(h_0)\\
    a_4^4=& -q^2 D_R -q^2 \frac{M_{\gamma,0}\rho_{0,R}}{\mu}\Gamma_R \frac{l}{h_0}- q^2\frac{M_{p,0}\rho_{0,R}}{\mu}\Gamma_R\frac{l}{h_0}f(h_0) -\frac{\omega\rho_0}{h_0}-\sigma_R^\uparrow \frac{l}{h_0}.
\end{align}
\end{subequations}
This is the desired matrix $A$.

\subsection{Eigenvalues}
\begin{figure}[hh]
    \centering
    \includegraphics[width=\columnwidth]{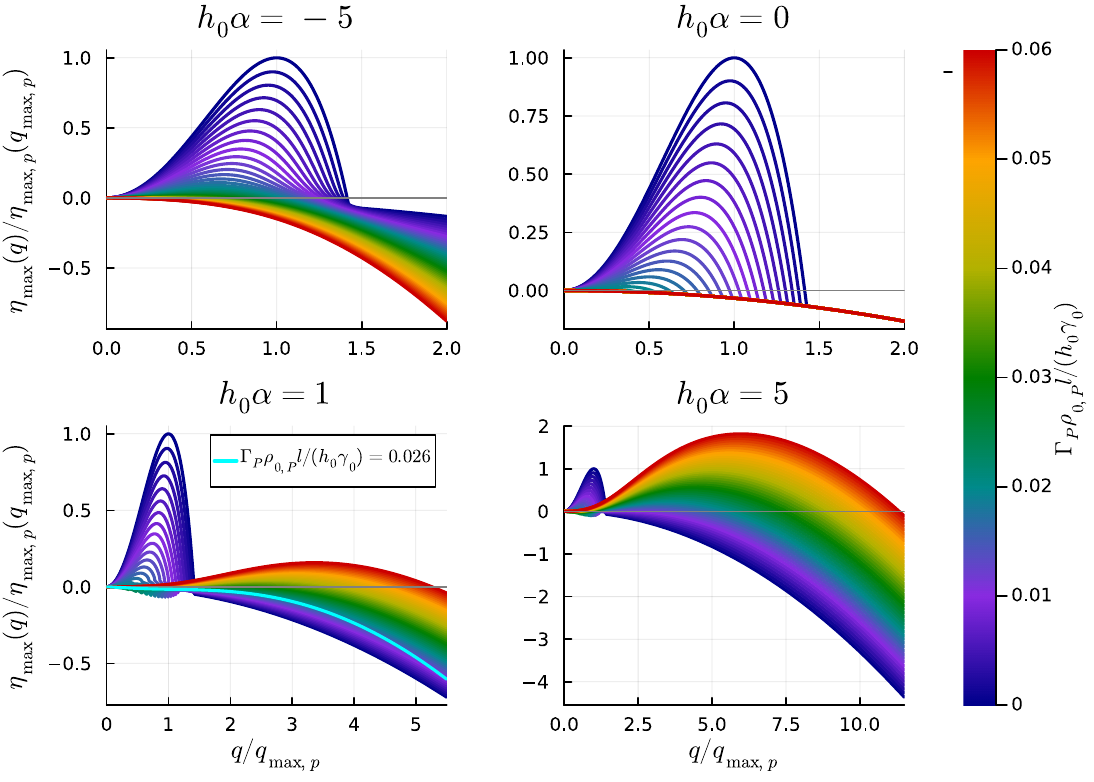}
    \caption{The largest eigenvalue $\eta_{m}(q)$ of $A(q)$ for different values of $\Gamma$ and $\alpha$. Pay attention that the y-axis (arbitrary but consistent units of dimensions $[1/T]$) is not fixed. In the third subplot an eigenvalue $\eta_{m}<0$ for an intermediate value of $\Gamma$ is highlighted in cyan, while there are both smaller and larger values of $\Gamma$ where $\eta_{m}>0$. We always choose $\Gamma_R\rho_{0,R}l/(h_0\gamma_0)=\frac{1}{2}\Gamma_P\rho_{0,P}l/(h_0\gamma_0)$. }
    \label{fig:eigenvalues}
\end{figure}

Here we want to discuss a few features of the eigenvalues of the matrix $A$. This can give further insight into the stability of the homogeneous state of the system and helps in understanding the features of the stability portrait in Figure 2.
Some exemplary eigenvalues for selected values of $\alpha$ and a range of values of $\Gamma_P$ are shown in Figure \ref{fig:eigenvalues}

The system is as we have seen well characterized by the largest eigenvalue $\eta_{m}$. When $\exists q : \eta_{m}(q)>0$ the homogeneous state  is unstable while when $\forall q: \eta_{m}(q)\leq0$ the homogeneous state is a stable steady state. Further, we can obtain the dominant wavelength of the system by $\lambda_{\max}=1/q_{\max}$ for the fastest growing mode $q_{\max}:=\text{argmax}\eta_{m}$, and the time-scale of the system by via $\tau\propto 1/\max \eta_{m}$. 

Figure $\ref{fig:eigenvalues}$ shows examples of the complex behavior of $\eta_{m}$, that stems from the four coupled differential equations \eqref{eq:TFE}, \eqref{eq:fick}, \eqref{eq:product} and \eqref{eq:reactant}.  The value of $\Gamma_R\rho_{0,R}l/(h_0\gamma_0)$ is not given in the plot but always chosen $\Gamma_R\rho_{0,R}l/(h_0\gamma_0)=\frac{1}{2}\Gamma_P\rho_{0,P}l/(h_0\gamma_0).$ 

For $\alpha\leq 0$ as seen in the top two subplots $\eta_{m}$ decreases with increasing $\Gamma_P$. Both $q_{\max}$ and $\max \eta_{m}$ continuously approach $0$. 

For $h_0\alpha=1$ (the bottom left subplot), we see  a transition  from $q_{\max}>0$ to $q_{\max}=0$ upon raising $\Gamma_P$. For some in-between values e.g. $\Gamma_P\rho_{0,P} l /(h_0\gamma_0)=0.026$ we have $q_{\max}=0$. When further raising $\Gamma_P$ an instability arises at high wave numbers leading to a jump discontinuity in $q_{\max}$ from $q_{\max}=0$ to $q_{\max}>0$. The absolute values $\max \eta_{m}$ are small as compared to $\Gamma=0$ and thus the timescales here are longer than in the passive case. That can also be seen in Figure 5. 

For $\alpha \gg 0$ (the bottom right subplot) the two instabilities bypass each other without ever allowing for $q_{\max}=0$ leading to another jump discontinuity in $q_{\max}$ between two positive finite values.

\subsection{Varying parameters}

     \begin{figure}[hh]
         \centering
         \includegraphics[width=\columnwidth]{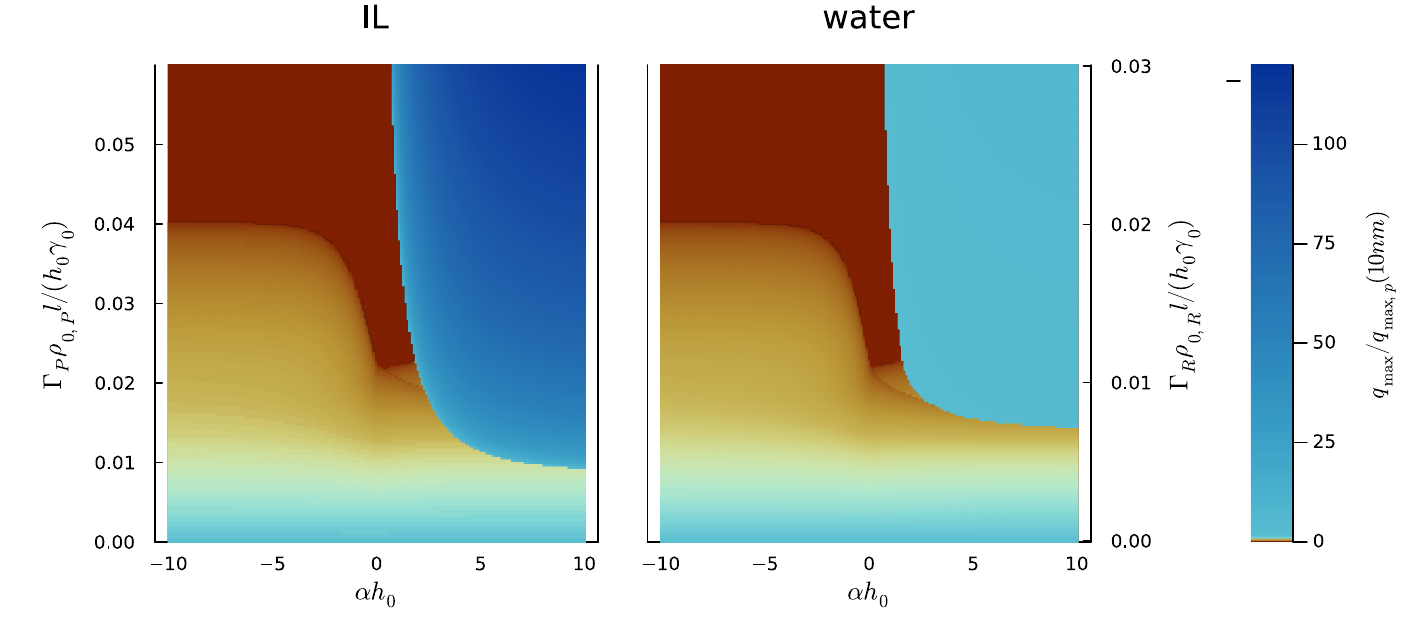}
         \caption{The fastest growing mode for  water
         $[(mPEG2)2IM]I$ at a temperature of $T=303.15 K$. In the right subplot the same values as reported in Figure \ref{fig:fourier} are used. For the 
         left subplot the parameters are chosen $R_c=1[\text{nm}]$, $R_A=1[\text{\AA}]$, $h_0=10[\text{nm}]$, $\nu=4.058\cdot 10^{-4}[\text{m}^2/\text{s}]$, $\rho_l=1.4098 [\text{kg}/\text{l}^3]$ , $\gamma_0=0.04596 [\text{N}/\text{m}]$, $\theta_0=\pi/9$, $\omega=1.6\cdot10^{7} [1/\text{s}] $, $\sigma_2=0.1\omega$, $h^*=1.44 h_0$. Viscosity and surface tension have been measured by \cite{zhai_influence_2023} for $[(mPEG2)2IM]I$ at a temperature of $T=303.15 K$, the density has been taken for the same ionic liquid and temperature form \cite{seidl_bis-polyethylene_2022}. The rates have been chosen to match the reaction rate per surface area reported by \cite{brown_ionic_2014} for the reaction of $2H_2O_2 \rightleftharpoons 2H_2O + O_2$ in aqueous solution catalyzed by Platinum particles. 
         }
         \label{fig:stability_portrait_il}
     \end{figure}


\begin{figure}[hh]
    \centering
    \includegraphics[width=\columnwidth]{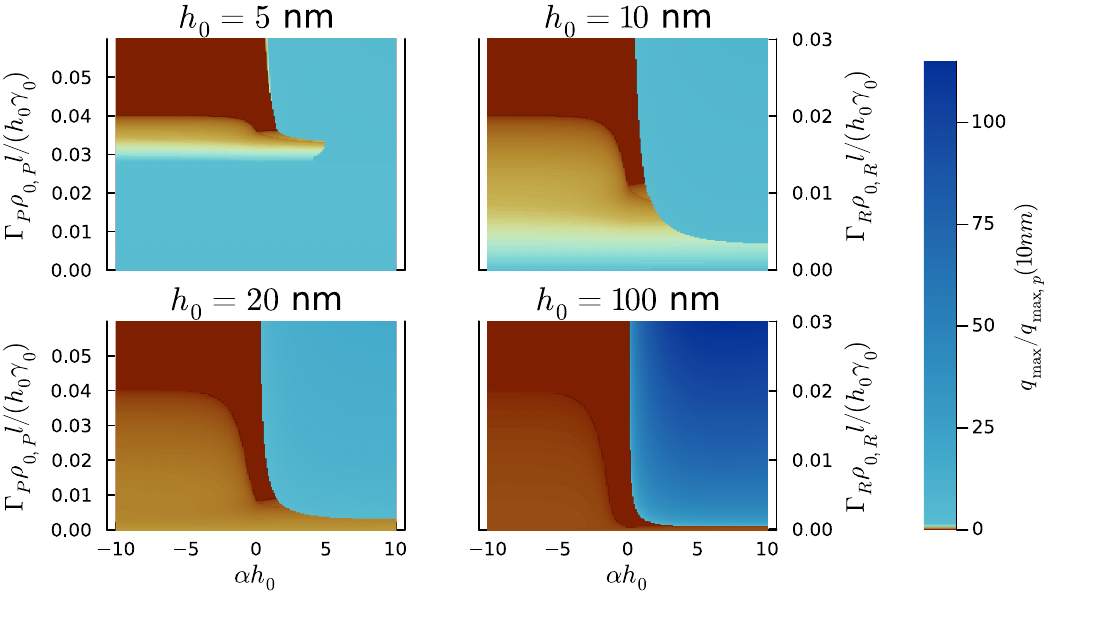}
    \caption{Stability diagrams upon varying the initial film height $h_0$. We normalize all modes by the fastest growing mode of a $10 \text{nm}$ film to make them better comparable. Notice that the color bar is not fixed. Apart from $h_0$ all values are chosen as reported in the Material and Methods section.}
    \label{fig:different_heights}
\end{figure}

So far we discussed only the stability portrait for parameters, that corresponds to a $10 [\text{nm}]$ film of water with a contact angle of $\theta=\frac{\pi}{9}$. Plugging in parameters for different liquids as done in Figure \ref{fig:stability_portrait_il} gives diagrams of similar shapes but different absolute values.

The only difference upon changing $\theta$ is that we move the threshold value for the onset of stability $\Gamma_c$ given by  $\Gamma_{c}(\rho_{0,P}+\rho_{0,R})l/(h_0\gamma_0) =1-\cos\theta_0$.


Upon changing the height of the initial film height $h_0$ the most striking difference is that the fastest growing mode of the passive case $q_{\max,p}$ changes dramatically. It varies as $q_{\max,p}\sim h_0^{-(n-1)/2}$, thus we can extremely stabilize a liquid film by raising its film height. Spinoidal rupture becomes irrelevant at film heights of at the very most $100 [\text{nm}]$. Upon increasing film height the differences between the mobility's $M_\gamma(h_0), M_p(h_0) , M(h_0)$ and $(h_0^2 + 2bh_0)/2$ become larger increasing the effect that leads to collapse. In Figure \ref{fig:different_heights} we report the stability portrait for different initial heights of the liquid film fixing all other parameters. We normalize all by the wave number of the fastest growing mode of a passive $10[\text{nm}]$ film to better compare the values. We see that upon raising the film height the fastest growing modes of the spinoidal rupture are weakened. The collapse on the other hand is enforced. At $h_0=100[\text{nm}]$ the film becomes stable for $\Gamma>0,\alpha<0$. For $\Gamma,\alpha\gtrless 0$ we have on the other hand excited wavelengths from the collapse.

\subsection{Influence of dissolved gases on surface tension of ionic liquids}

\begin{figure}[hh]
    \centering
    \includegraphics{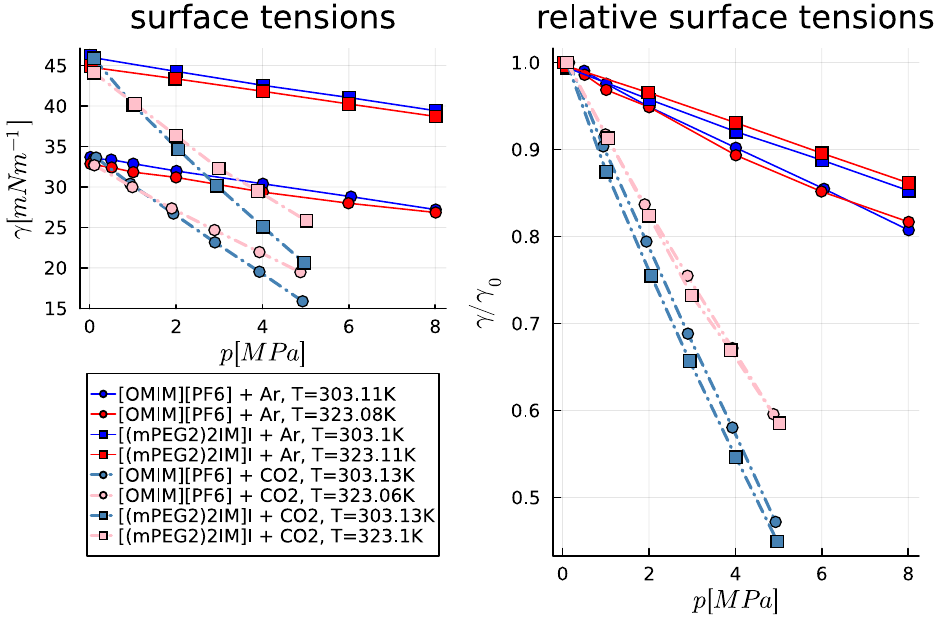}
    \caption{Experimentally determined surface tension of binary mixtures of an IL and dissolved gases as a function of pressure \cite{zhai_influence_2023}.}
    \label{fig:experimental_data}
\end{figure}

As measured by \cite{zhai_influence_2023} the liquid-gas surface tension of various ionic liquids is affected by the concentration of dissolved chemicals such as gases. Upon applying a pressure of different gases the experimentally determined surface tension is found to decrease quite linearly with increasing pressure, which is more pronounced for the more soluble \ce{CO2} than for the less soluble \ce{Ar}.  The data used to create Figure \ref{fig:experimental_data} is taken from \cite{zhai_influence_2023}.

\end{document}